\documentclass[10pt,journal,compsoc]{IEEEtran}
\usepackage{graphicx}
\usepackage{balance}
\usepackage{amssymb}
\usepackage{amsmath}
\usepackage{graphicx}
\usepackage{subfigure}
\usepackage{float}
\usepackage{color}
\usepackage{fmtcount}
\usepackage{multirow}
\usepackage{booktabs}
\usepackage{makecell}
\usepackage{array}

\makeatletter
\newif\if@restonecol
\makeatother

\usepackage[linesnumbered,ruled,vlined]{algorithm2e}
\usepackage{algpseudocode}
\usepackage{amsmath}

\newtheorem{definition}{Definition}

\newcommand{\eat}[1]{}

\graphicspath{{figures/}}

\begin{document}


\title{Revisiting $k$-Nearest Neighbor Graph Construction on High-Dimensional Data : Experiments and Analyses}



\author{Yingfan Liu,
        Hong Cheng
        and~Jiangtao Cui
\IEEEcompsocitemizethanks{
\IEEEcompsocthanksitem Yingfan Liu and Jiangtao Cui are with the School of Computer Science and Technology, Xidian University, Xi'an, China. E-mail: \{liuyingfan, cuijt\}@xidian.edu.cn.
\IEEEcompsocthanksitem  Hong Cheng is with the Chinese University of Hong Kong, Hong Kong SAR, China. E-mail: hcheng@se.cuhk.edu.hk.
}
}

\IEEEtitleabstractindextext{%

\begin{abstract}

The $k$-nearest neighbor graph (KNNG) on high-dimensional data is a data structure widely used in many applications such as similarity search, dimension reduction and clustering.  Due to its increasing popularity, several methods under the same framework have been proposed in the past decade. This framework contains two steps, i.e. building an initial KNNG (denoted as \texttt{INIT}) and then refining it by neighborhood propagation (denoted as \texttt{NBPG}). However, there remain several questions to be answered. First, it lacks a comprehensive experimental comparison among representative solutions in the literature. Second, some recently proposed indexing structures, e.g., SW and HNSW, have not been used or tested for building an initial KNNG. Third, the relationship between the data property and the effectiveness of \texttt{NBPG} is still not clear. To address these issues, we comprehensively compare the representative approaches on real-world high-dimensional data sets to provide practical and insightful suggestions for users. As the first attempt, we take SW and HNSW as the alternatives of \texttt{INIT} in our experiments. Moreover, we investigate the effectiveness of \texttt{NBPG} and find the strong correlation between the huness phenomenon and the performance of \texttt{NBPG}. 



\end{abstract}

}

\maketitle

\IEEEraisesectionheading{\section{Introduction}}
\label{sec:intro}

The $k$-nearest neighbor graph (KNNG) on high-dimensional data is a useful data structure for many applications in various domains such as computer vision, data mining and machine learning.  Given a $d$-dimensional data set $D \subset \mathbb{R}^{d}$ and a positive integer $k$, a KNNG on $D$ treats each point $u \in D$ as a graph node and creates directed edges from $u$ to its $k$ nearest neighbors (KNNs) in $D \setminus \{u\}$.  In this paper, we use \textit{point, vector and node} exchangeably.

KNNG is widely used as a building brick for several fundamental operations, such as approximate $k$-nearest neighbor (ANN) search, mainfold learning and clustering.  The state-of-the-art ANN search methods (e.g., DPG~\cite{Li2019TKDE} and NSG~\cite{Fu2019VLDB}) first construct an accurate KNNG on $D$ and then adjust the neighborhood for each node on top of the KNNG. In order to generate a low-dimensional and compact description of $D$, many manifold learning methods~(e.g., LLE~\cite{LLE}, Isomap~\cite{Tenenbaum2000Science}, t-SNE~\cite{Maaten2014JMLR} and LargeVis \cite{TangJ2016WWW}) requires its KNNG, which captures the local geometric properties of $D$.  Besides, some clustering methods~\cite{Zhang2012ECCV, Wang2015MDMA, Deng2018ICDE} improve their performance following the idea that each point and its KNNs should be assigned into the same cluster.

\begin{figure}[t]
\begin{center}
\subfigure[\textbf{Sift}]
{
\includegraphics[width=0.22\textwidth]{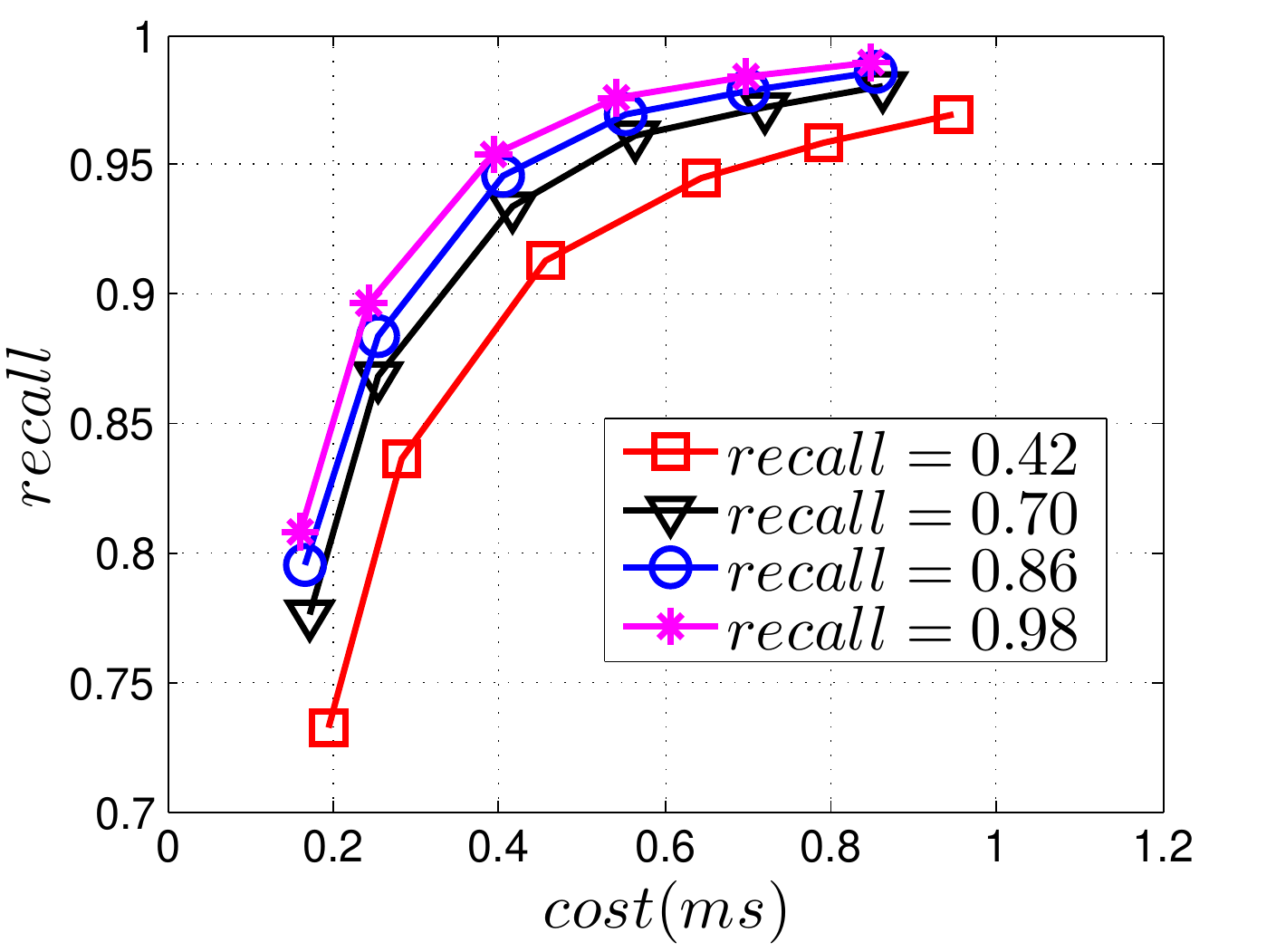}
\label{fig:motivation_dpg_sift}
}
\subfigure[\textbf{Gist}]
{
\includegraphics[width=0.22\textwidth]{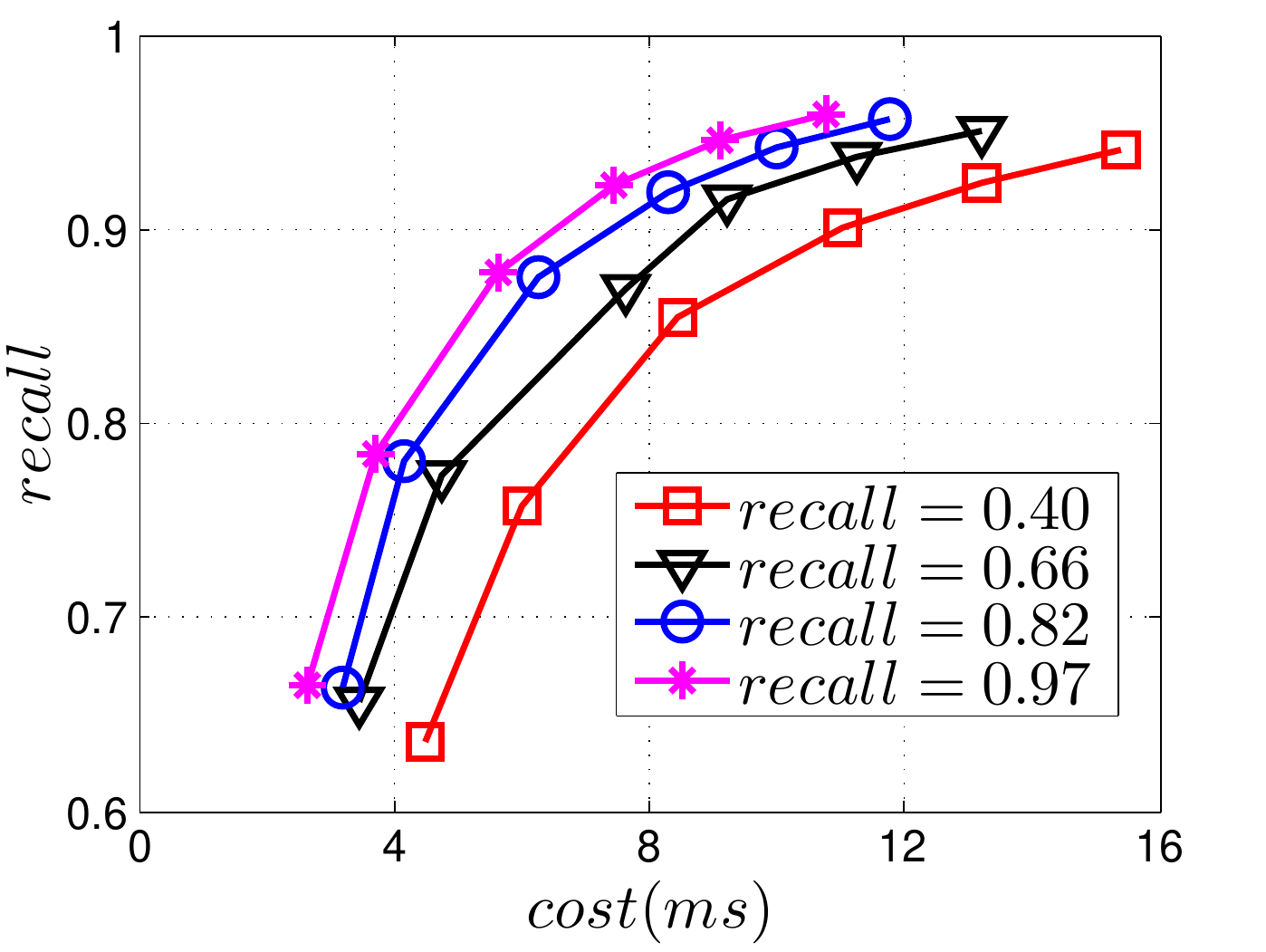}
\label{fig:motivation_dpg_gist}
}
\end{center}\vspace{-3ex}
\caption{The effect of KNNG's accuracy on DPG's performance. Each curve corresponds to a KNNG built for DPG.}\vspace{-2ex}
\label{fig:motivation_dpg}
\end{figure}

Since computing the exact KNNG requires huge cost in practice, researchers focused on the approximate KNNG to balance efficiency and accuracy. In practice, an accurate KNNG is urgent, because its accuracy will affect the downstream applications. Let us take DPG~\cite{Li2019TKDE}, an ANN search method, as an example. We conduct ANN search with DPG on two widely-used data sets, i.e., Sift and Gist, whose details could be found in Section~\ref{ssec:exp_init}, and show the results in Figure~\ref{fig:motivation_dpg}.  As KNNG's accuracy increases, DPG's performance is obviously enhanced, where less cost leads to more accurate results.


During the past decade, a few approximate methods~\cite{ChenJ2009JMLR, WangJD2012CVPR, ZhangYM2013ECML, TangJ2016WWW} have been proposed.  In general, they follow the same framework that quickly constructs an initial KNNG with low or moderate accuracy and then refines it by neighborhood propagation.  Existing methods usually employ some index structures to construt the initial KNNG, including binary partition tree~\cite{ChenJ2009JMLR, WangJD2012CVPR}, random projection tree~\cite{TangJ2016WWW} and locality sensitive hashing~\cite{ZhangYM2013ECML}.  For neighborhood propagation, existing methods follow the same principle ``a neighbor of my neighbor is also likely to be my neighbor''~\cite{Dong2011WWW}, despite some variations in technical details.  In the rest of this paper, we denote the first step of constructing an \textbf{init}ial KNNG as \texttt{INIT} and the second step of \textbf{n}eigh\textbf{b}orhood \textbf{p}ropa\textbf{g}ation as \texttt{NBPG}.

Given the current research outcomes on KNNG construction in the literature, there remain several questions to be answered.  First, it lacks a comprehensive experimental comparison among representative solutions in the literature.  This may be caused by the fact that these authors come from different research communities (e.g., data mining, computer vision, machine learning, etc.) and are not familiar with some other works.  Second, some related techniques in the literature have not been considered for the purpose of KNNG construction, such as SW graph~\cite{Malkov2014IS} and HNSW ~\cite{Malkov2018PAMI}, which are the state-of-the-art methods for ANN search.  It remains a question whether they are potentially suitable methods for KNNG construction.  Third, \texttt{NBPG} methods are effective on some data but not so effective on some other data.  The relationship between the effectiveness of \texttt{NBPG} and data property is not clear yet.

To address the above issues, we revisit existing methods and make a comprehensive experimental study for both \texttt{INIT} and \texttt{NBPG} steps, to understand the strengths and limitations of different methods and provide deeper insights for users.  In addition to existing methods for the \texttt{INIT} step, we adopt two related techniques for ANN search, i.e., SW and HNSW, as alternative methods for \texttt{INIT}, denoted as \texttt{SW KNNG} and \texttt{HNSW KNNG} respectively.  Through our extensive experiments, we first recommend \texttt{HNSW KNNG} for \texttt{INIT}.

We categorize \texttt{NBPG} methods into three categories, i.e., \texttt{UniProp}~\cite{ChenJ2009JMLR, ZhangYM2013ECML, TangJ2016WWW}, \texttt{BiProp}~\cite{Dong2011WWW} and \texttt{DeepSearch}~\cite{WangJD2012CVPR}.  Both \texttt{UniProp} and \texttt{BiProp} adopt the iterative approach to refine the initial KNNG step by step.  But \texttt{UniProp} only considers nearest neighbors in \texttt{NBPG}, while \texttt{BiProp} considers both nearest neighbors and reverse nearest neighbors.  In contrast, \texttt{DeepSearch} runs for only one iteration, which conducts ANN search on an online rapidly-built proximity graph, e.g., the initial KNNG in~\cite{WangJD2012CVPR} or the HNSW graph as pointed out in this paper.  According to our experimental results, there is no dominator in all cases. But, we do not recommend the popular \texttt{UniProp}, because it usually cannot achieve a high-accuracy KNNG and is sensitive to $k$. \texttt{KGraph}~\footnote{As an improved implementation of \cite{Dong2011WWW}, it does not appear in the experiments of several existing works~\cite{ChenJ2009JMLR, WangJD2012CVPR, ZhangYM2013ECML, TangJ2016WWW}}, a variant of \texttt{BiProp}, has the best performance to construct a high-accuracy KNNG, but requires much more memory than other methods.  \texttt{Deep HNSW}, a new combination under the framework, presents the best balance among efficiency, accuracy and memory requirement in most cases. 


Besides the experimental study, we further investigate the effectiveness of \texttt{NBPG} versus data property.  We employ the definition of \textbf{node hubness} in~\cite{Radovanovic2010} to characterize each node, where the node hubness of a node is defined as the number of its exact reverse KNNs. Further, we extend this definition and define the \textbf{data hubness} of a data set as the normalized sum of the largest node hubness of its top nodes up to a specified percentage, in order to characterize the data set. With those two definitions in our analyses, we have two interesting findings. First, a high hub data usually has worse \texttt{NBPG} performance than a low hub data.  Second, a high hub node usually has higher accuracy of their KNNs than a low hub ndoe. 

We summarize our contributions as follows.
\vspace{-0.5ex}
\begin{itemize}
	\item We revisit the framework of constructing KNNG and introduce new methods or combinations following it.
	\item We conduct a comprehensive experimental comparison among representative solutions and provide our practical suggestions.  
	\item We investigate the effectiveness of \texttt{NBPG} and discover the strong correlation between its performance and the hubness phenomenon.
\end{itemize}
\vspace{-0.5ex}

The rest of this paper is organized as follows. In Section~\ref{sec:preli}, we define the problem and show the framework. In Section~\ref{sec:initg}, we present \texttt{INIT} methods and conduct experimental evaluations on \texttt{INIT}. In Section~\ref{sec:nbpg}, we describe various \texttt{NBPG} methods and show our experimental results.  In Section~\ref{sec:explo}, we investigate the effectiveness of \texttt{NBPG} mechanism and present our insights.  In Section~\ref{sec:rework}, we discuss related work. Finally, we conclude this paper in Section~\ref{sec:clu}.



\section{Overview of KNNG Construction}
\label{sec:preli}

\subsection{Problem Definition}
\label{ssec:def}

Let $D \subset \mathbb{R}^{d}$ be a data set that consists of $n$ $d$-dimensional real vectors.  The KNNG on $D$ is defined as follows.

\begin{definition}
\textbf{[KNNG].} Given $D \subset \mathbb{R}^d$ and a positive integer $k$, let $G = (V, E)$ be the KNNG of $D$, where $V$ is the node set and $E$ is the edge set.  Each node $u \in V$ uniquely represents a vector in $D$.  A directed edge $(u, v)\in E$ exists iff $v$ is one of $u$'s KNNs in $D \setminus \{u\}$.
\end{definition}

\begin{figure}[t]
\begin{center}
\includegraphics[width=0.2\textwidth]{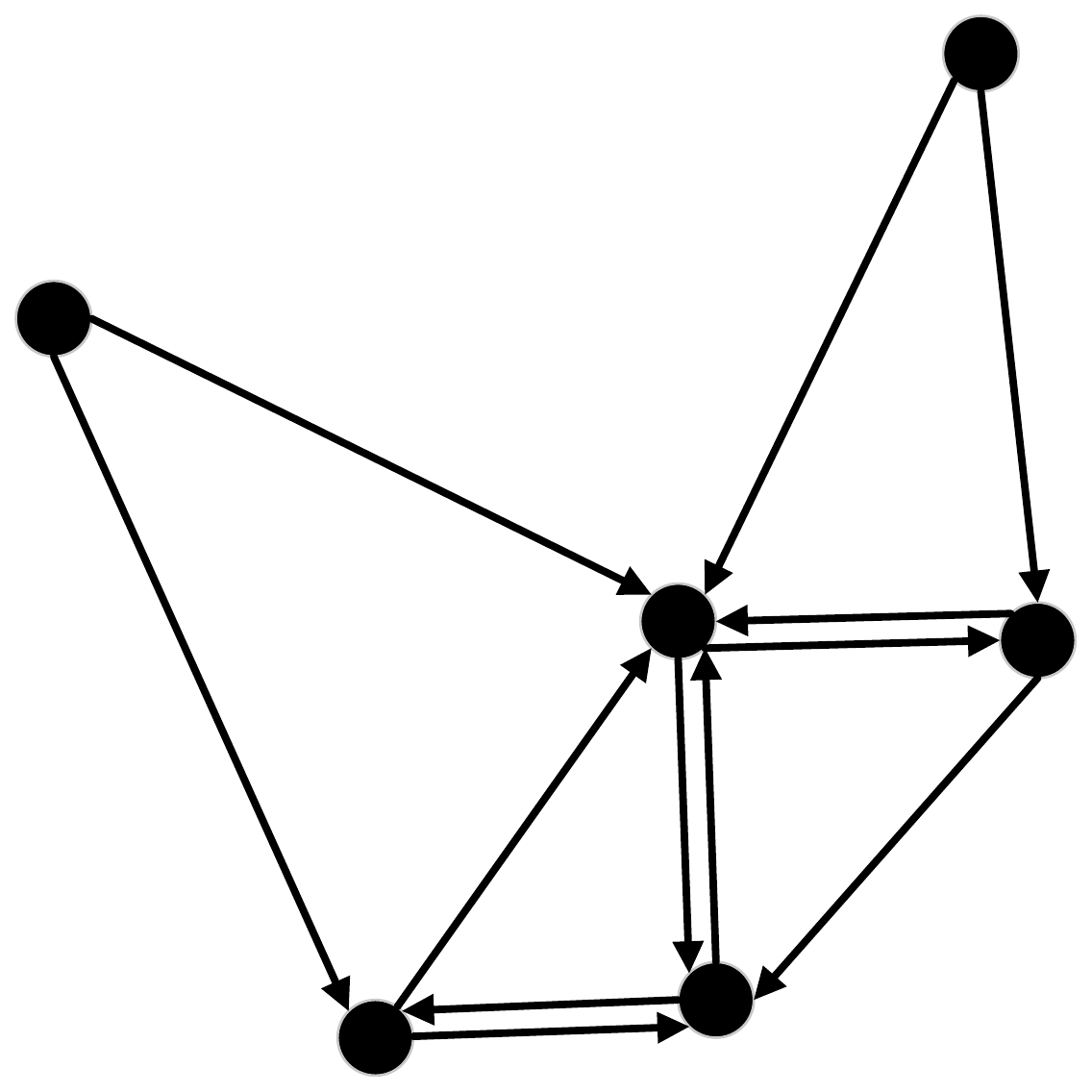}
\end{center}\vspace{-3ex}
\caption{An illustrative KNNG with $n = 6$ and $k = 2$.}\vspace{-2ex}
\label{fig:knng} 
\end{figure}

The KNNG of a data set with 6 data points is shown in Figure~\ref{fig:knng}.  For each node, it has two out-going edges pointing to its 2NNs.  \emph{For simplicity, we refer to \textbf{approximate KNNG} as KNNG and \textbf{approximate KNNs} as KNNs in the rest of the paper.  When we refer to \textbf{exact KNNG} and \textbf{exact KNNs}, we will mention them explicitly.}

In this paper, we focus on \textbf{in-memory solutions on high-dimensional dense vectors with the Euclidean distance as the distance measure}, which is the most popular setting in the literature. This setting makes us focus on the comprehensive comparison on existing works. We would solve this problem with other settings, e.g., disk-based solutions, sparse data and other distance measures, in future works. 

\subsection{The Framework of KNNG Construction}
\label{ssec:framework}


Algorithm~\ref{alg:framework} shows the framework of KNNG construction, adopted by many methods in the literature~\cite{ChenJ2009JMLR, WangJD2012CVPR, ZhangYM2013ECML, TangJ2016WWW}.  This framework contains two steps, i.e., \texttt{INIT} and \texttt{NBPG}.  The philosophy of this framework is that \texttt{INIT} geneartes an initial KNNG quickly, and then \texttt{NBPG} efficiently refines it, which follows the principle ``a neighbor of my neighbor is also likely to be my neighbor''~\cite{Dong2011WWW}.  Note that we do not expect \texttt{INIT} costs a lot for high accuracy, because the refinement will be done in \texttt{NBPG}. 

\begin{algorithm}[h]
\caption{Framework of KNNG Construction}
\label{alg:framework}
\KwIn{Data set $D$ and a positive integer $k$}
\KwOut{KNNG $G$}
\texttt{INIT} : generate an initial graph $G$ quickly\;
\texttt{NBPG} : refine $G$ by neighborhood propagation\;
\textbf{return} $G$\;
\end{algorithm}

In this paper, we will introduce a few representative approaches for \texttt{INIT} and \texttt{NBPG} respectively.  For each method, we also analyze its \textbf{memory requirement} and \textbf{time complexity}. For memory requirement, we only count the \textit{auxiliary data structures} of each method, ignoring the common structures, i.e., the data $D$ and the KNNG $G$.

\subsection{KNNG vs Proximity Graph}
\label{ssec:proxg}

Proximity graphs (e.g., SW~\cite{Malkov2014IS}, HNSW~\cite{Malkov2018PAMI}, DPG~\cite{Li2019TKDE} and NSG~\cite{Fu2019VLDB}) are the state-of-the-art methods for ANN search. Like KNNG, they treat each point $u$ as a graph node, but have distinct strategies to define $u$'s edges.  The relationship between KNNG and proximity graph is twisted. On one hand, KNNG could be treated as a special type of proximity graph. Hence, KNNG is used for ANN search in recent benchmarks~\cite{Annoy, Nmslib, Li2019TKDE}. In those benchmarks, \emph{KNNG is usually denoted as KGraph}, which is actually a KNNG construction method~\cite{Dong2011WWW}. On the other hand, some proximity graphs such as DPG and NSG require an accurate KNNG as the premise, so that their construction cost is higher than KNNG. However, SW and HNSW are not built on top of KNNG and thus may be built faster than KNNG with adequate settings. As a new perspective, we show that SW and HNSW could also be used to build KNNG.  

Moreover, we show ANN search on a proximity graph in Algorithm~\ref{alg:anns}, which will be discussed more than once in this paper. Its core idea is similar to \texttt{NBPG}, i.e., ``a neighbor of my neighbor is also likely to be my neighbor''. Let $H$ be a proximity graph and $q$ be a query. The search starts on specified or randomly-selected node(s) $ep$, which are first pushed into candidate set $pool$~(a sorted node list). Let $L = \max (k, efSearch)$ be the size of $pool$. In each iteration in Line 4-11, we find the first unexpanded node $u$ in $pool$ and then expand $u$ in Lines 6-10, denoted as $expand(q, u, H)$. This expansion treats each neighbor $v$ of $u$ on $H$ as a candidate to refine $pool$ in Line 8-10, which is denoted as $update(pool, v)$ . Once the first $efSearch$ nodes in $pool$ have been expanded, the search process terminates and returns the first $k$ nodes in $pool$.  Obviously, $efSearch$ is key to the search performance. A large $efSearch$ increases both the cost and the accuracy. 
 
\begin{algorithm}[t]
\caption{$Search\_on\_Graph(H, q, k, efSearch, ep)$}
\label{alg:anns}
\KwIn{$H$, $q$, $k$, $efSearch$ and $ep$}
\KwOut{KNNs of $q$}
let $pool$ be the candidate set and push $ep$ in $pool$\;
$L = max(k, efSearch)$ and $i = 0$\;
\While {$i < efSearch$} {
	$u=pool[i]$ and mark $u$ as expanded\;
	/* \textbf{Procedure:} $expand(q, u, H)$*/\\
	\For {each neighbor $v$ of $u$ on $H$} {
		/* \textbf{Procedure:} $update(pool, v)$*/\\
		$pool.add(v, dist(q, v))$\;
		sort $pool$ in ascending order of $dist(q, \cdot)$\;
		\textbf{if} $pool.size() > L$, \textbf{then} $pool.resize(L)$\;
	}
	$i = $ index of the first unexpanded node $u$ in $pool$\;
}
\textbf{return} the first $k$ nodes in $pool$\;
\end{algorithm}



\section{Constructing Initial Graphs}
\label{sec:initg}

In this section, we introduce several representative methods for \texttt{INIT}.  They can be classified into two broad categories: the partition-based approach and the small world based approach.  The main idea of the former is to partition the data points into sufficiently small but intra-similar groups.  Then each point finds its KNNs in the most promising group(s) and an initial KNNG is generated.  The partition-based approach includes \texttt{Multiple Division} \cite{WangJD2012CVPR}, \texttt{LSH KNNG} \cite{ZhangYM2013ECML}, and \texttt{LargeVis} \cite{TangJ2016WWW}.  These methods adopt different partitioning techniques.  The main idea of the small world based approach is to create a navigable small world graph among the data points in a greedy manner, the structure of which is helpful for identifying the KNNs of any data point.  The small world based approach was proposed to process ANN search in the literature, but not for KNNG construction.  But we find it is trivial to extend them for KNNG construction.  So we include these techniques in our study, which are denoted as \texttt{SW KNNG} \cite{Malkov2014IS} and \texttt{HNSW KNNG} \cite{Malkov2018PAMI}.  We conduct extensive experiments to test and compare these methods for \texttt{INIT}.

\subsection{Multiple Random Division}
\label{ssec:multidiv}

\begin{figure}[t]
\begin{center}
\includegraphics[width=0.45\textwidth]{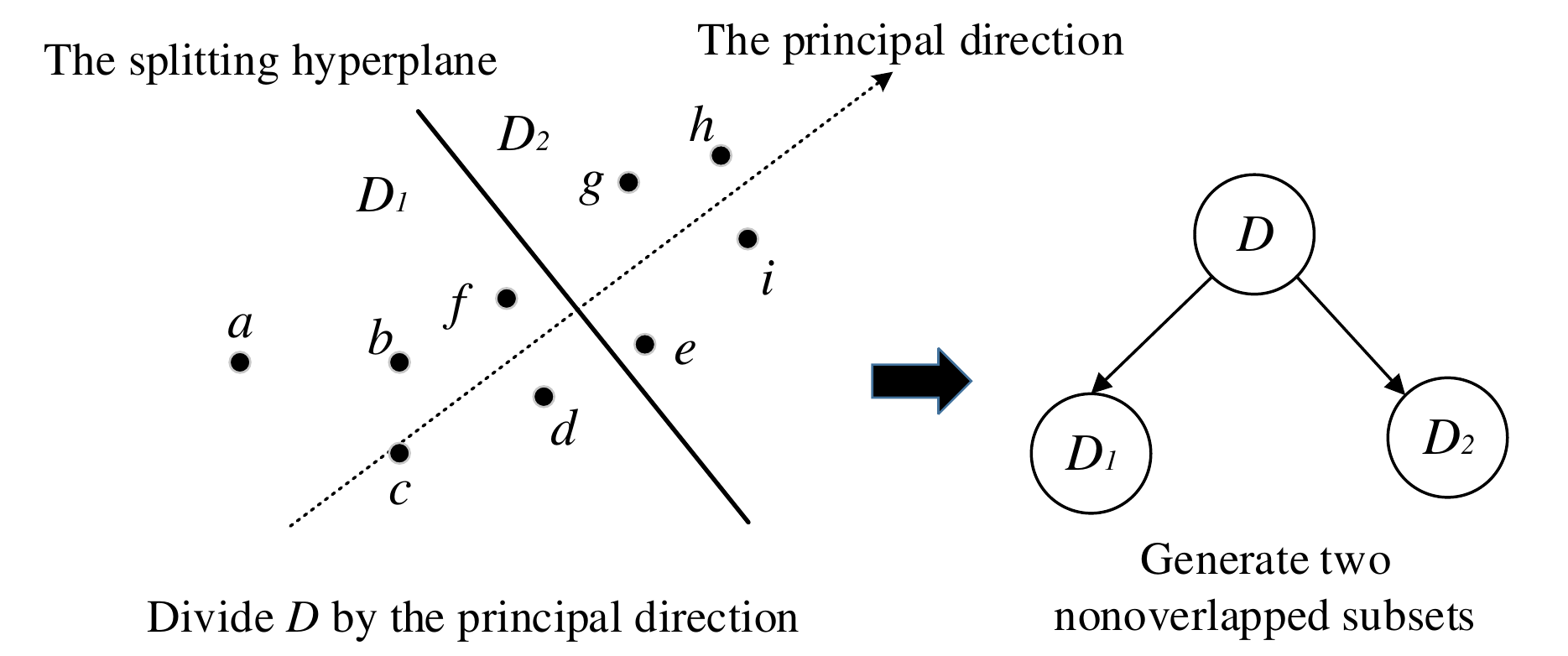}
\end{center}\vspace{-3ex}
\caption{Illustrate the division of $D$ by principal direction (represented by the solid arrow).  As shown by the dashed line, the splitting hyperplane that is perpendicular to the principal direction divides $D$ into two disjoint groups, $D_1$ and $D_2$, which are further partitioned recursively.{\color{blue}the figure should re-created.}}\vspace{-2ex}
\label{fig:randomdiv}
\end{figure}

\begin{figure*}[t]
\begin{center}
\includegraphics[width=0.8\textwidth]{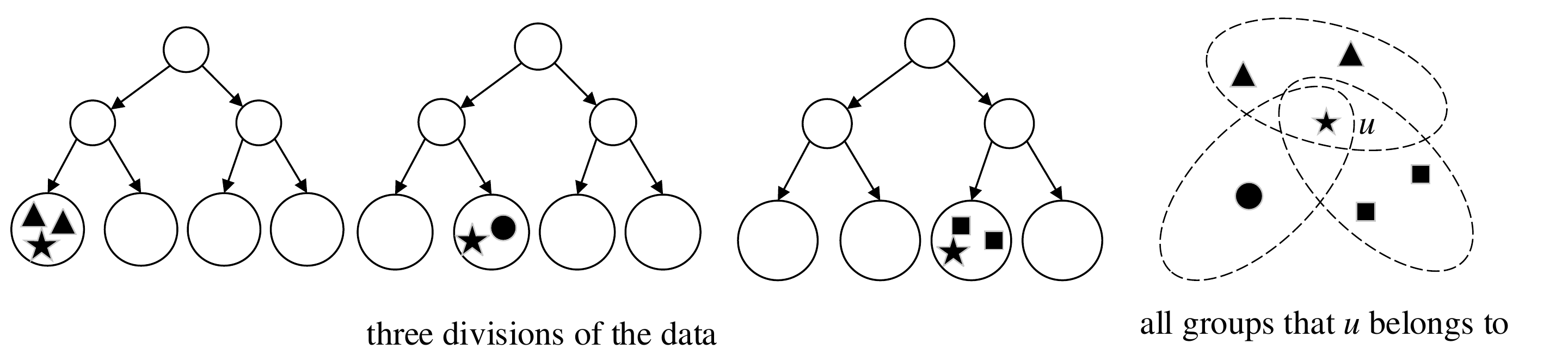}
\end{center}\vspace{-3ex}
\caption{This figure illustrates three random divisions of a data set $D$.  For data point $u$ denoted as a star, its KNNs are computed in the group of each division, and then combined to return the best KNNs.}\vspace{-2ex}
\label{fig:multidiv}
\end{figure*}

\begin{figure*}[t]
\begin{center}
\includegraphics[width=0.8\textwidth]{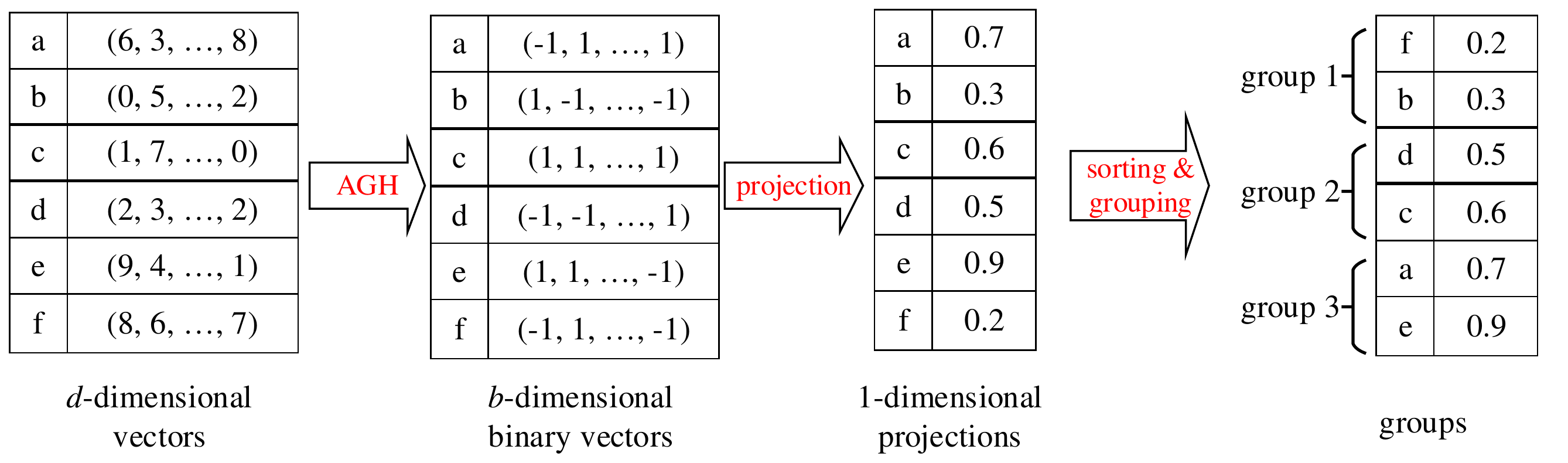}
\end{center}\vspace{-3ex}
\caption{The process of creating a division by LSH in \texttt{LSH KNNG}.}\vspace{-2ex}
\label{fig:lshknng}
\end{figure*}

Wang et al.\ \cite{WangJD2012CVPR} proposed a multiple random divide-and-conquer approach to build base approximate neighborhood graphs and then ensemble them together to obtain an initial KNNG.  Their method, which we denote as \texttt{Multiple Division}, recursively partitions the data points in $D$ into two disjoint groups.  To assign nearby points to the same group, the partitioning is done along the principal direction, which could be rapidly generated~\cite{WangJD2012CVPR}.  The data set $D$ is partitioned into two disjoint groups $D_1$ and $D_2$ (as illustrating in Figure \ref{fig:randomdiv}), which are further split in a recursive manner until the group to be split contains at most $T_{div}$ points.  Then a neighborhood subgraph for each finally-generated group is built in a brute-force manner, which costs $O(d * T_{div}^2)$.

A single division as described above yields a base approximate neighborhood graph containing a series of isolated subgraphs, and it is unable to connect a point with its closer neighbors lying in different subgraphs.  Thus, \texttt{Multiple Division} uses $L_{div}$ random divisions.  Rather than computing the principal direction from the whole group, \texttt{Multiple Division} computes it over a random sample of the group.  For each point $u \in D$, it has a set of KNNs from each division and thus totally $L_{div}$ sets of KNNs, from which the best KNNs are generated.  We illustrate this process in Figure \ref{fig:multidiv}.

\textbf{Cost Analysis:} Since those $L_{div}$ divisions could be generated independently, the memory requirement is $O(n)$ to store the current division.  For each division, the recursive partition costs $O(n * d * \log n)$ and the brute-force procedure on all finally-generated groups requires $O(n * d * T_{div})$. With $L_{div}$ divisions, the cost of \texttt{Multiple Division is $O(L_{div} * n * d * (\log n + T_{div}))$}. We can see that $L_{div}$ and $T_{div}$ are key to the cost of \texttt{Multiple Division}.

\subsection{LSH KNNG}
\label{ssec:lshknng}

Zhang et al.\ \cite{ZhangYM2013ECML} proposed a method, denoted as \texttt{LSH KNNG}, which uses the locality sensitive hashing (LSH) technique to divide $D$ into groups with equal size, and then builds a neighborhood graph on each group using the brute-force method.  To enhance the accuracy, similar to \texttt{Multiple Division}, $L_{hash}$ divisions are created to build $L_{hash}$ base approximate neighborhood graphs, which are then combined to generate an initial KNNG of higher accuracy.

The details of such a division are illustrated in Figure \ref{fig:lshknng}.  For each $d$-dimensional vector, \texttt{LSH KNNG} uses Anchor Graph Hashing (AGH)~\cite{LiuW2011ICML}, an LSH technique, to embed it into a binary vector in $\{-1, 1\}^b$. Then, the binary vector is transferred to a one-dimensional projection on a random vector in $\mathbb{R}^b$.  All points in $D$ are sorted in the ascending order of their one-dimensional projections and then divided into equal-size disjoint groups accordingly.  Let $T_{hash}$ be the size of such a group. Then a neighborhood graph is built on each group in the brute-force manner.

\textbf{Cost Analysis: } Like \texttt{Multiple Division}, each LSH division could be built independently. Hence, its memory requirement is caused by the binary codes and the division result. In total, its memory requirement is $O(n * b)$. In each division, generating binary codes costs $O(n * d * b)$ and the brute-force procedure on all groups costs $O(n * d * T_{hash})$. In total, the cost of \texttt{LSH KNNG} is $O(L_{hash} * n * d * (b + T_{hash}))$. We can see that the cost of \texttt{LSH KNNG} is sensitive to $L_{hash}$ and $T_{hash}$, since $b$ is usually a fixed value. 

\subsection{LargeVis}
\label{ssec:largevis}

Tang et al.\ \cite{TangJ2016WWW} proposed a method, denoted as \texttt{LargeVis}, to construct a KNNG for the purpose of visualizing large-scale and high-dimensional data.  Like \texttt{Multiple Division}, \texttt{LargeVis} recursively partitions the data points in $D$ into two small groups, but the partitioning is along a random projection (RP) that connects two randomly-sampled points from the group to be split.  The partitioning results are organized as a tree called RP tree~\cite{Dasgupta2008SOTC}.  Similar to \texttt{Multiple Division}, $L_{tree}$ RP trees will be built to enhance the accuracy.  An initial KNNG is thus generated by conducting a KNN search procedure for each point on those trees. Each search process will be terminated after retrieving $L_{tree} * k$ candidates~\footnote{Summarized according to the public codes from the authors~\textit{https://github.com/lferry007/LargeVis}. }.

\textbf{Cost Analysis: } Unlike \texttt{Multiple Division} and \texttt{LSH KNNG}, we have to store all $L_{tree}$ RP trees in the memory, since ANN search for each point is done on those trees simultaneously. Each tree requires $O(n*d)$ space, since each inner node stores a random projection vector. Hence, the total memory requirement is $O(L_{tree} * n * d)$. For each query, its main cost is to reach the leaf node from the root node in each tree and then verify $L_{tree} * k$ candidates. Hence, the total cost of \texttt{LargeVis} is $O(L_{tree} * n * d * (\log n+ k))$. We can see that $L_{tree}$ and $k$ are key to the cost of \texttt{LargeVis}. 

\subsection{Small World based Approach}
\label{ssec:hnswknng}

SW graph~\cite{Malkov2014IS} and HNSW graph~\cite{Malkov2018PAMI} are the state-of-the-art methods for ANN search~\cite{Nmslib, Li2019TKDE}.  In the literature, neither technique (SW or HNSW) has been used for \texttt{INIT}.  According to their working mechanism, it is trivial to extend them for \texttt{INIT}, while it needs a thorough comparison with other \texttt{INIT} methods in the literature.

\begin{figure}
\begin{center}
\includegraphics[width=0.333\textwidth]{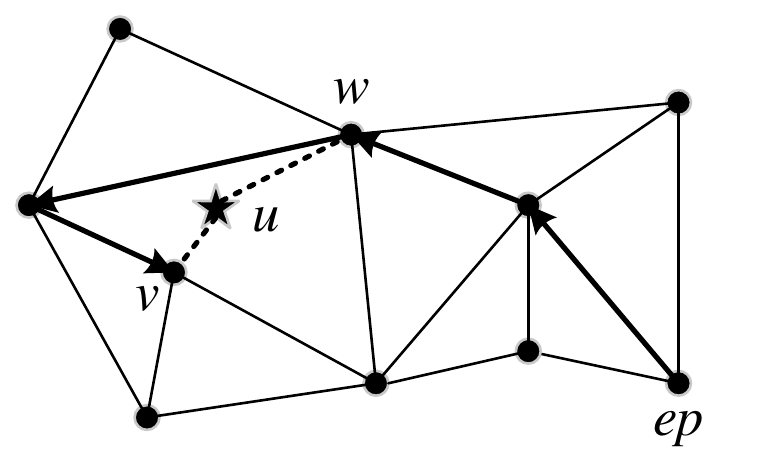}
\end{center}\vspace{-3ex}
\caption{The insertion of data point $u$ in SW.  $v$ and $w$ are neighbors found by ANN search, then undirected edges (dashed lines) are created. Bold arrows represents the search parth.}\vspace{-2ex}
\label{fig:sw}
\end{figure}


The SW graph $G_{sw}$ construction process starts from an empty graph, and iteratively inserts each data point $u$ into $G_{sw}$.  In each insertion, we first treat $u$ as a new node and then conduct ANN search $\mathcal{A} = Search\_on\_Graph(G_{sw}, u, M_{sw}, efConstruction, ep)$ as in Algorithm~\ref{alg:anns}, where $G_{sw}$ is the current SW graph and contains only a part of $D$.  Afterwards, undirected edges are created between $u$ and $M_{sw}$ neighbors in $\mathcal{A}$.  Figure \ref{fig:sw} illustrates the insertion process for a data point $u$.  It is easy to generate an initial KNNG during the construction of $G_{sw}$.  Before building $G_{sw}$, each point's KNN set is initialized as empty. When inserting $u$, its similar candidates will be retrieved by the search process and a distance is computed between $u$ and each candidate $v$. We use $v$ (resp.\ $u$) to refine the KNN set of $u$ (resp.\ $v$). Once the SW graph is constructed, an initial KNNG is also generated. For simplicity, this method is denoted as \texttt{SW KNNG}.

\textbf{Cost Analysis :} The main memory requirement of \texttt{SW KNNG} is to store the SW graph, which needs $O(n * M_{sw})$ space. Its main cost is to insert $n$ nodes. By experiments, we find that each insertion expands $O(efConstruction)$ nodes, as demonstrated in Appendix~\ref{sec:append_cost_sw}. But, there is no explicit upper bound on the neighborhood size. Let $M_{sw}^{max}$~\footnote{$M_{sw}^{max}$ is specific w.r.t. the data set, which is obviously affected by the hubness phenomenon as discussed in Section~\ref{ssec:explo_def}} be maximum neighborhood size in $G_{sw}$. In the worst case, its time complexity is $O(n * d * efConstruction * M_{sw}^{max})$.

\begin{figure}
\begin{center}
\includegraphics[width=0.3\textwidth]{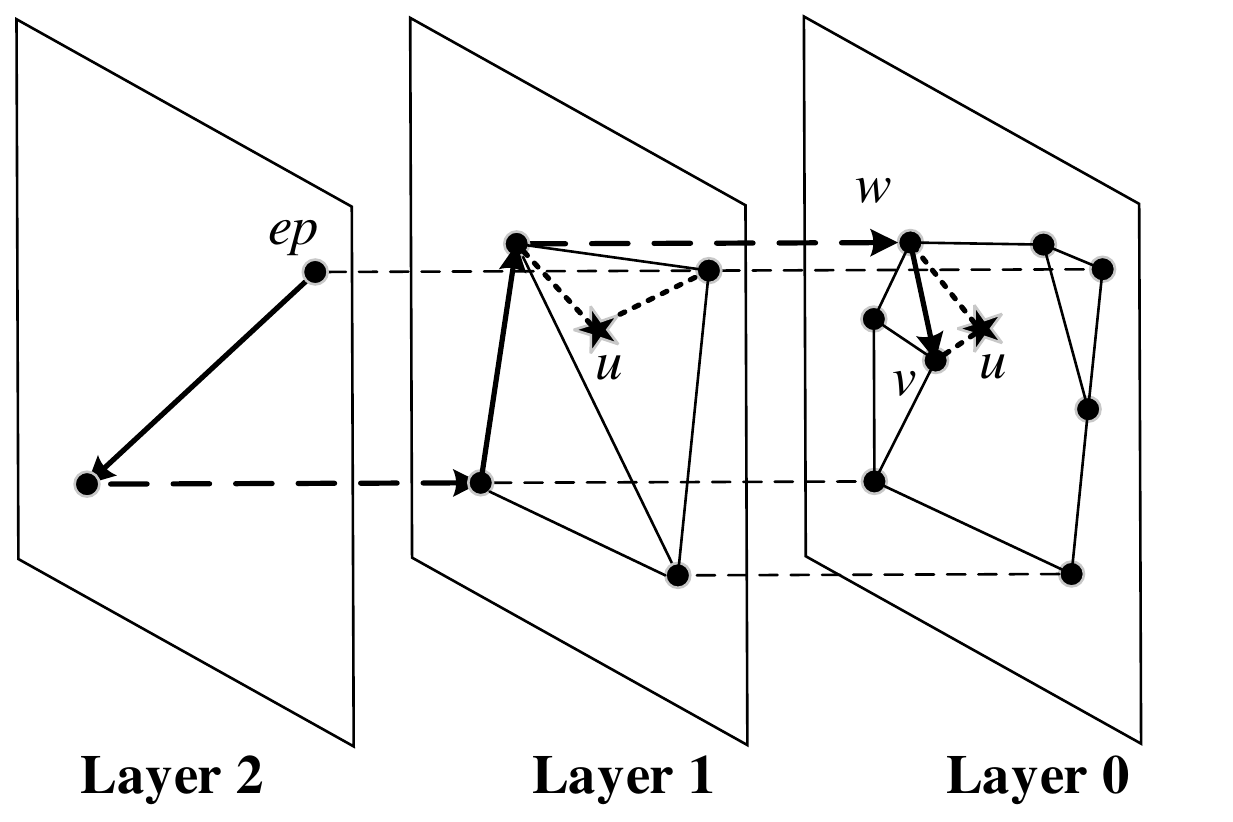}
\end{center}\vspace{-3ex}
\caption{The insertion of data point $u$ in HNSW. $l_h(u) = 1$. After the search, undirected edges (dashed lines) are added between $u$ and its neighbors in  both layer 0 and layer 1.}\vspace{-2ex}
\label{fig:hnsw}
\end{figure}

The HNSW garph $G_{hnsw}$ is a hierarchical structure of multiple layers.  A higher layer contains fewer data points and layer 0 contains all data points in $D$.  To insert a new point $u$, an integer maximum layer $l_h(u)$ is randomly selected so that $u$ will appear in layers 0 to $l_h(u)$.  To insert $u$, the search process starts from the entry point $ep$ in the highest layer.  The nearest node found in the current layer will serve as the entry point of search in the next layer, and this search process repeats in the next layer till layer 0.  Note that the search in each layer follows Algorithm~\ref{alg:anns}.    After the search, HNSW creates undirected edges between $u$ and $M_{hnsw}$ neighbors selected from $efConstruction$ returned ones, which is done in layers from $l_h(u)$ to layer 0 respectively. Unlike SW, HNSW imposes a threshold $M_{hnsw}$ on the neighborhood size.   Once the edges of a node exceeds $M_{hnsw}$, HNSW prunes them by a link diversification strategy.  Figure~\ref{fig:hnsw} illustrates an insertion in HNSW. By the same process in \texttt{SW KNNG}, we build the initial KNNG by tracing each pair of distance computation when building the HNSW graph.  We denote this method as \texttt{HNSW KNNG}.

\textbf{Cost Analysis: } The main requirement of \texttt{HNSW KNNG} is caused by the HNSW graph, which contains $O(n)$ nodes and each node has up to $M_{hnsw}$ neighbors. Hence, its memory requirement is $O(n * M_{hnsw})$. The cost of \texttt{HNSW KNNG} mainly contains two parts, i.e., ANN search for each node and pruning operations for oversize nodes. We find that each search expands $O(efConstruction)$ nodes as demonstrated in Appendix~\ref{sec:append_cost_hnsw} and thus the total cost for ANN search is $O(n * d * M_{hnsw} * efConstruction)$. Each pruning operation costs $O(M_{hnsw}^2 * d)$, but the exact number of those operations is unknown in advance. In the worst case, adding each undirected edge will lead to a pruning operation. In this case, there are totally $n*M_{hnsw}$ pruning operations. However, the practical number of pruning operations on real data is far smaller than $n*M_{hnsw}$, as shown in Appendix~\ref{sec:append_cost_hnsw}.  Overall, the total cost of \texttt{HNSW KNNG} is $O(n * d * M_{hnsw} * (efConstruction + M_{hnsw}^2))$ in the worst case.

\subsection{Experimental Evaluation of INIT}
\label{ssec:exp_init}

\subsubsection{Experimental Settings}
\label{sssec:settings}

We first describe the experimental settings in this paper, for both \texttt{INIT} methods and \texttt{NBPG} methods.

\noindent\textbf{Data Sets}. We use 8 real data sets with various sizes and dimensions from different domains, as listed in Table~\ref{tb:data}.  \textbf{Sift}\footnote{\textit{http://corpus-texmex.irisa.fr}} contains 1,000,000 128-dimensional SIFT vectors.  \textbf{Nuscm}\footnote{\textit{http://lms.comp.nus.edu.sg/research/NUS-WIDE.htm}} consists of 269,648 225-dimensional block-wise color moments. \textbf{Nusbow}\footnote{\textit{http://lms.comp.nus.edu.sg/research/NUS-WIDE.htm}} has 269,648 500-dimensional bag of words based on SIFT descriptors. \textbf{Gist}\footnote{\textit{http://corpus-texmex.irisa.fr}} contains 1,000,000 960-dimensional Gist vectors. \textbf{Glove}\footnote{\textit{http://nlp.stanford.edu/projects/glove/}} consists of 1,193,514 100-dimensional word features extracted from Tweets. \textbf{Msdrp}\footnote{\textit{http://www.ifs.tuwien.ac.at/mir/msd/download.html}} contains 994,185 1440-dimensional Rhythm Patterns extracted from the same number of contemporary popular music tracks. \textbf{Msdtrh}\footnote{\textit{http://www.ifs.tuwien.ac.at/mir/msd/download.html}} contains 994,185 420-dimensional Temporal Rhythm Histograms extracted from the same number of contemporary popular music tracks. \textbf{Uqv}\footnote{\textit{http://staff.itee.uq.edu.au/shenht/UQ\_VIDEO/index.htm}} contains 3,305,525 256-dimensional feature vectors extracted from the keyframes of a video data set.

\begin{table}[t]
\centering
\caption{Data Summary : the space is measured in MB and the cost of the baseline method is measured in seconds.}
\begin{tabular}{|l|l|l|m{0.8cm}|m{1.2cm}|l|}
\hline
Data &  $n$ &  $d$ & Space & Baseline & Domain \\
\hline
Sift & 1,000,000 & 128 & 488 & 6,115 & Image \\
\hline
Nuscm & 269,648 & 225 & 231 & 803 & Image \\
\hline
Nusbow & 269,648 & 500 & 514 & 1,850 & Image \\
\hline
Gist & 1,000,000 & 960 & 3,662 & 63,430 & Image \\
\hline
Glove & 1,193,514 & 100 & 455 & 6,648 & Text \\
\hline
Msdrp & 994,185 & 1,440 & 5,461 & 87,630 & Audio \\
\hline
Msdrth & 994,185 & 420 & 1,593 & 24,053 & Audio \\
\hline
Uqv & 3,305,525 & 256 & 3,228 & 178,089 & Video \\
\hline
\end{tabular}
\label{tb:data}
\end{table}

\noindent\textbf{Performance Indicators}. We estimate the performance each method in three factors, i.e., cost,  accuracy and memory requirement.  We use the execution time to measure the construction cost, denoted as $cost$. The accuracy is measured by $recall$.  Given a KNNG $G$, let $N_k(u)$ be the KNNs of $u$ from $G$, and $N_k^*(u)$ be its exact KNNs. Then $recall(u) = |N_k^{*}(u) \cap N_k(u)| / k$.  The $recall$ of $G$ is averaged over all nodes in $G$.  The memory requirement is estimated by the memory space occupied by the auxiliary structures of each method. By default, $k$ is set to be 20 in this paper.   

\noindent\textbf{Computing Environments}. All experiments are conducted on a workstation with Intel(R) Xeon(R) E5-2697 CPU and 512 GB memory. The codes are implemented by C++ and compiled by \texttt{g++4.8} with \texttt{-O3} option.  We shut down the SIMD instructions to make a fair competition, since some public codes are implemented without opening those instructions. We run those methods in parallel~\footnote{We do this due to two reasons. First, all the methods are easy to be executed in parallel. Second, there are many experiments in this paper.} and the number of threads is set to be 16 by default.

\subsubsection{Compared Methods and Parameter Selection}
\label{sssec:comp_mehtod_init}

We evaluate the five methods described in this section.  As listed below, their parameters are carefully selected following the goal, i.e., generating rapidly an initial KNNG with low or moderate accuracy in \texttt{INIT}. Note that the refinement will be done in \texttt{NBPG}.  

\begin{itemize}
	\item \texttt{Multiple Division :} $T_{div} = 500$, as in~\cite{WangJD2012CVPR}, and $L_{div}$ (10 by default)  is varied in $\{2, 4, 6, 8, 10\}$.
	\item \texttt{LSH KNNG :}  $T_{hash} = 200$, optimized by experiments, and $L_{hash}$ is varied in $\{5, 10, 20, 40\}$. 
	\item \texttt{LargeVis :} we vary $L_{tree}$ (50 by default) in $\{25, 50, 75, 100\}$ and use default settings for others. 
	\item \texttt{SW KNNG :} $M_{sw} = k$ and $efConstruction$ is varied in $\{5, 10, 20, 40, 80\}$. 
	\item \texttt{HNSW KNNG :} $M_{hnsw} = 20$, optimized by experiments, and $efConstruction$ (80 by default) is varied in $\{20, 40, 80, 160\}$. 
\end{itemize}


Moreover, we use the brute-force method as the baseline method, which leads to $n*(n-1)/2$ distance computations. We list the cost of the baseline method in Table~\ref{tb:data}.

\subsubsection{Experimental Results}
\label{ssec:exp_ig}

\begin{figure*}
\begin{center}
\subfigure
{
\includegraphics[width=0.75\textwidth, trim=50 388 45 205,clip]{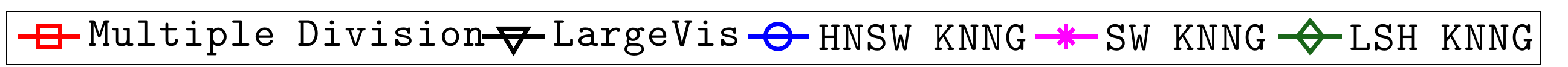}
} \vspace{-2ex}
\\
\subfigure[\textbf{Sift}]
{
\includegraphics[width=0.22\textwidth]{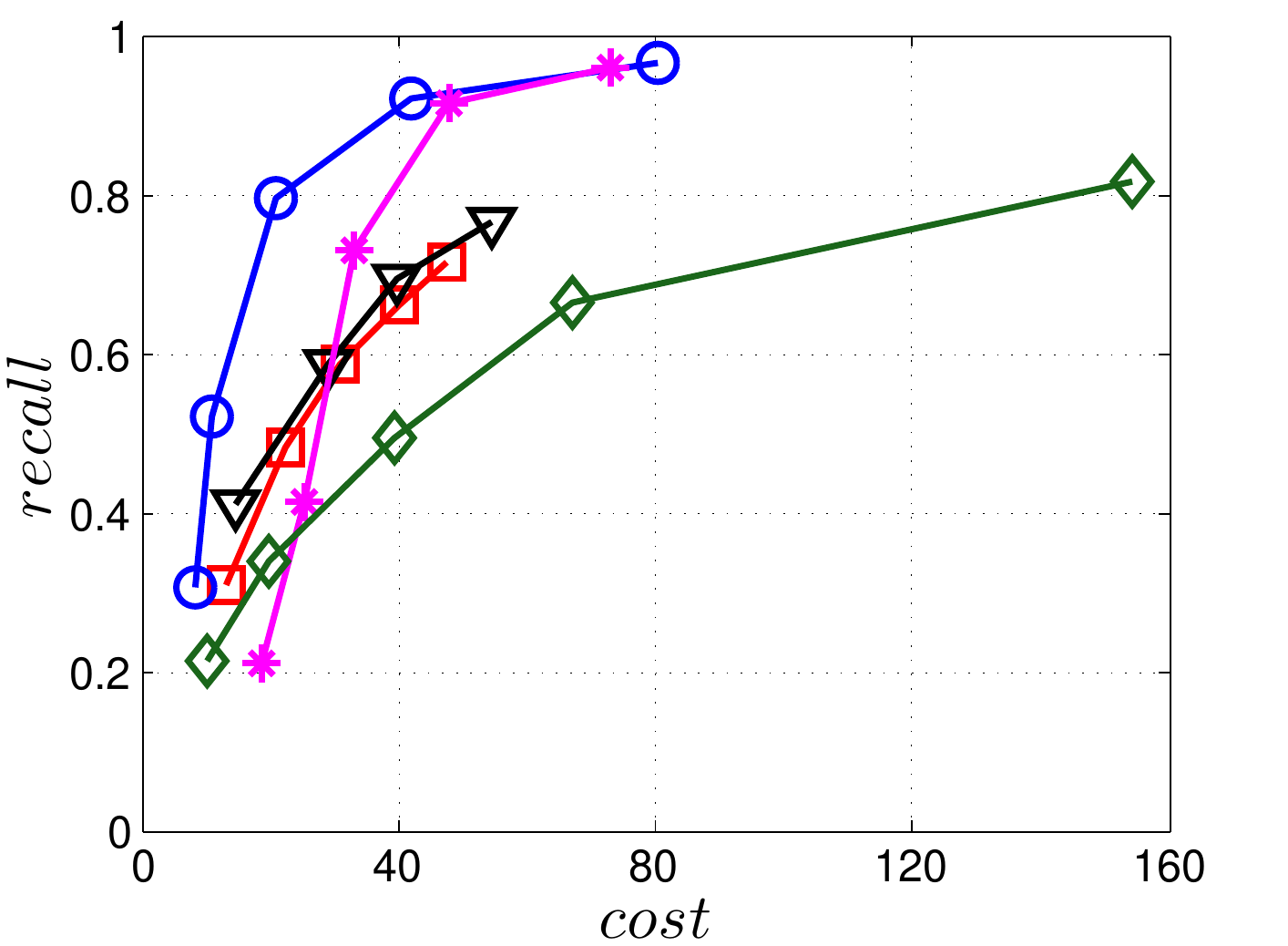}
\label{fig:initg_sift}
}
\subfigure[\textbf{Nuscm}]
{
\includegraphics[width=0.22\textwidth]{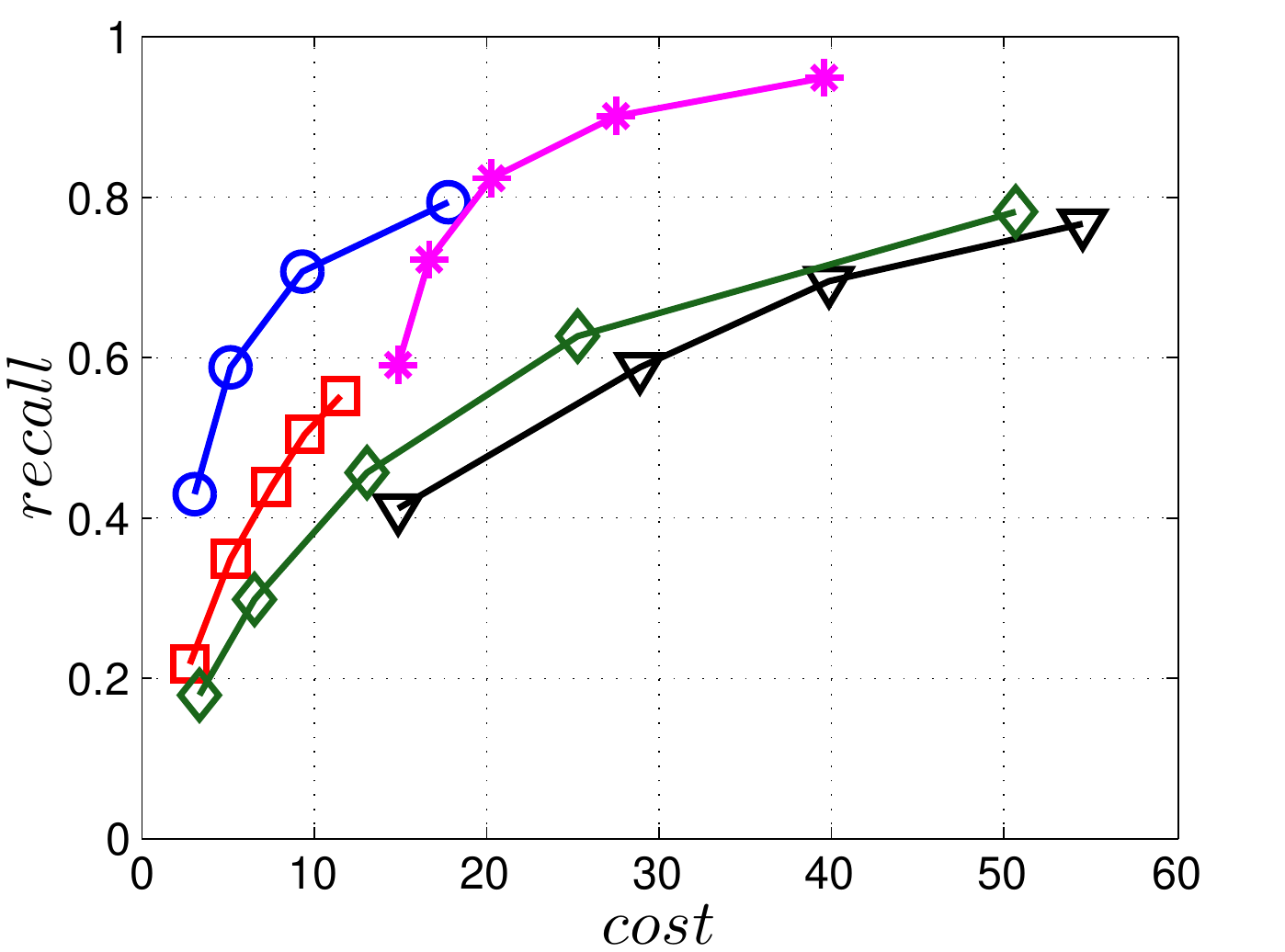}
\label{fig:initg_nuscm}
}
\subfigure[\textbf{Nusbow}]
{
\includegraphics[width=0.22\textwidth]{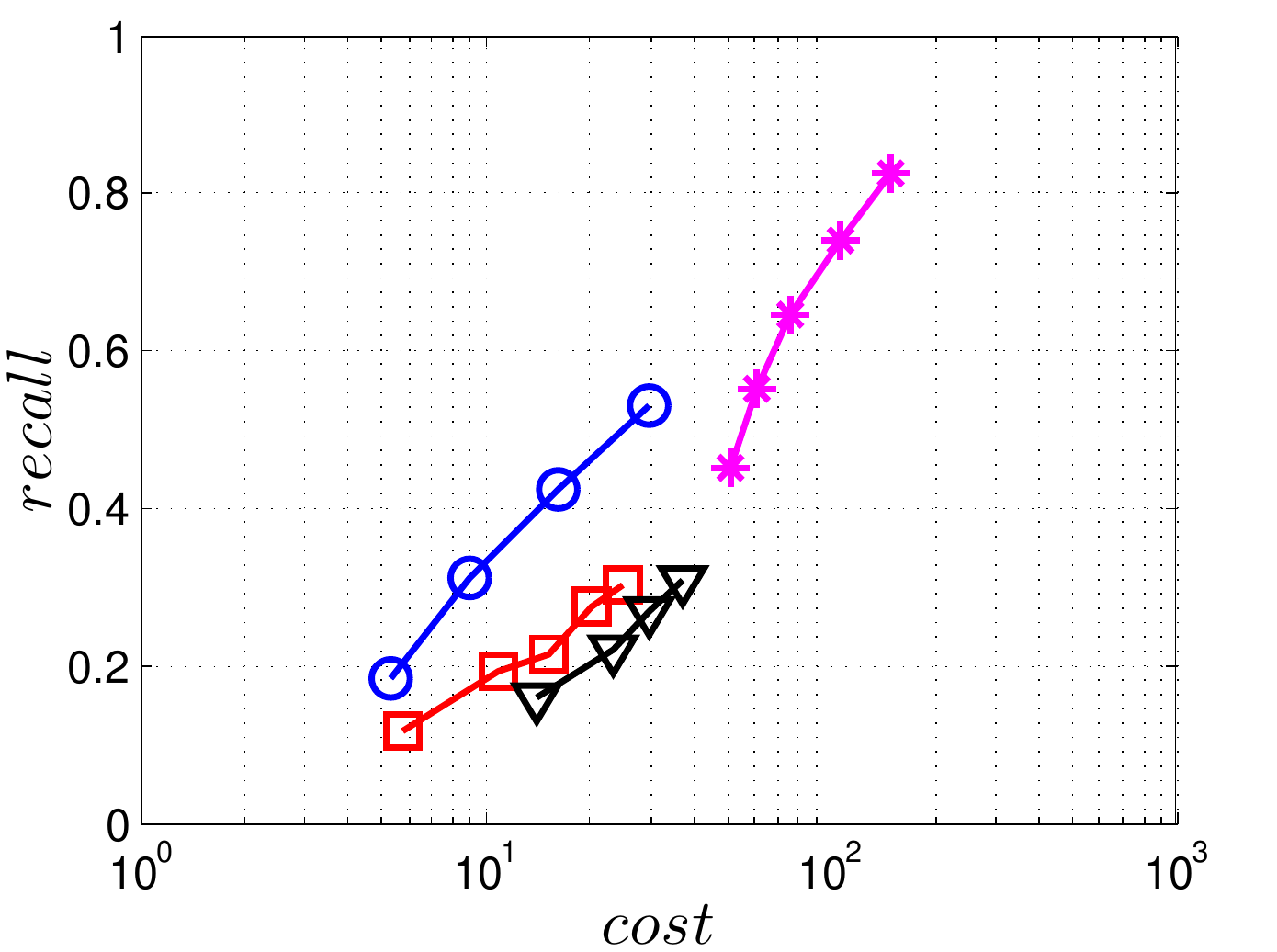}
\label{fig:initg_nusbow}
}
\subfigure[\textbf{Gist}]
{
\includegraphics[width=0.22\textwidth]{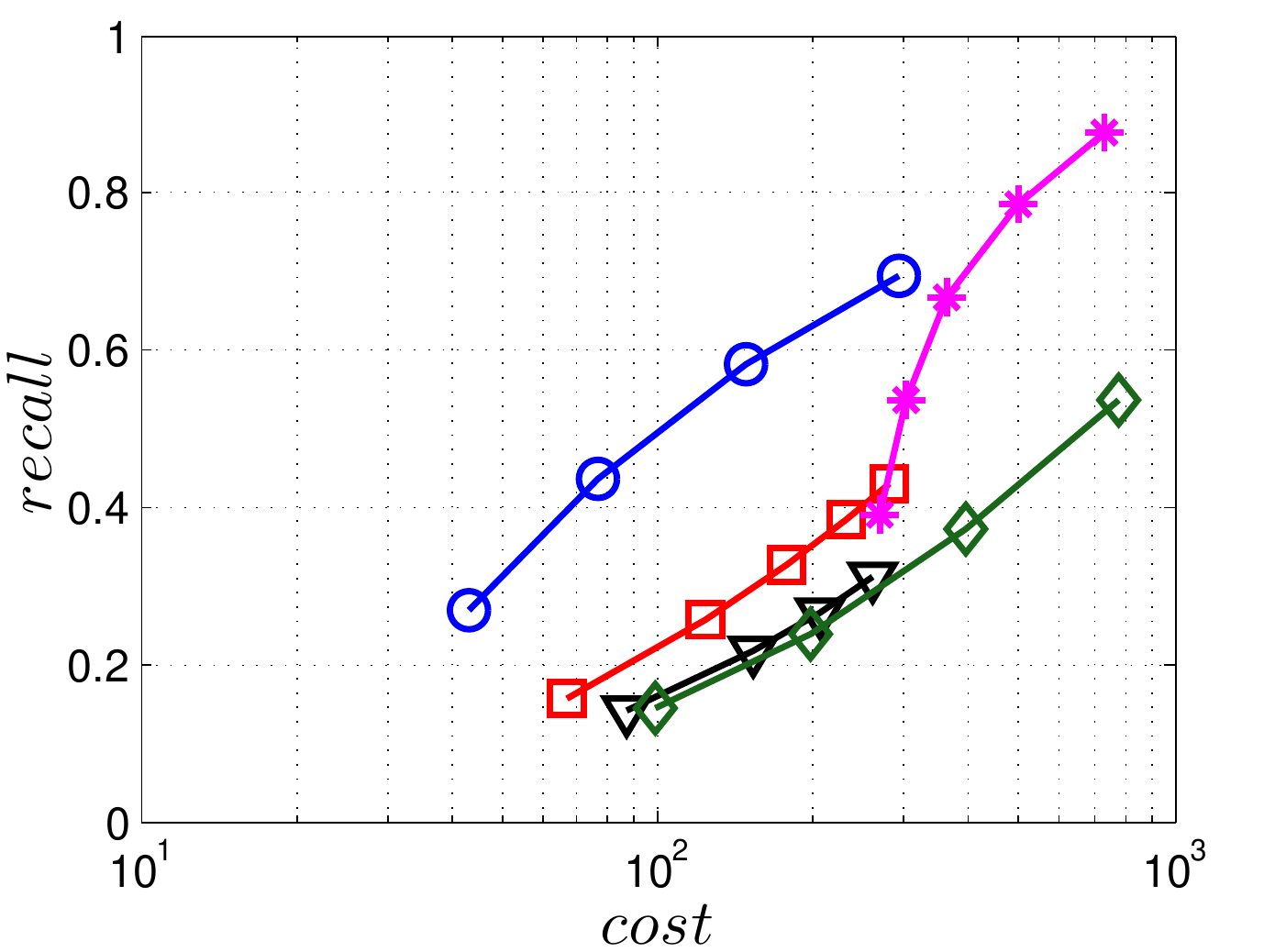}
\label{fig:initg_gist}
}
\subfigure[\textbf{Glove}]
{
\includegraphics[width=0.22\textwidth]{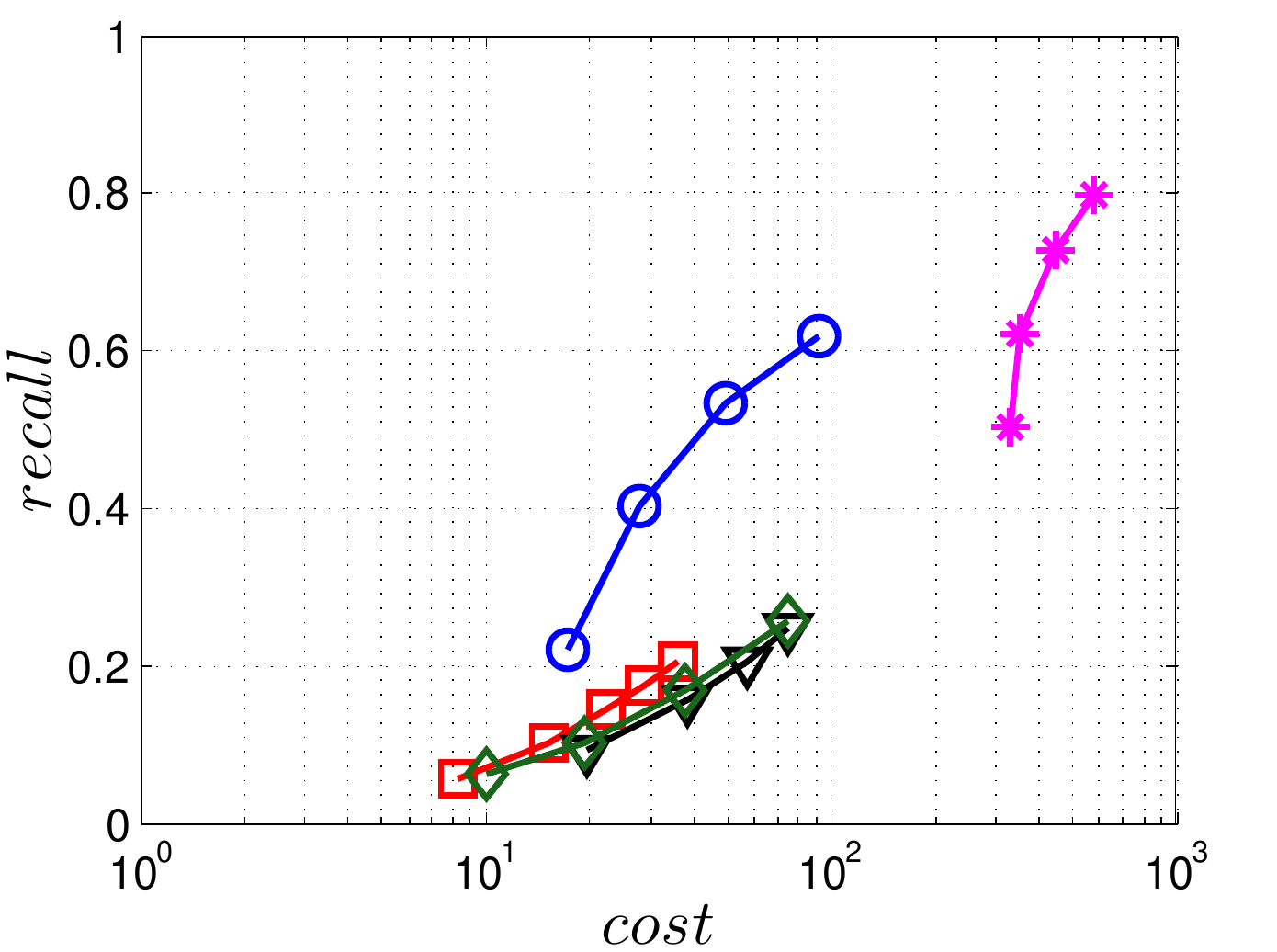}
\label{fig:initg_glove}
}
\subfigure[\textbf{Msdrp}]
{
\includegraphics[width=0.22\textwidth]{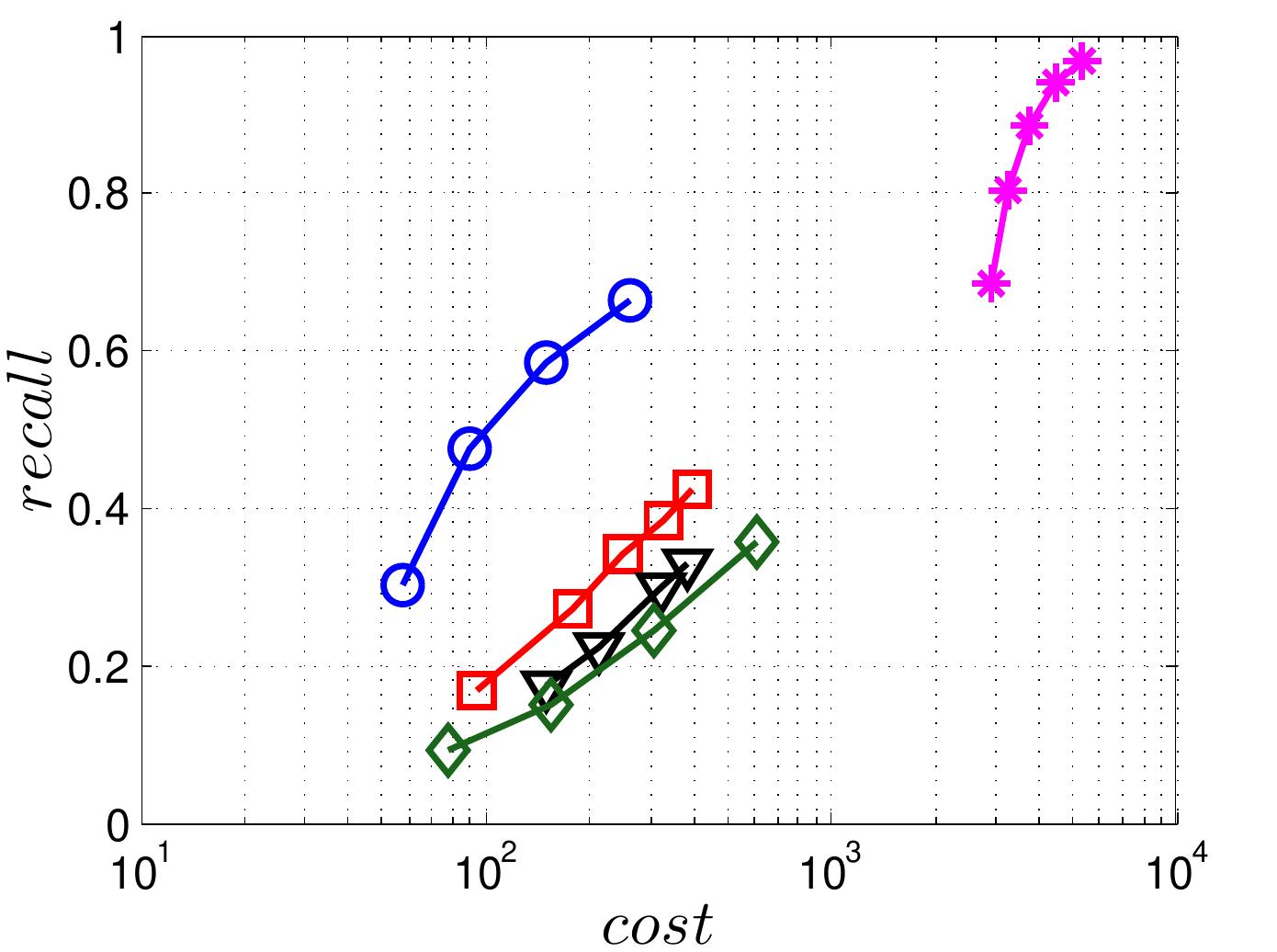}
\label{fig:initg_msdrp}
}
\subfigure[\textbf{Msdtrh}]
{
\includegraphics[width=0.22\textwidth]{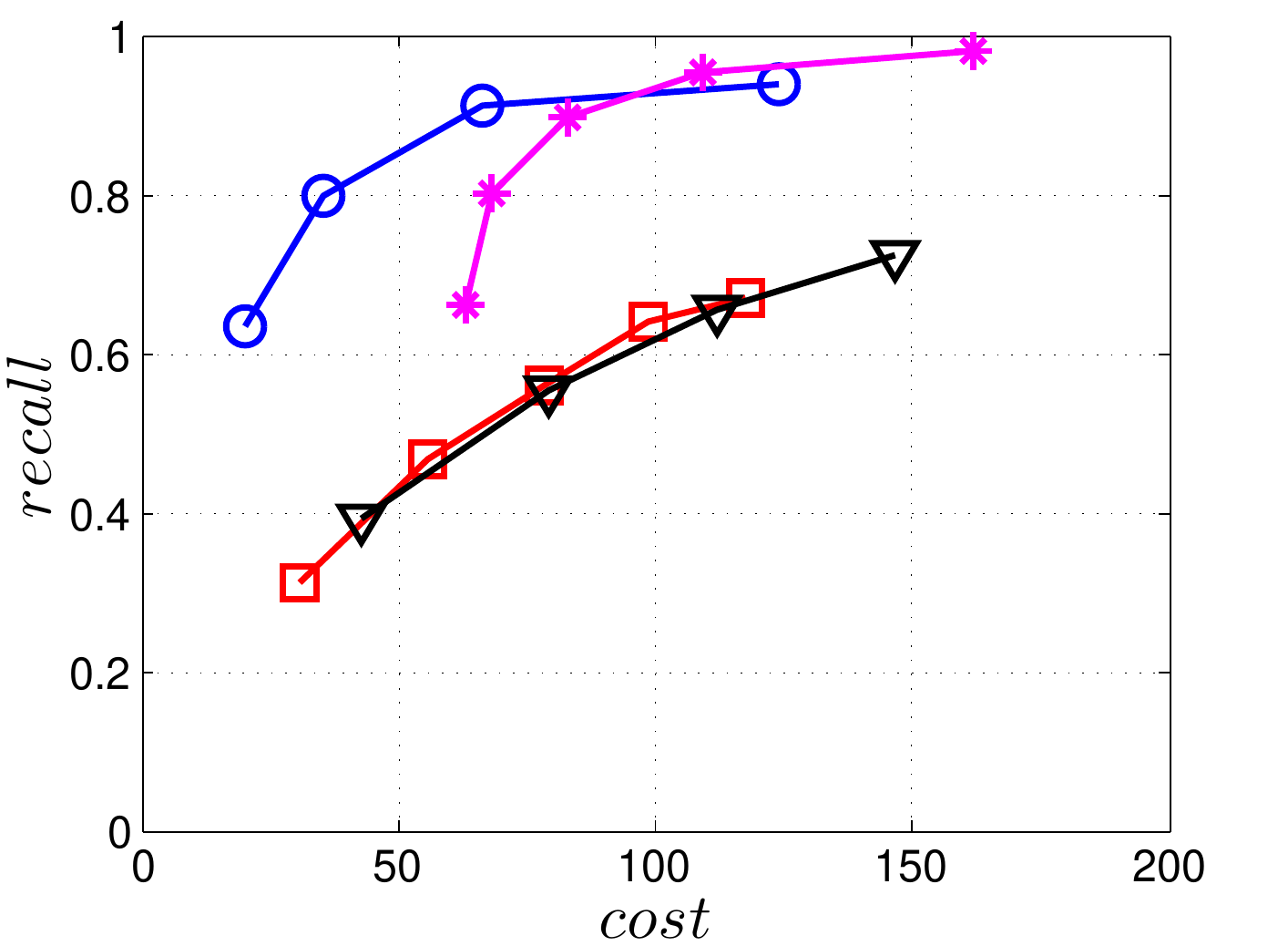}
\label{fig:initg_msdtrh}
}
\subfigure[\textbf{Uqv}]
{
\includegraphics[width=0.22\textwidth]{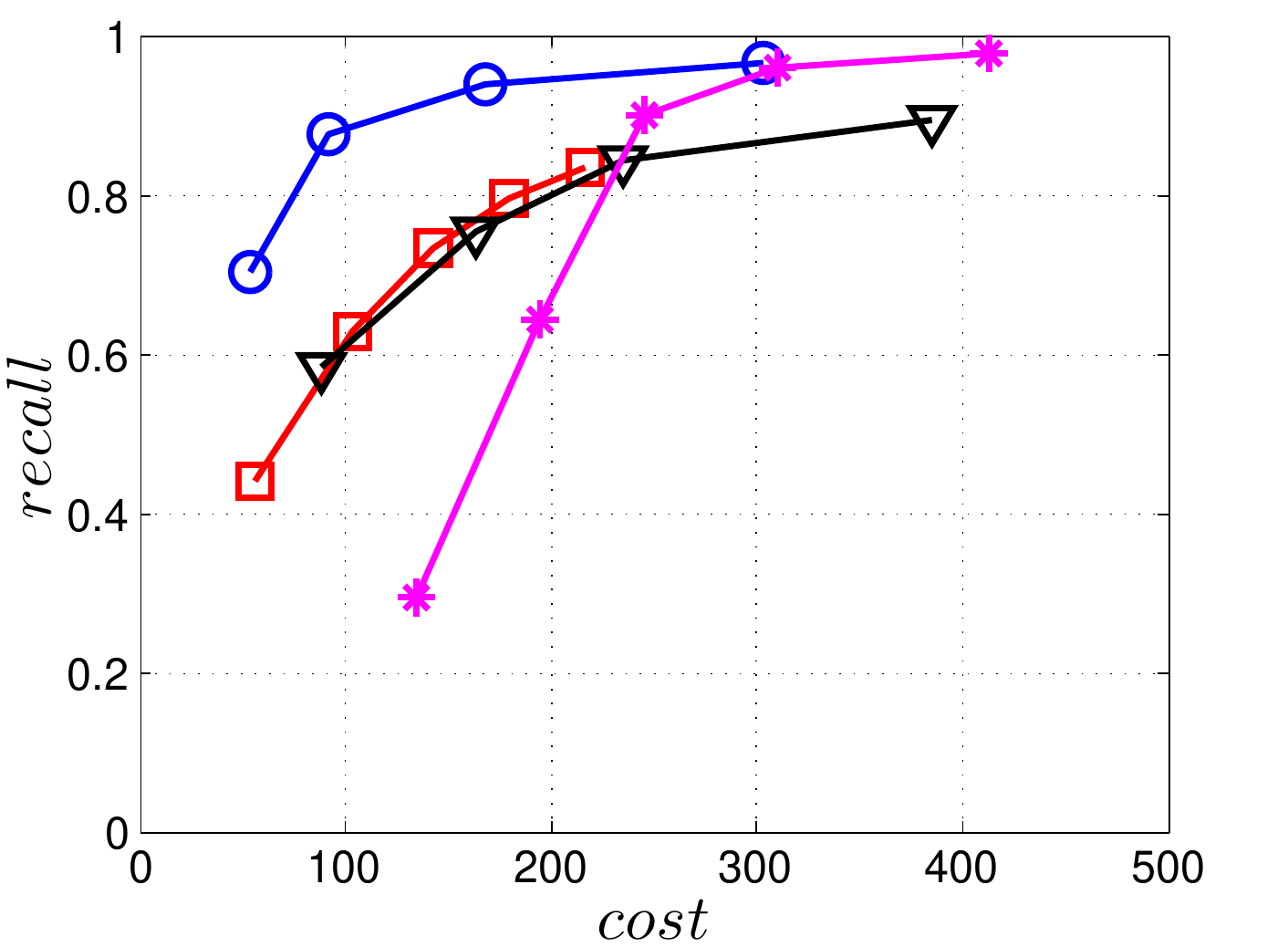}
\label{fig:initg_uqv}
}
\end{center}\vspace{-3ex}
\caption{Comparing \texttt{INIT} Methods. $cost$ is measured in seconds.} \vspace{-2ex}
\label{fig:initg}
\end{figure*}

\eat{
We show the experimental results in Figure~\ref{fig:initg}. \texttt{HNSW KNNG} obviously outperforms the other methods in almost all cases.  Its superior performance is brought by the search power of the HNSW graph.  In contrast, \texttt{SW KNNG} shows unstable performance.  On data sets Sift and Msdtrh, its performance is close to that of \texttt{HNSW KNNG}, but on data sets Msdrp and Glove, its performance is much worse as indicated by the high cost.  This can be explained by the existence of super nodes with many neighbors in the SW graph, which is caused by the \emph{hubness phenomenon} as discussed later in Section~\ref{ssec:explo_def}.  \texttt{HNSW KNNG} avoids such super nodes by imposing a threshold on the edge number for each node.

\texttt{Multiple Division} performs better than \texttt{LargeVis} and \texttt{LSH KNNG}. This superior performance may be brought by the principal direction based projection, which is more effective than LSH functions and random projections for data partitioning.  On several data sets, \texttt{LSH KNNG} fails to construct a KNNG due to runtime errors in computing the eigenvectors when training the AGH model.
}

A good \texttt{INIT} method should have a good balance between $cost$  and $recall$. We show the experimental results of $cost$ vs $recall$ in Figure~\ref{fig:initg}.  We can see that small world based approaches obviously outperform existing methods that employ traditional index structures. Compared with \texttt{SW KNNG}, \texttt{HNSW KNNG} presents stable balance bwetween $cost$ and $recall$.  We notice that on data such as Nusbow, Glove and Msdrp, \texttt{SW KNNG} presents high accuracy and high cost. As mentioned previously, this is not the purpose of \texttt{INIT}. By the way, we believe that it is caused by the \emph{hubness phenomenon} as discussed later in Section~\ref{ssec:explo_def}. 

\texttt{Multiple Division} performs better than \texttt{LargeVis} and \texttt{LSH KNNG}. This superior performance may be brought by the principal direction based projection, which is more effective than LSH functions and random projections for data partitioning.  On Nusbow, Msdtrh and Uqv, \texttt{LSH KNNG} fails to construct a KNNG due to runtime errors in computing the eigenvectors when training the AGH model.

We show the memory requirement of auxiliary data structures of \texttt{INIT} methods with their default settings in Table~\ref{tb:memory_init}.  We can see that \texttt{LargeVis} requires the most memory space due to storing 50 RP trees, followed by \texttt{HNSW KNNG} and \texttt{SW KNNG} that need to store an HNSW graph and SW graph repressively.  Due to the independence of multiple divisions, \texttt{Multiple Division} and \texttt{LSH KNNG} only store the structures for one division in the memory and thus require obviously less memory. 
	
\begin{table}
     \centering
     \caption{Memory Requirement (MB) of Auxiliary Structures for \texttt{INIT} Methods} \vspace{-0.45ex}
     \begin{tabular}{|l|m{1.2cm}|m{0.7cm}|m{0.8cm}|m{0.7cm}|m{0.7cm}|}
          \hline
           Data & \texttt{Multiple Division} & \texttt{LSH KNNG} & \texttt{Large Vis} & \texttt{SW KNNG} & \texttt{HNSW KNNG}\\
          \hline
          Sift & 7.7 & 30.5 & 722 & 164 & 180 \\
          \hline
          Nuscm & 2.3 & 8.2 & 243 & 44 & 49 \\
          \hline 
          Nusbow & 3.2 & 8.2 & 329 & 44 & 49 \\
          \hline 
          Gist & 11.3 & 30.5 & 874 & 164 & 179 \\
          \hline 
          Glvoe & 9.2 & 36.4 & 1,208 & 196 & 215 \\
          \hline
          Msdrp & 16 & 30.3 & 1,013 & 163 & 179 \\
          \hline
          Msdtrh & 8.4 & 30.3 & 881 & 163 & 179 \\
          \hline
          Uqv & 25.6 & 100.9 & 2,488 & 542 & 654 \\
          \hline
     \end{tabular}\vspace{0.2ex}
     \label{tb:memory_init}
\end{table}

\noindent\textbf{Summary}.  For \texttt{INIT}, we first recommend \texttt{HNSW KNNG}. This is because \texttt{HNSW KNNG} presents consistently superior performance over other competitors. Even though its memory requirement is the second largest, it is still acceptable, especially compared to the data space as shown in Table~\ref{tb:data}.  Our second recommendation is \texttt{Multiple Division}, which has good performance and requires little memory space. Without the information of the data hubness, we do not recommend \texttt{SW KNNG}, since the user may encounter huge cost to complete the task.



\section{Neighborhood Propagation}
\label{sec:nbpg}

In this section, we study the \texttt{NBPG} technique in the framework of KNNG construction, which is used to refine the initial KNNG.  In general, all methods in the literature follow the same principle: ``a neighbor of my neighbor is also likely to be my neighbor''~\cite{Dong2011WWW}, but differ in the technical details.  We categorize them into three categories: \texttt{UniProp}~\cite{ZhangYM2013ECML, TangJ2016WWW}, \texttt{BiProp}~\cite{Dong2011WWW} and \texttt{DeepSearch}~\cite{WangJD2012CVPR}, and describe their core ideas as follows.


\subsection{UniProp}
\label{ssec:uni_prop}

As a popular \texttt{NBPG} method, {\texttt{UniProp} stands for uni-directional propagation, which iteratively refines a node's KNNs from the KNNs of its KNNs~\cite{ZhangYM2013ECML, TangJ2016WWW}.   Let $N_k^t (u)$ be the KNNs of $u$ found in iteration $t$.  We can obtain $N_k^0(u)$ from the initial KNNG constructed in the \texttt{INIT} step.  Algorithm \ref{alg:uni_prop} shows how to iteratively refine the KNNs of each node.  In iteration $t$, we initially set $N_k^t(u)$ as $N_k^{t-1}(u)$ and then conduct $update(N_k^{t}(u), w)$  (as in Algorithm~\ref{alg:anns}) for each $w \in \cup_{v\in N_k^{t-1}(u)} N_k^{t-1}(v)$. \texttt{UniProp} repeats for $nIter$ times and terminates. 

\textbf{Cost Analysis:} \texttt{UniProp} needs to store the last-iteration KNNG. Hence its space requirement is $O(n*k)$. Its time complexity is $O(nIter * n * d * k^2)$ in the worst case. 

\begin{algorithm}[t]
\caption{\texttt{UniProp}}
\label{alg:uni_prop}
\KwIn{$D$, $k$, $nIter$, initial KNNG $G$}
\KwOut{$G$}
\For {$t : 1 \to nIter$} {
	\For {each $u$ in $D$}
	{	
		$N_k^{t} (u) = N_k^{t-1} (u)$\;
		\For {each $v \in N_k^{t-1}(u)$}
		{
			\For {each $w \in N_k^{t-1}(v)$}
			{
				$update(N_k^{t}(u), w)$\;			
			}
		}
	}
}
\textbf{return} $G$\;
\end{algorithm}

\subsection{BiProp}
\label{ssec:bi_prop}

In contrast to \texttt{UniProp}, another type of method considers both the \emph{nearest neighbors} and the \emph{reverse nearest neighbors} of a node for neighborhood propagation ~\cite{Dong2011WWW}.  Thus we denote it as \texttt{BiProp} which stands for bi-directional propagation.  For each node $u$, \texttt{BiProp} maintains a larger neighborhood $u.pool$ with size $L \geq k$ , which contains $u$'s $L$ nearest neighbors found so far. \texttt{BiProp} also takes an iterative strategy.  Let $N_m^t (u)$ be the $m$ nearest neighbors of $u$ in iteration $t$, where $m$ is a variable and does not necessarily equal to $k$. Note that $N_m^t (u) \subset u.pool$.  We use $R_m^t (u)$ to denote $u$'s reverse neighbor set $\{ v | u \in N_m^t (v) \}$ and thus $\sum_{u \in D} |R_m^t(u)| = \sum_{v \in D} |N_m^t(v)|$.  Then we denote the candidate set for neighborhood propagation as:
\[B_m^t (u) = N_m^t (u) \cup R_m^t (u).\]

We show the procedure of \texttt{BiProp} in Algorithm~\ref{alg:bi_prop}. $u.pool$ is initialized randomly, and so is $B_m^0(u)$. In iteration $t$, a brute-force search is conducted on $B_m^{t-1} (u)$ for each $u$, as in Lines 4 to 9. At the end of each iteration, \texttt{BiProp} will update $N_m^t(u)$ and $B_m^t(u)$ in Lines 10 and 11. 

In practice, the reverse neighbor set $R_m^t (u)$ may have a large cardinality (e.g., tens of thousands or even more) for some $u \in D$.  Processing such a super node in one iteration costs as high as $O(d*(m + |R_m^t (u)|)^2)$.  To address this issue, \texttt{BiProp} imposes a threshold $T$ on $|R_m^t (u)|$ and the oversize sets will be reduced by random shuffle in each iteration.

Both \texttt{NNDes}\footnote{As a well-known KNNG construction method, it is publicly available on \textit{https://code.google.com/archive/p/nndes}.} and \texttt{KGraph}\footnote{It is publicly available on \textit{https://github.com/aaalgo/kgraph. It is well known as a method for ANN search, but ignored for KNNG construction}.} follows the idea of \texttt{BiProp}.  Their key difference is how to set $m$ and $L$.  In \texttt{NNDes}, $L=k$ and $m \leq k$ for all nodes, which is specified in advance.  As in \texttt{KGraph}, $L$ is a specified value, but $m$ varies with node $u$ and the iteration $t$, denoted as $m^t(u)$.  To determine $m^t(u)$, three rules must be satisfied simultaneously, summarized from the source codes. Let $\delta$ be a small value~(e.g., 10). 

\begin{itemize}

\item \textbf{Rule 1 :} $m^0(u) = \delta$ and $m^t(u) - m^{t-1}(u) \leq \delta$.

\item \textbf{Rule 2 :} $| N_m^t (u) \setminus N_m^{t-1}(u)| \leq \delta$.

\item \textbf{Rule 3 :} $m^t(u)$ increases, iff $\exists v \in B_m^{t-1} (u)$ such that the brute-force search on $B_m^{t-1} (u)$ refines $N_k^t(v)$.

\end{itemize}
}

\textbf{Cost Analysis:} In \texttt{NNDes}, we need to store $N_m^t(\cdot)$ and $B_m^t(\cdot)$. Since $m \leq k$ and $T=k$, the required memory is $O(n*k)$. Its time complexity is $O(nIter * n * d * k^2$) in the worst case. In \texttt{KGraph}, we need to store $u.pool$ and $B_m^t(u)$ for each $u$. $|N_m^t(u)|$ is bounded by $L$ and thus $|B_m^t (u)| \leq L + T$. Its memory requirement is $O(n * L)$, which could be large enough since $L$ is usually set as a large value~(e.g., 100). Its time complexity is $O(nIter * n *  d * (L + T)^2)$ in the worst case, but it is very fast in practice due to the filters as discussed in Section~\ref{ssec:opt}.

\begin{algorithm}[t]
\caption{\texttt{BiProp}}
\label{alg:bi_prop}
\KwIn{$D$, $k$, $nIter$}
\KwOut{$G$}
\For {each $u \in D$} 
{
	randomly initialize $u.pool$ and $B_m^0(u)$\;
}
\For {$t : 1 \to nIter$} {
	\For {each $u$ in $D$}
	{	
		/* \textbf{Procedure: } $brute\_force(B_m^{t-1}(u))$ */ \\
		\For {each $v$ in $B_m^{t-1}(u)$}
		{
			\For {each $w$ in $B_m^{t-1}(u)$ and $v < w$}
			{
				$update(v.pool, w)$\;
				$update(w.pool, v)$\;			
			}
		}
	}
	\For {each $u$ in $D$}
	{
		update $N_m^t(u)$ and $B_m^t(u)$\;
	}
}
\For {each $u$ in $D$}
{
	$N_k(u)=$ the first $k$ nodes in $u.pool$\;
}
\textbf{return} $G$\;
\end{algorithm}

\subsection{DeepSearch}
\label{ssec:deep_search}

\eat{
Wang et al.\ introduced the idea of \texttt{DeepSearch}~\cite{WangJD2012CVPR}, which follows the idea of ANN search on a proximity graph~\cite{WangJD2012CVPR, Malkov2018PAMI}.  A proximity graph on $D$ is determined in advance, which is not necessarily the initial KNNG.  Then we treat each point $u$ as a query and perform ANN search on the proximity graph, in order to refine $u$'s KNNs.  The proximity graph in~\cite{WangJD2012CVPR} is the initial KNNG constructed by \texttt{Multiple Division}.  In our experimental study, we try using an HNSW graph as another option of the proximity graph for \texttt{DeepSearch}, and show that it achieves better performance.

A greedy strategy is used in the ANN search process \cite{Li2019TKDE, Malkov2018PAMI}. For each $u \in D$, a node list $pool$ is maintained to store the best $efSerach$ neighbors found so far.  At start, $pool$ is filled by $u$'s KNNs from the initial KNNG.  The search process iteratively fetches the closest unexplored neighbor $v$ from $pool$ and then explores $v$'s neighbor list on the proximity graph to refine both $u$'s KNNs and $pool$. Once all nodes in $pool$ have been explored, the search process terminates.

Unlike \texttt{UniProp} and \texttt{BiProp}, \texttt{DeepSearch} runs only for one iteration. The quality of returned KNNG is controlled by $efSearch$ instead of the number $nIter$ of iterations.
}

Wang et al.\ introduced the initial idea of \texttt{DeepSearch}~\cite{WangJD2012CVPR}, as shown in Algorithm~\ref{alg:deep_search}. It follows the idea of ANN search on a proximity graph $H$, created online before neighborhood propagation. For $u \in D$, it refines $N_k(u)$ by conducting ANN search with staring nodes from the initial KNNG. Unlike \texttt{UniProp} and \texttt{BiProp}, \texttt{DeepSearch} runs only for one iteration.

The proximity graph $H$ is key to the performance of \texttt{DeepSearch}. An appropriate $H$ should satisfy two conditions. First, $H$ should be rapidly built, since it is built online. Second, $H$ should have good ANN search performance. $H$ is the initial KNNG built by \texttt{Multiple Division} in~\cite{WangJD2012CVPR}, denoted as \texttt{DeepMdiv}, which satisfies the first condition but fails for the second. Among existing proximity graphs, HNSW is a good choice of \texttt{DeepSearch}, which satisfies both conditions. We denote this method as \texttt{Deep HNSW}. Others such as DPG~\cite{Li2019TKDE} and NSG~\cite{Fu2019VLDB} require an expensive modification on an online-built KNNG, which violates the first condition. Thus they are not suitable for \texttt{DeepSearch}. The construction cost of SW is unstable as shown in Section~\ref{ssec:exp_init} and its ANN search performance is usually worse than other proximity graphs~\cite{Li2019TKDE}. Hence SW is not suitable.  


\textbf{Cost Analysis: } \texttt{DeepSearch} needs to store the proximity graph. Thus \texttt{DeepMdiv} requires $O(n*k)$ and \texttt{Deep HNSW} needs $O(n*M_{hnsw})$. The cost of \texttt{DeepSearch} is determined by two factors, i.e. the number of expanded nodes and their neighborhood sizes. We find that both \texttt{DeepMdiv} and \texttt{Deep HNSW} expand $O(efSearch)$ nodes, as demonstrated in Appendix~\ref{sec:append_cost_deep}. Each node on KNNG has $k$ neighbors and thus the cost of \texttt{DeepMdiv} is $O(n * d * efSearch * k)$. Each node on the HNSW graph has at most $M_{hnsw}$ edges and thus the cost of \texttt{Deep HNSW} is $O(n * d * efSearch * M_{hnsw})$. 

\begin{algorithm}[t]
\caption{\texttt{DeepSearch}}
\label{alg:deep_search}
\KwIn{$k$, proximity graph $H$ and $efSearch$}
\KwOut{$G$}
\For {each $u$ in $D$} {
	$ep = N_k(u)$ from the initial KNNG\;
	$N_k(u) = Search\_on\_Graph(H, u, k, efSearch, ep)$\;
}
\textbf{return} $G$\;
\end{algorithm}

\subsection{Optimizations}
\label{ssec:opt}

In \texttt{NBPG}, there exist repeated distance calculations for the same pair of points, which increases the computational cost.  In this section, we discuss how to reduce such repeated computations which can be categorized into two sources: \textbf{intra-iteration} and \textbf{inter-iteration}.

\begin{figure}[h]
\begin{center}
\includegraphics[width=0.48\textwidth]{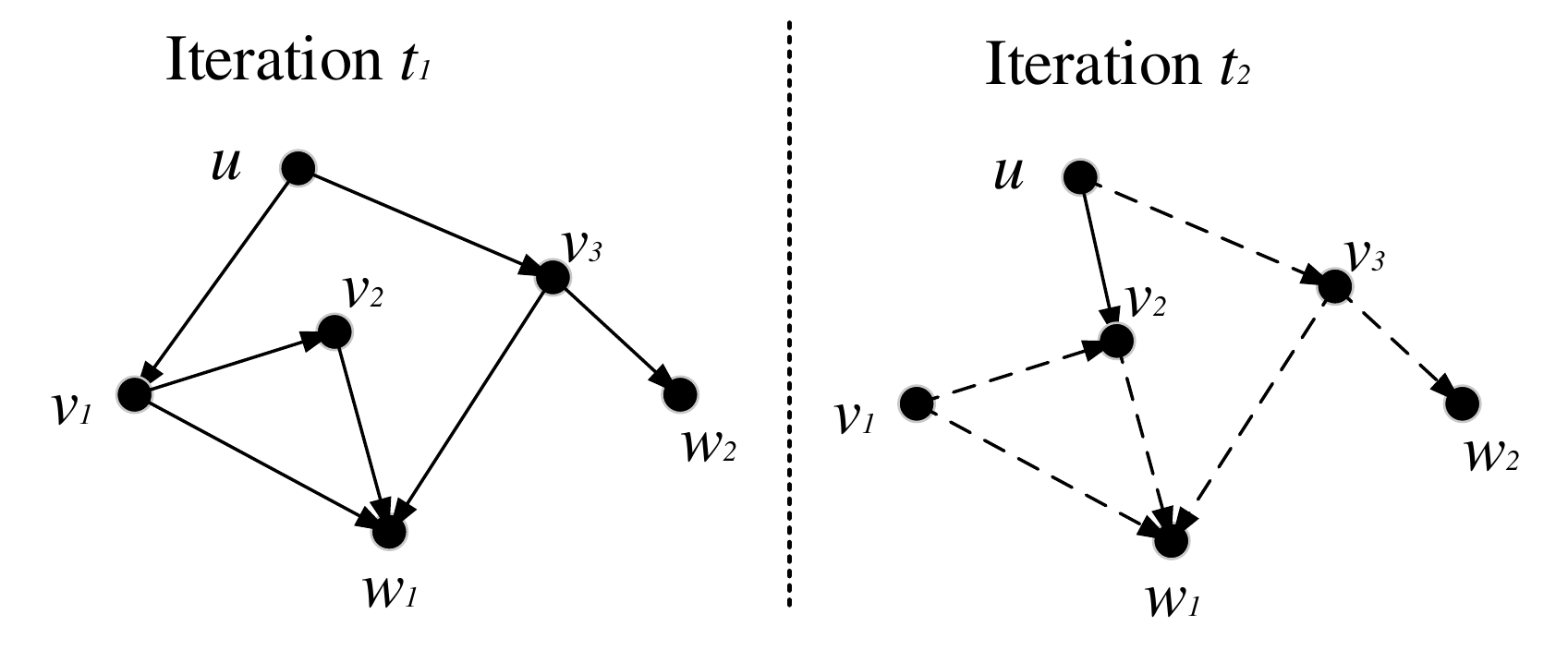}
\end{center}\vspace{-3ex}
\caption{Illustrations of repeated pairs with $u$ as the query in \texttt{UniProp}.} \vspace{-2ex}
\label{fig:rep}
\end{figure}

In the left graph in Figure~\ref{fig:rep}, $w_1$ is a neighbor of both $v_1$ and $v_3$, which are the neighbors of $u$.  In the neighborhood propagation, the pair $(u, w_1)$ will be checked twice in the same iteration.  Thus we call this case intra-iteration repetition.  To avoid the intra-iteration repetition, we use a global filter, a bitmap $visited$ of size $n$ in each iteration.  $visited[u]$ indicates whether $u$ has been visited or not.  Let us take \texttt{UniProp} as an example. Before line 4 in Algorithm~\ref{alg:uni_prop}, we initialize all $visited$ elements as $false$. Then we check $visited[w]$ for candidate $w$ before line 6.  If $w$ is a new candidate, we conduct $update(N_k^t(u), w)$ and mark $visited[w]$ as $true$. Otherwise, $w$ is just ignored.  The global filter is also applicable to \texttt{BiProp} and \texttt{DeepSearch}. 

Inter-iteration repetitions refer to the repeated computation in different iterations.  Consider the two iterations $t_1, t_2$ in Figure \ref{fig:rep} in which the pair $(u, w_2)$ is considered twice.  To avoid the inter-iteration repetition, we can use a local filter.  Let us take \texttt{UniProp} as an example.  For each neighbor $v\in N_k^{t}(u)$, we assign it an attribute $isOld$, which is set as $false$ at start. If $v$ is firstly added to $N_k^t (u)$ as a new neighbor, its $isOld$ is $false$, otherwise it is $true$ as an old neighbor.  In Figure~\ref{fig:rep}, we use the solid arrows to represent new KNN pairs and the dashed arrows for old KNN pairs.  In iteration $t_2$, $w_2$ is an old neighbor of $v_3$ and $v_3$ is an old neighbor of $u$.  Thus the pair $(u, w_2)$ has been considered in a previous iteration and $w_2$ should be ignored by $u$.  Note that the inter-iteration repetition cannot be completely avoided by the local filter. Consider $u$ finds $w_1$ via a new neighbor $v_2$ in iteration $t_2$, which is a repeated computation from iteration $t_1$.  The local filter can also be used for \texttt{BiProp}, but is not necessary for \texttt{DeepSearch} with only one iteration. 

According to our knowledge, the global filter has been widely used in existing studies, while the local filter has not. The filters will accelerate \texttt{NBPG} methods, but will not change the cost analysis in the worst case. On the other hand, they introduce new data structures and require more memory. The space required by the global filter is $O(n)$, while that by the local filter is determined by the neighborhood sizes. To be specific, \texttt{UniProp} and \texttt{NNDes} require $O(n * k)$, while \texttt{KGraph} needs $O(n*L)$. Since the local filter is attached to the neighbors, it will not change the final space complexity.


\subsection{Experimental Evaluation of NBPG}
\label{ssec:exp_nbpg}

We use the same experimental settings in Section~\ref{sssec:settings}. 

\begin{figure*}[!htbp]
\begin{center}
\includegraphics[width=0.75\textwidth, trim=80 342 0 253,clip]{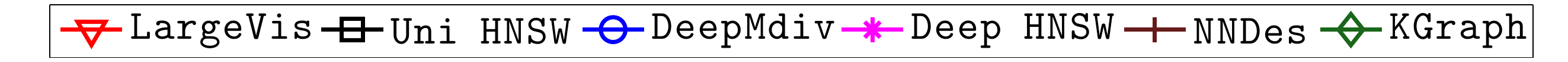}
 \vspace{-1ex}
\\
\subfigure[\textbf{Sift}]
{
\includegraphics[width=0.22\textwidth]{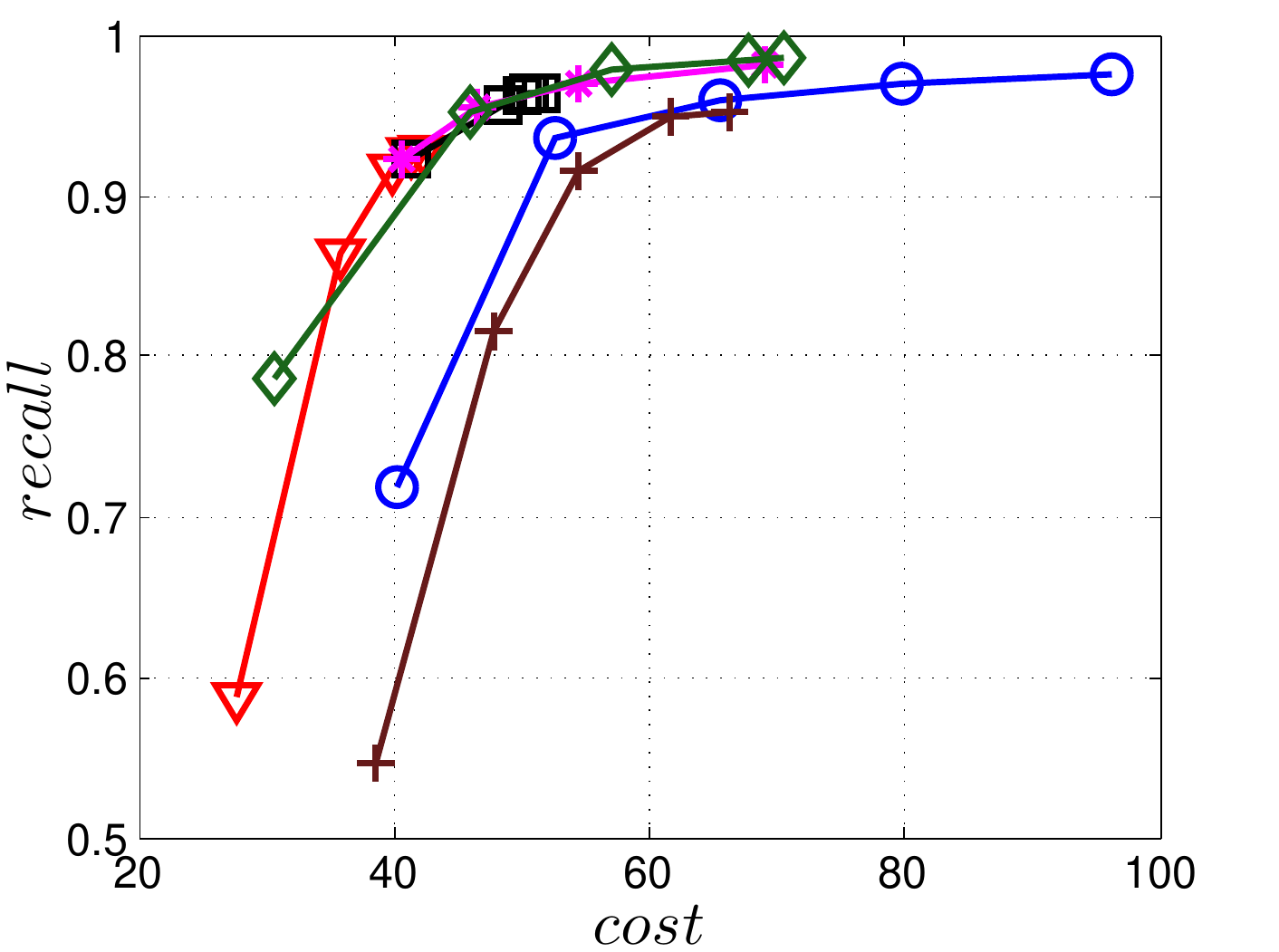}
\label{fig:nbpg_sift}
}
\subfigure[\textbf{Nuscm}]
{
\includegraphics[width=0.22\textwidth]{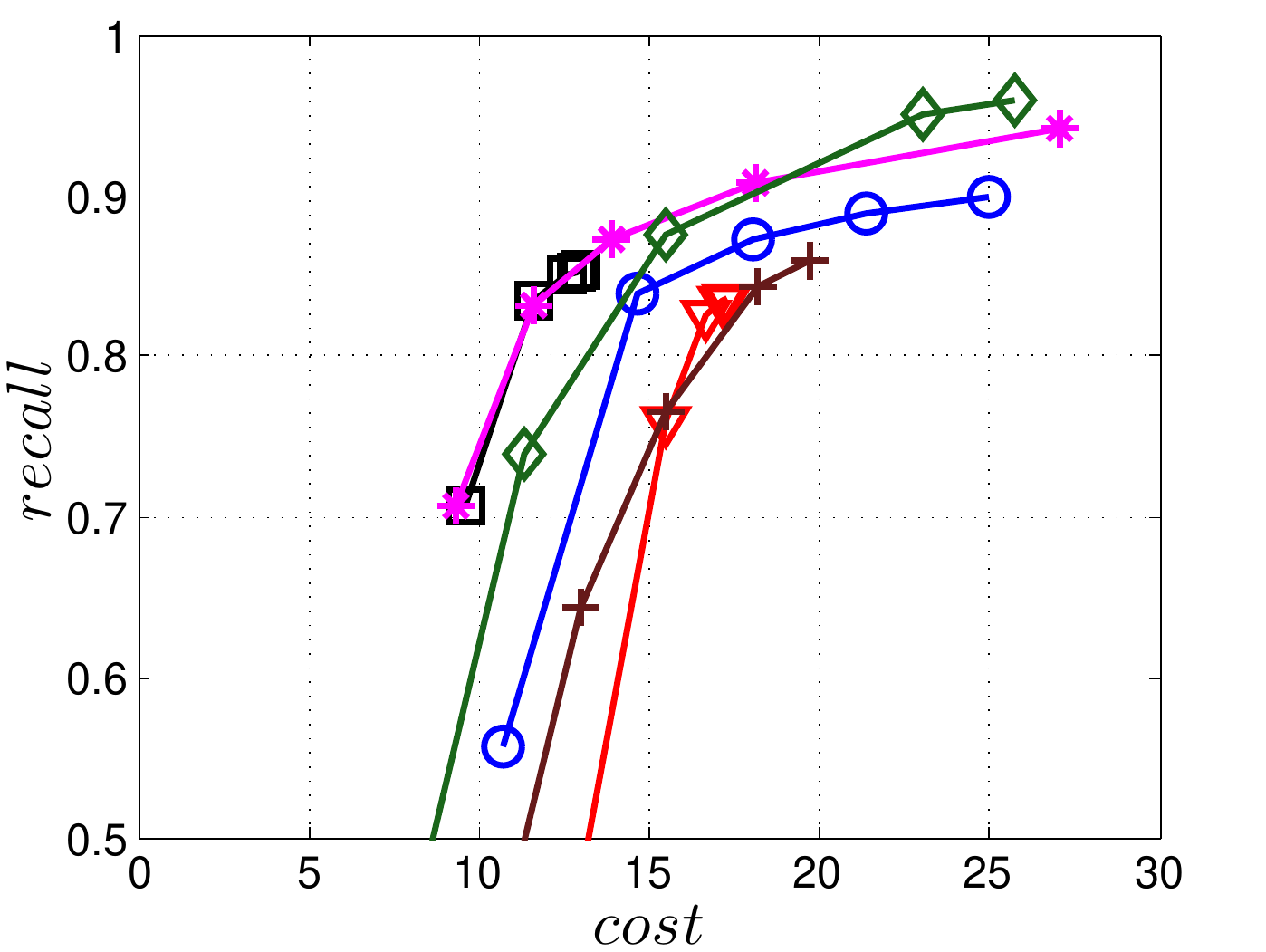}
\label{fig:nbpg_nuscm}
}
\subfigure[\textbf{Nusbow}]
{
\includegraphics[width=0.22\textwidth]{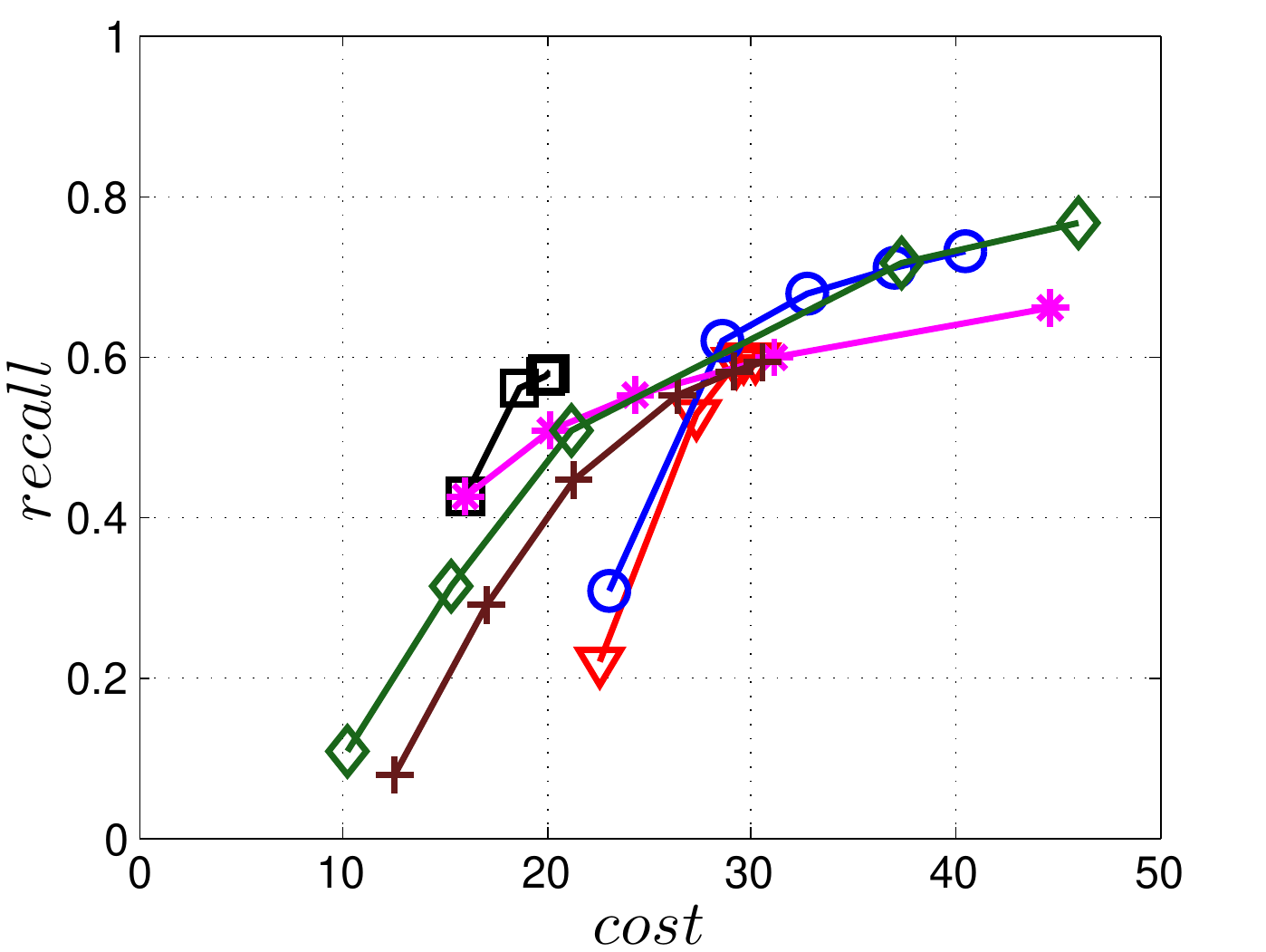}
\label{fig:nbpg_nusbow}
}
\subfigure[\textbf{Gist}]
{
\includegraphics[width=0.22\textwidth]{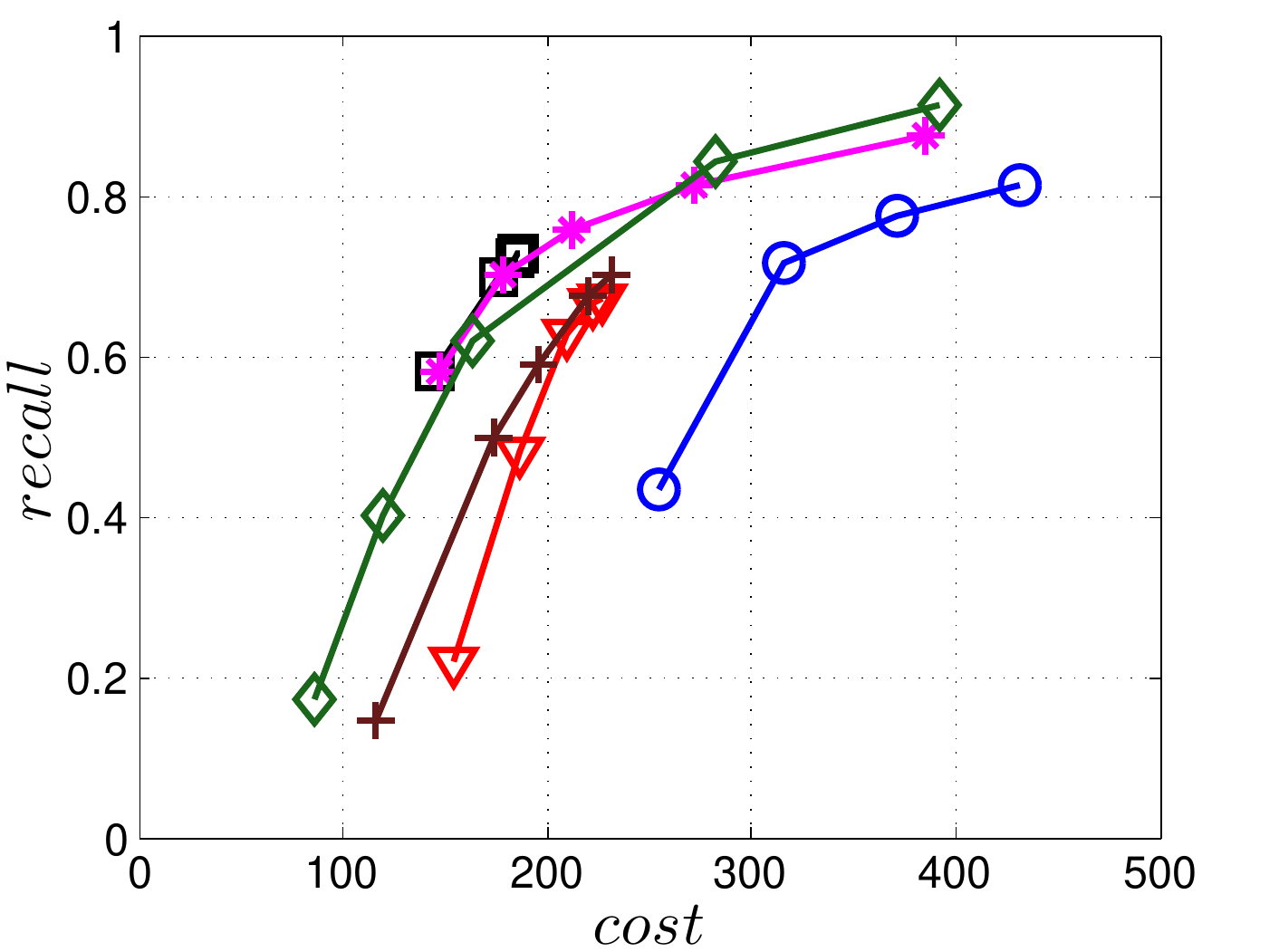}
\label{fig:nbpg_gist}
}
\subfigure[\textbf{Glove}]
{
\includegraphics[width=0.22\textwidth]{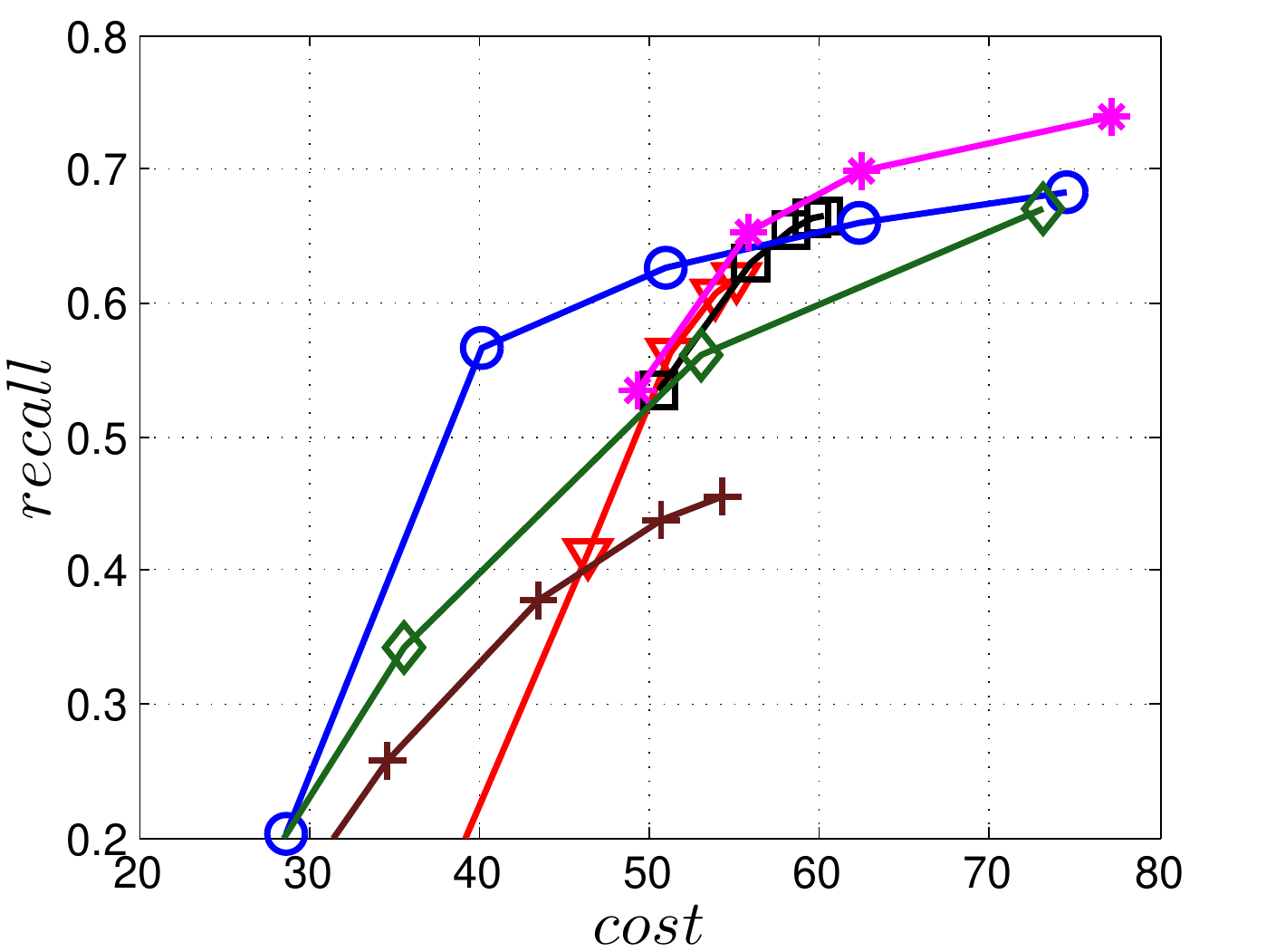}
\label{fig:nbpg_glove}
}
\subfigure[\textbf{Msdrp}]
{
\includegraphics[width=0.22\textwidth]{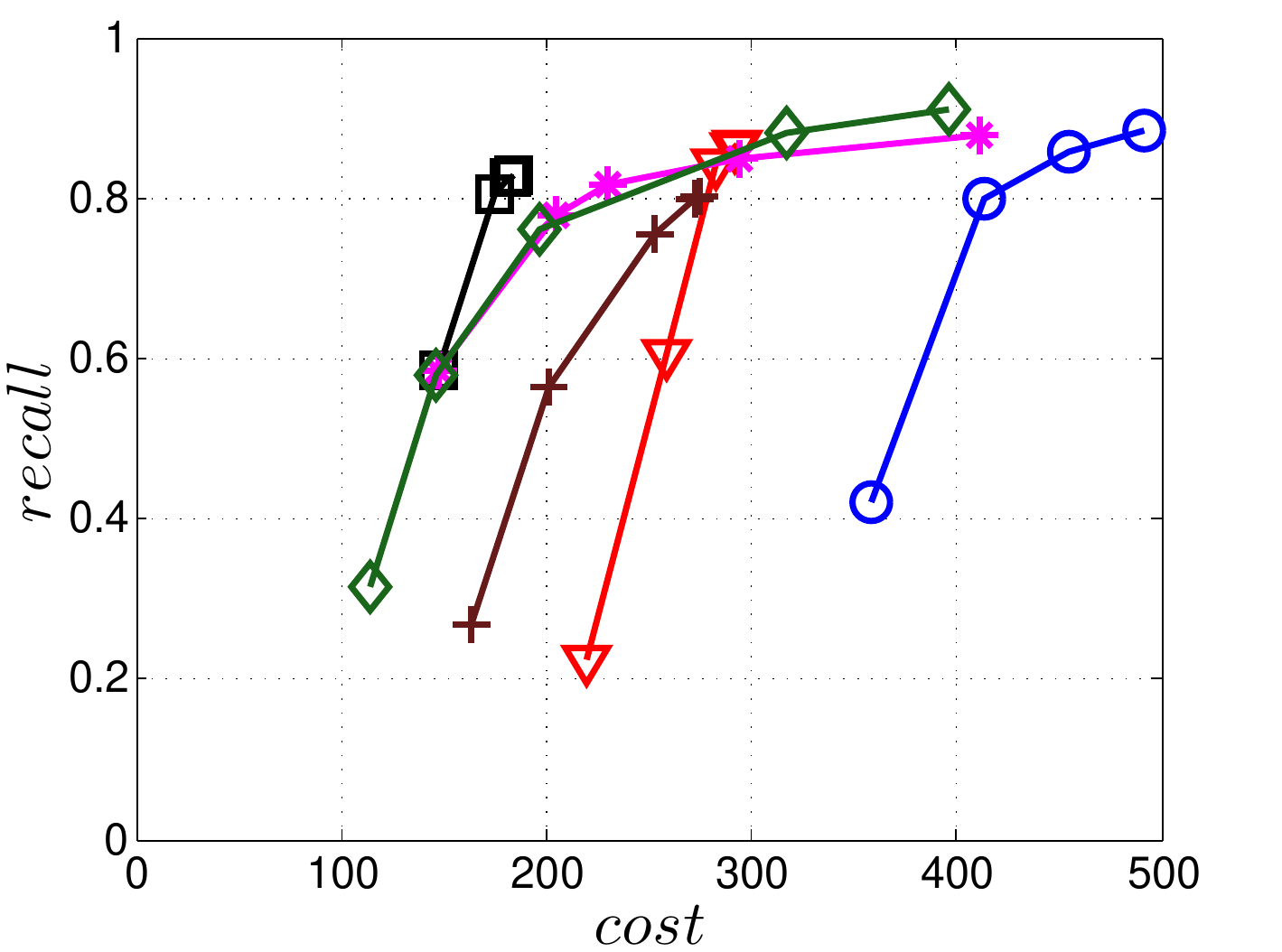}
\label{fig:nbpg_msdrp}
}
\subfigure[\textbf{Msdtrh}]
{
\includegraphics[width=0.22\textwidth]{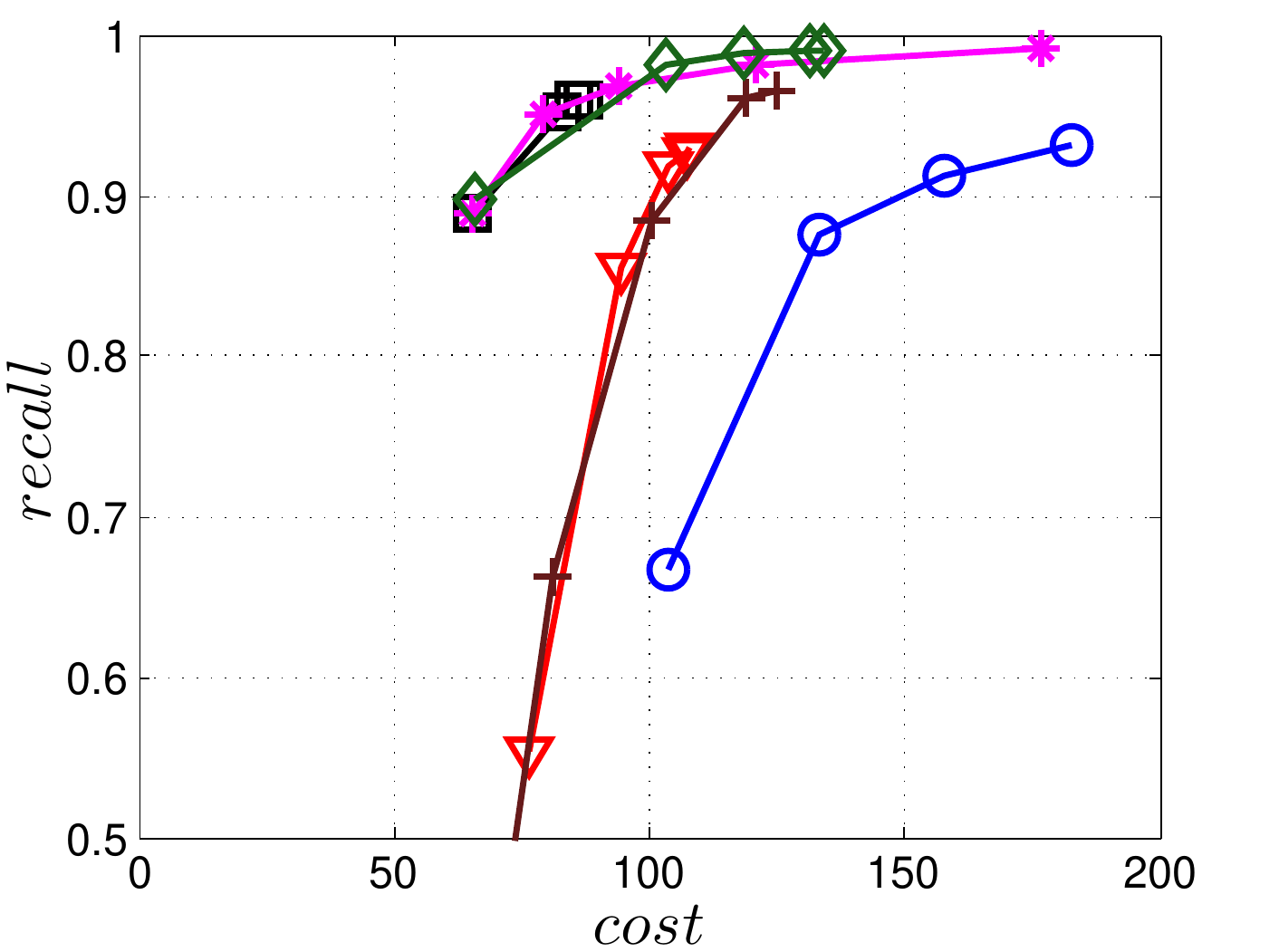}
\label{fig:nbpg_msdtrh}
}
\subfigure[\textbf{Uqv}]
{
\includegraphics[width=0.22\textwidth]{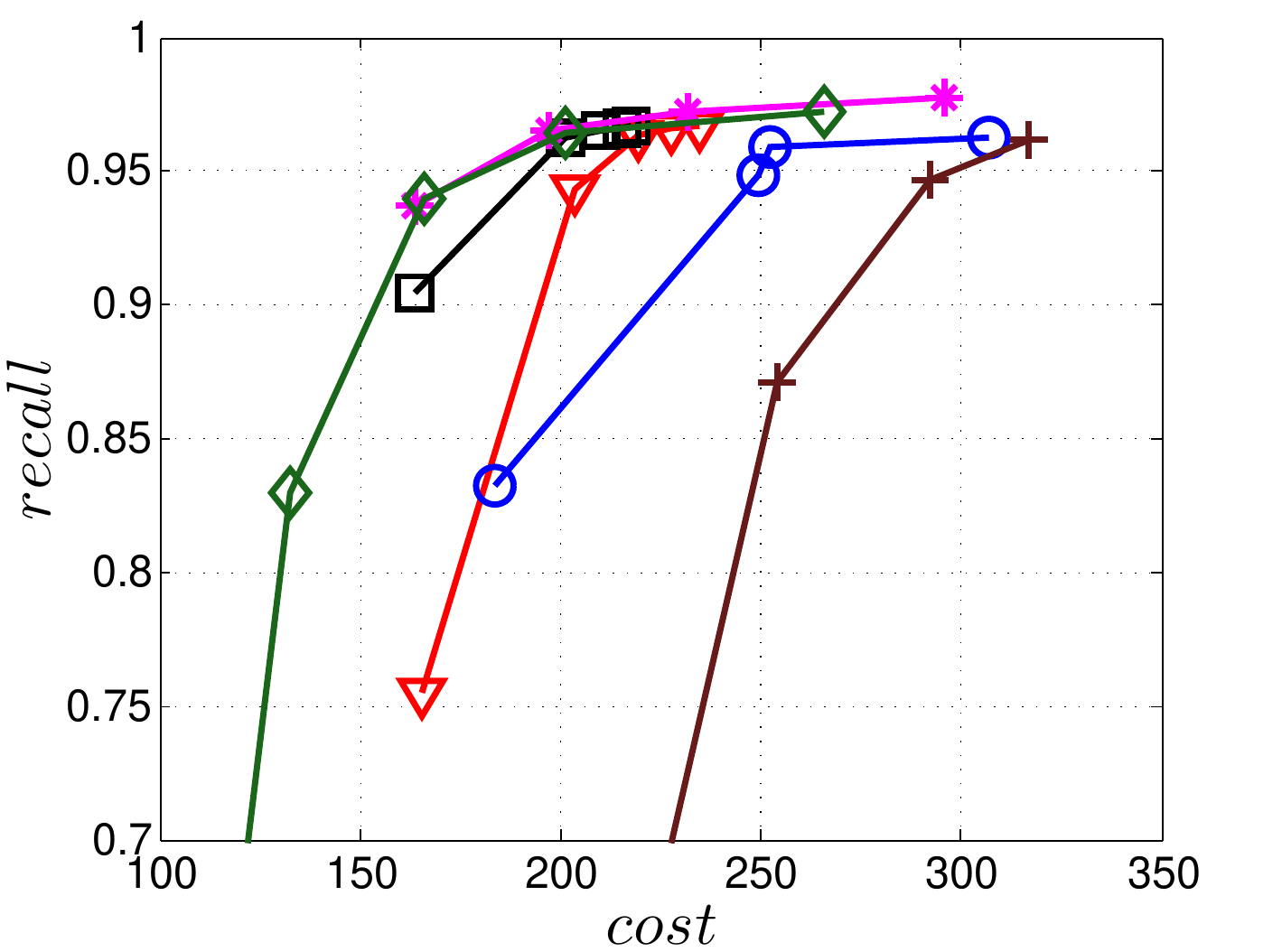}
\label{fig:nbpg_uqv}
}
\end{center} \vspace{-3ex}
\caption{Comparing \texttt{NBPG} Methods. $cost$ is measured in seconds.} \vspace{-2ex}
\label{fig:nbpg}
\end{figure*}

\subsubsection{Compared Methods and Parameter Selection}
\label{sssec:method_param}

We compare 6 \texttt{NBPG} methods as listed in the following and carefully optimize their parameters for their best performance. Following the framework, \texttt{NBPG} is also combined with our first choice \texttt{HNSW KNNG} in \texttt{INIT}, i.e.,  where $efConstruction = 80$ and $M_{hnsw}=20$ by default. We just list the settings of key parameters as follows. 


\begin{itemize}
\item \texttt{UniProp}: \texttt{LargeVis} \cite{TangJ2016WWW}~\footnote{We improve the codes from the authors by adding the local filter to reduce inter-iteration repeated distance computations in \texttt{UniProp}.} and \texttt{Uni HNSW}. \texttt{LargeVis} takes 50 RP trees to build the initial KNNG, while \texttt{Uni HNSW} uses \texttt{HNSW KNNG}. The number $nIter$ of iterations is varied from 0 to 4.

\item \texttt{BiProp}: \texttt{NNDes} \cite{Dong2011WWW} and \texttt{KGraph} \cite{Dong2011WWW}.  The number $nIter$ of iterations is set between 0 and 16. 

\item \texttt{DeepSearch}: \texttt{DeepMdiv} \cite{WangJD2012CVPR} and \texttt{Deep HNSW}. $efSearch$ in \texttt{DeepSearch} is set between 0 and 160.  The number $L_{div}$ of random divisions is 10 in \texttt{DeepMdiv} .
\end{itemize}

\subsubsection{Experimental Results}
\label{sssec:nbpg_er}

The experimental results of $cost$ vs $recall$ are plotted in Figure~\ref{fig:nbpg}.  We first compare the methods within each category.  For \texttt{UniProp}, we can see that \texttt{Uni HNSW} obviously outperforms \texttt{LargeVis} in most cases, because \texttt{Uni HNSW} performs \texttt{UniProp} based on a better initial KNNG generated by \texttt{HNSW KNNG}.  Notably, both methods have significant improvement in the first two iterations but rapidly converge afterwards.  This is because most KNNs have become old neighbors in subsequent iterations and fewer candidates will be found for KNN refinement. For \texttt{BiProp}, \texttt{KGraph} obviously defeats \texttt{NNDes}.  This is because \texttt{NNDes} fetches neighbors from a smaller neighbor pool in \texttt{NBPG} than \texttt{KGraph}.  Moreover, \texttt{KGraph} carefully determines the neighborhood size $m^t(u)$ and limits the size of $R_m^t(u)$ for each $u$ in each iteration, to balance efficiency and accuracy.  For \texttt{DeepSearch}, \texttt{Deep HNSW} is much better than \texttt{DeepMdiv} on all data sets except Nusbow, due to two reasons. First, \texttt{HNSW KNNG} outperforms \texttt{Multiple Division} in \texttt{INIT}.  Second, the HNSW graph has much better performance on ANN search than KNNG, as demonstrated in recent benchmarks ~\cite{Nmslib, Annoy, Li2019TKDE}.

We list the memory requirement of auxiliary data structures of \texttt{NBPG} methods in Table~\ref{tb:memory_nbpg}. We can see that \texttt{KGraph} requires the most memory for auxiliary structures, followed by \texttt{NNDes}. On the data with relatively small dimensions, such as Sift, Nuscm and Glove, \texttt{KGraph} even needs more space than the data itself. This is because \texttt{KGraph} maintains a large neighbor pool for each node and stores reverse neighbors. The space requirement of \texttt{UniProp} is mainly caused by the last-iteration KNNG, while \texttt{Deep HNSW} and \texttt{DeepMdiv} are due to their proximity graphs.

\begin{table}[t]
\centering
\caption{Memory Requirement (MB) of Auxiliary Structures for \texttt{NBPG} Methods}
    \begin{tabular}{|l|m{1cm}|m{0.8cm}|m{1cm}|m{0.6cm}|m{0.6cm}|}
    	\hline
    	Data & \texttt{UniProp} & \texttt{NNDes} & \texttt{KGraph} & \texttt{Deep HNSW} & \texttt{Deep Mdiv}\\
    	\hline
    	Sift & 97 & 332 & 1,158 & 181 & 78 \\
    	\hline
    	Nuscm & 26 & 90 & 309 & 49 & 21 \\
    	\hline
    	Nusbow & 26 & 90 & 318 & 49 & 21 \\
    	\hline 
    	Gist & 97 & 332 & 1,177 & 180 & 78 \\
    	\hline 
    	Glvoe & 115 & 396 & 1,346 & 216 & 93 \\
    	\hline
    	Msdrp & 96 & 330 & 1,101 & 180 & 77 \\
    	\hline
    	Msdtrh & 96 & 330 & 1,116 & 180 & 77 \\
    	\hline
    	Uqv & 319 & 1,098 & 3,644 & 657 & 256 \\
    	\hline
    \end{tabular}
\label{tb:memory_nbpg}
\end{table}

Let us consider and compare all three categories of \texttt{NBPG} together. Overall, There is no denominator in all cases.  Even with small memory requirement, \texttt{UniProp} usually cannot achieve high accuracy due to its fast convergence.   \texttt{KGraph} is the best choice for a high-recall KNNG, but not for a moderate-recall KNNG. Besides, \texttt{KGraph} requires obviously more memory and thus is not suitable when the memory is not enough. We can see that \texttt{Deep HNSW} is the most balanced method, considering efficiency, accuracy and memory requirement simultaneously.



\begin{figure}[h]
\begin{center}
\subfigure[\textbf{Nuscm}]
{
\includegraphics[width=0.22\textwidth]{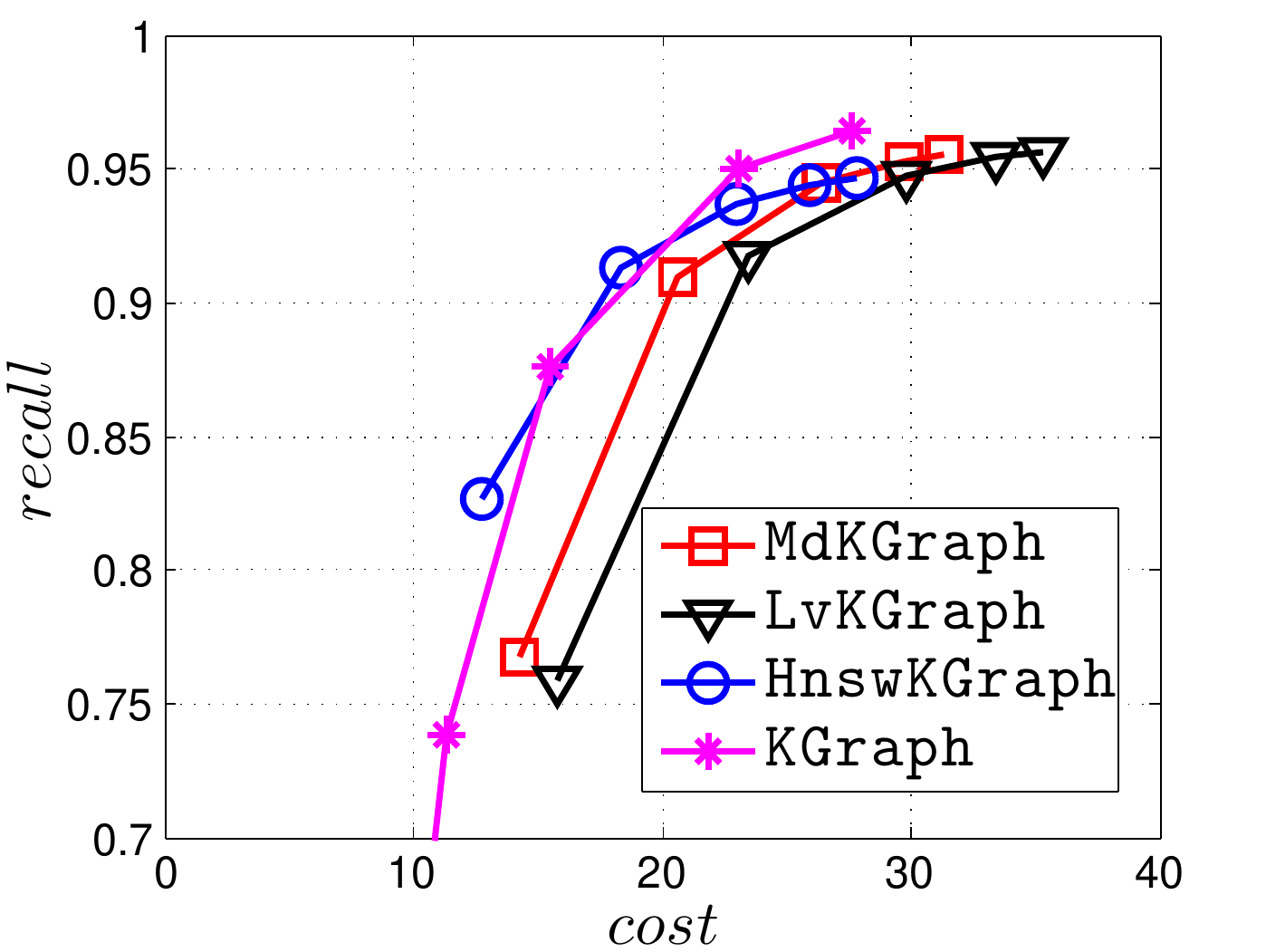}
\label{fig:influ_kgraph_nuscm}
}
\subfigure[\textbf{Msdrp}]
{
\includegraphics[width=0.22\textwidth]{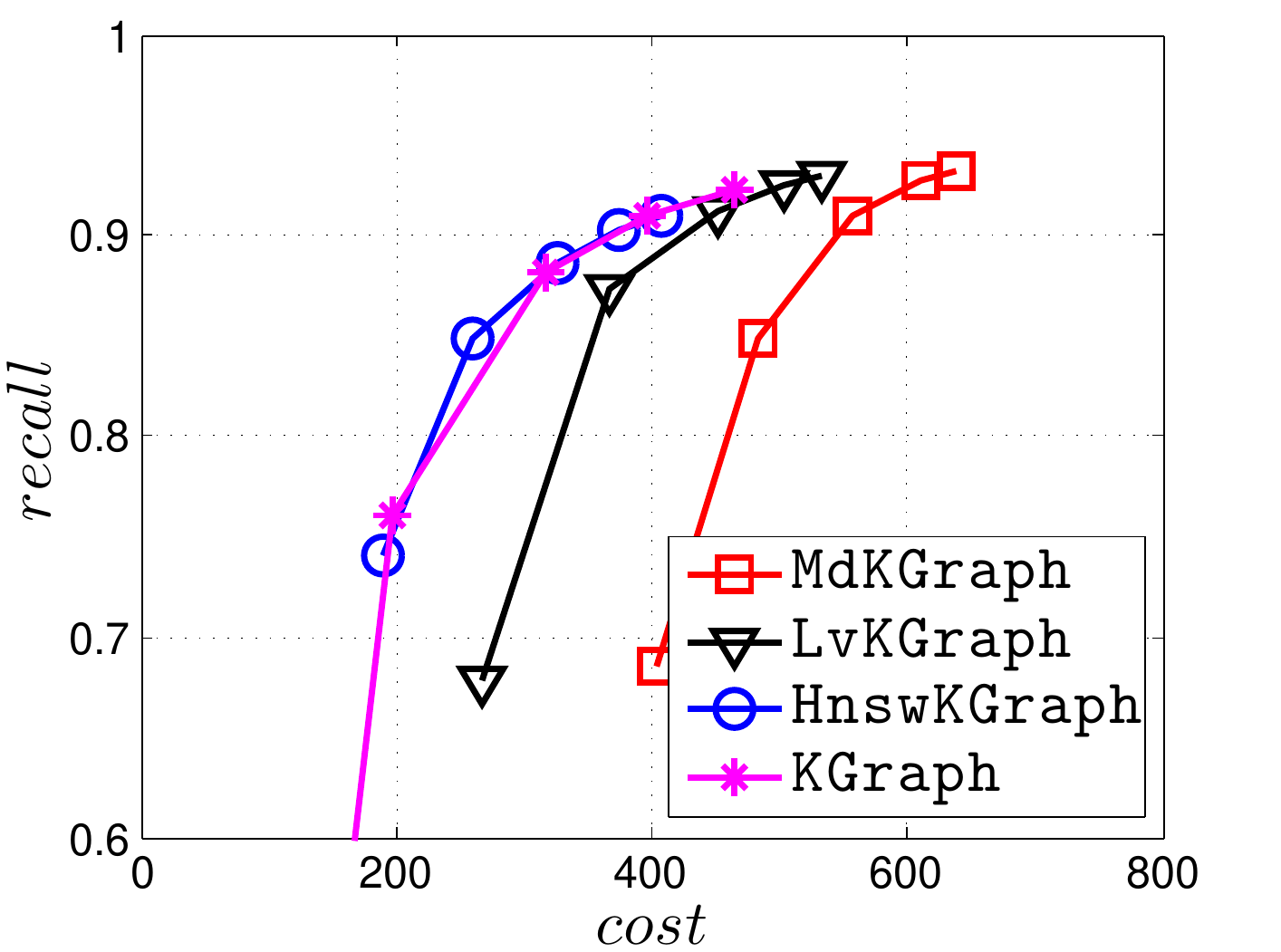}
\label{fig:influ_kgraph_msdrp}
}
\end{center}\vspace{-3ex}
\caption{Performance of \texttt{KGraph} with various initial KNNGs. $cost$ is measured in seconds.}
\label{fig:influ_kgraph} \vspace{-2ex}
\end{figure}

\begin{figure}[h]
\begin{center}
\subfigure[\textbf{Sift100M}]
{
\includegraphics[width=0.22\textwidth]{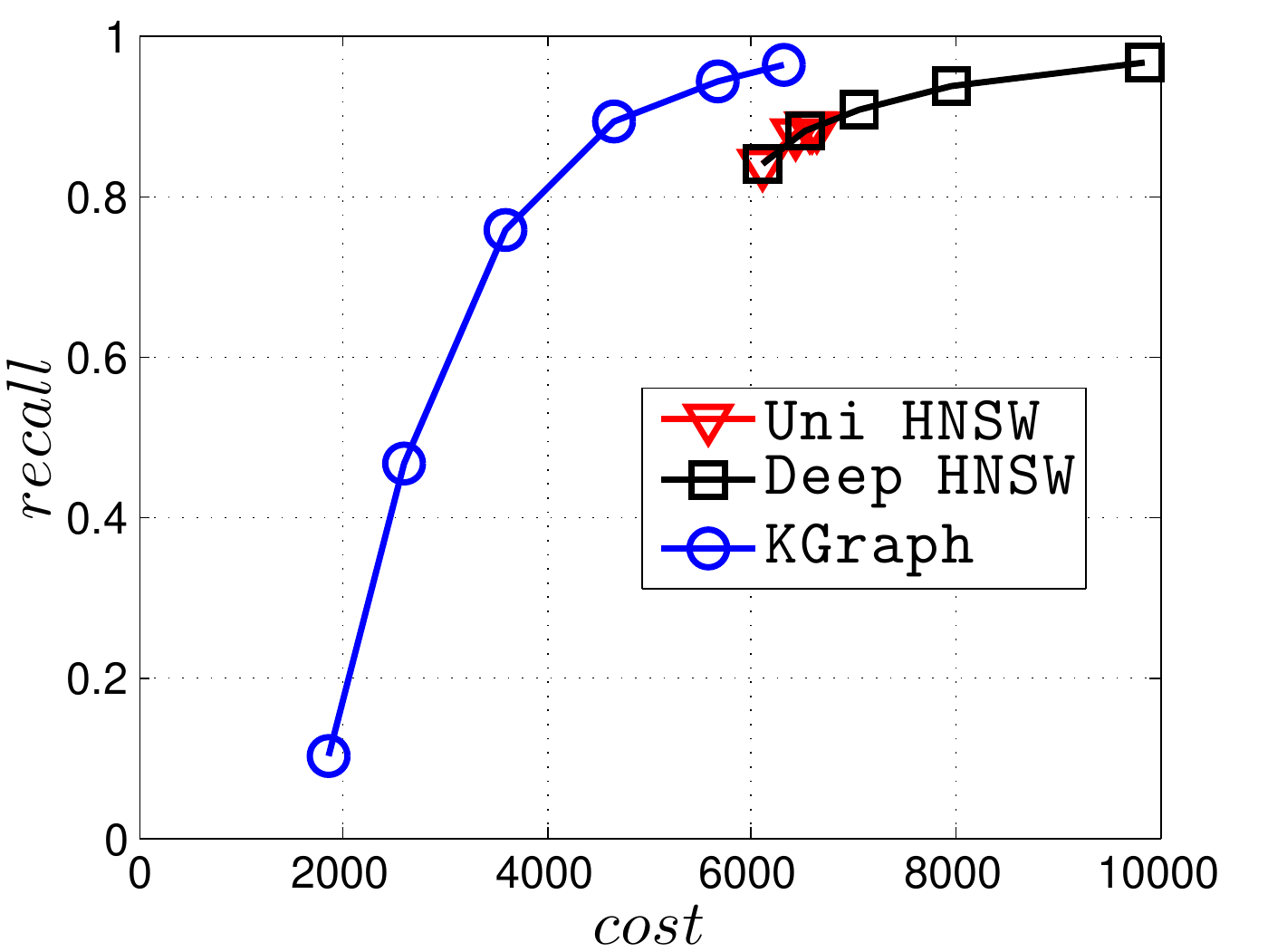}
\label{fig:large_sift100m}
}
\subfigure[\textbf{Deep100M}]
{
\includegraphics[width=0.22\textwidth]{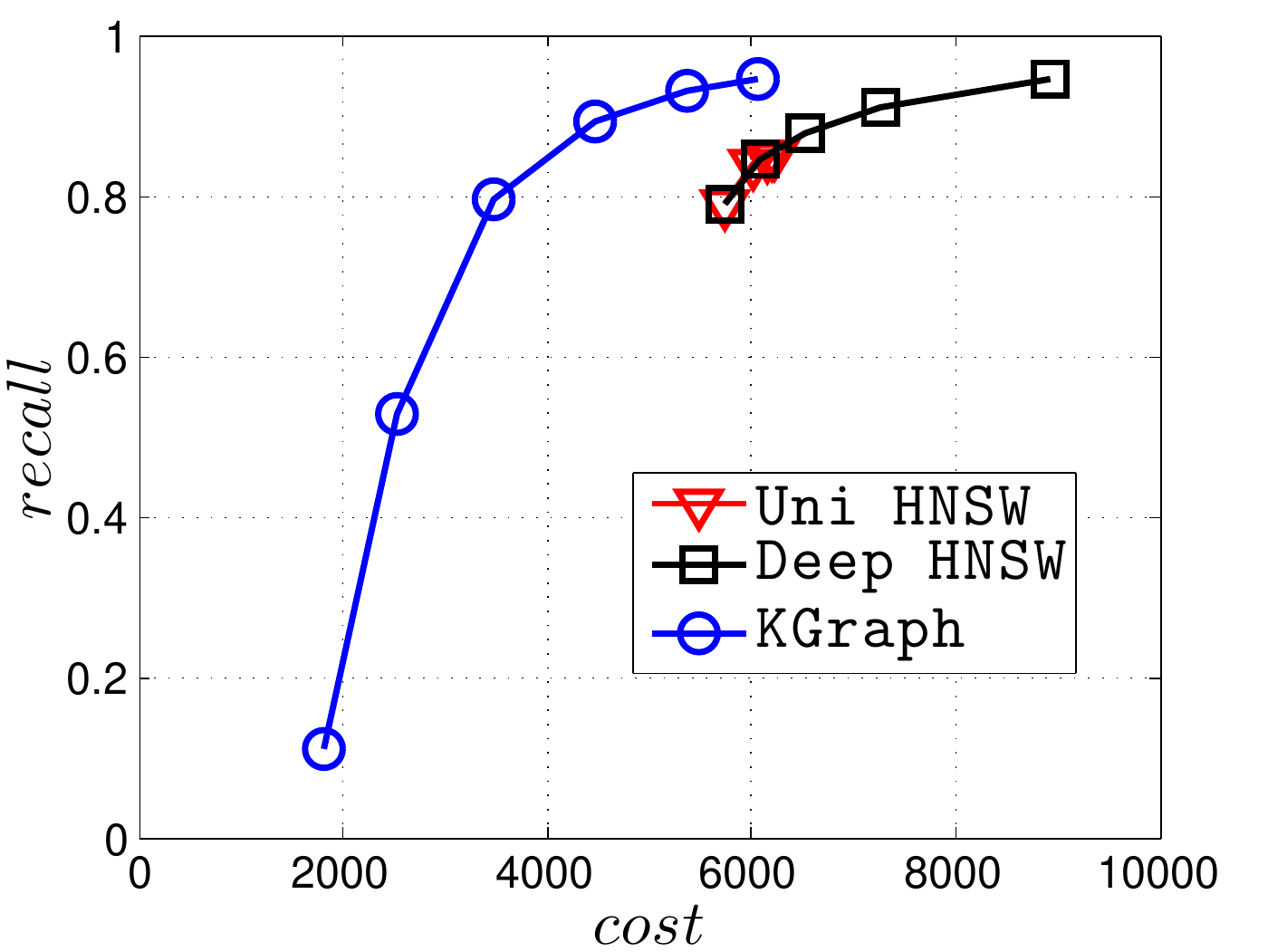}
\label{fig:large_deep100m}
}
\end{center}\vspace{-3ex}
\caption{\texttt{NBPG} Performance on large data, with $cost$ in seconds.} \vspace{-2ex}
\label{fig:large}
\end{figure}

\subsubsection{KGraph with an Initial KNNG}

\texttt{KGraph} demonstrates impressive performance in the above experiments, but it starts from a random initial KNNG.  Thus it is interesting to study the following question: \emph{How well can \texttt{KGraph} perform if we provide a good-quality initial KNNG?}  To answer this question, we create new combinations, \texttt{LvKGraph}, \texttt{MdKGraph} and \texttt{HnswKGraph} with three different initial KNNGs (i.e., \texttt{LargeVis}, \texttt{Multiple Division} and \texttt{HNSW KNNG} respectively) and then perform \texttt{KGraph} style of neighborhood propagation on top of these initial KNNGs.  We show the results in Figure~\ref{fig:influ_kgraph}.

\begin{figure*}[h]
\begin{center}
\includegraphics[width=0.75\textwidth, trim=50 287 45 155,clip]{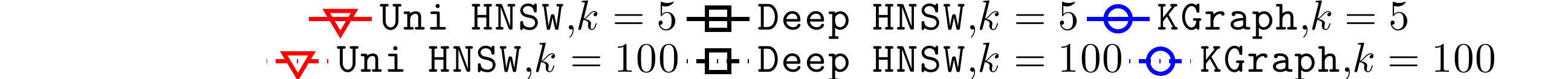}
\vspace{-1ex}
\\
\subfigure[\textbf{Sift}]
{
\includegraphics[width=0.22\textwidth]{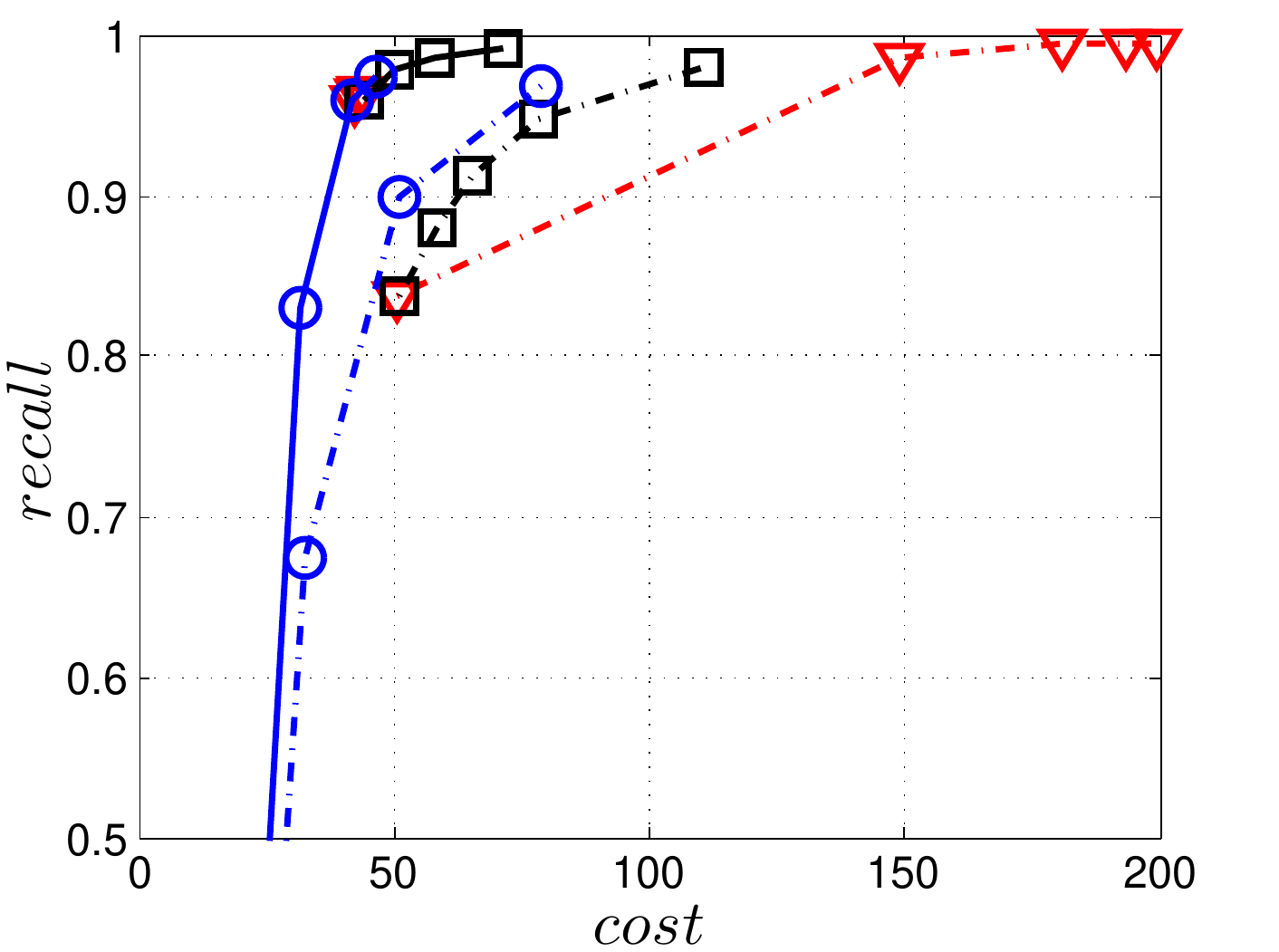}
\label{fig:kinflu_sift}
}
\subfigure[\textbf{Nuscm}]
{
\includegraphics[width=0.22\textwidth]{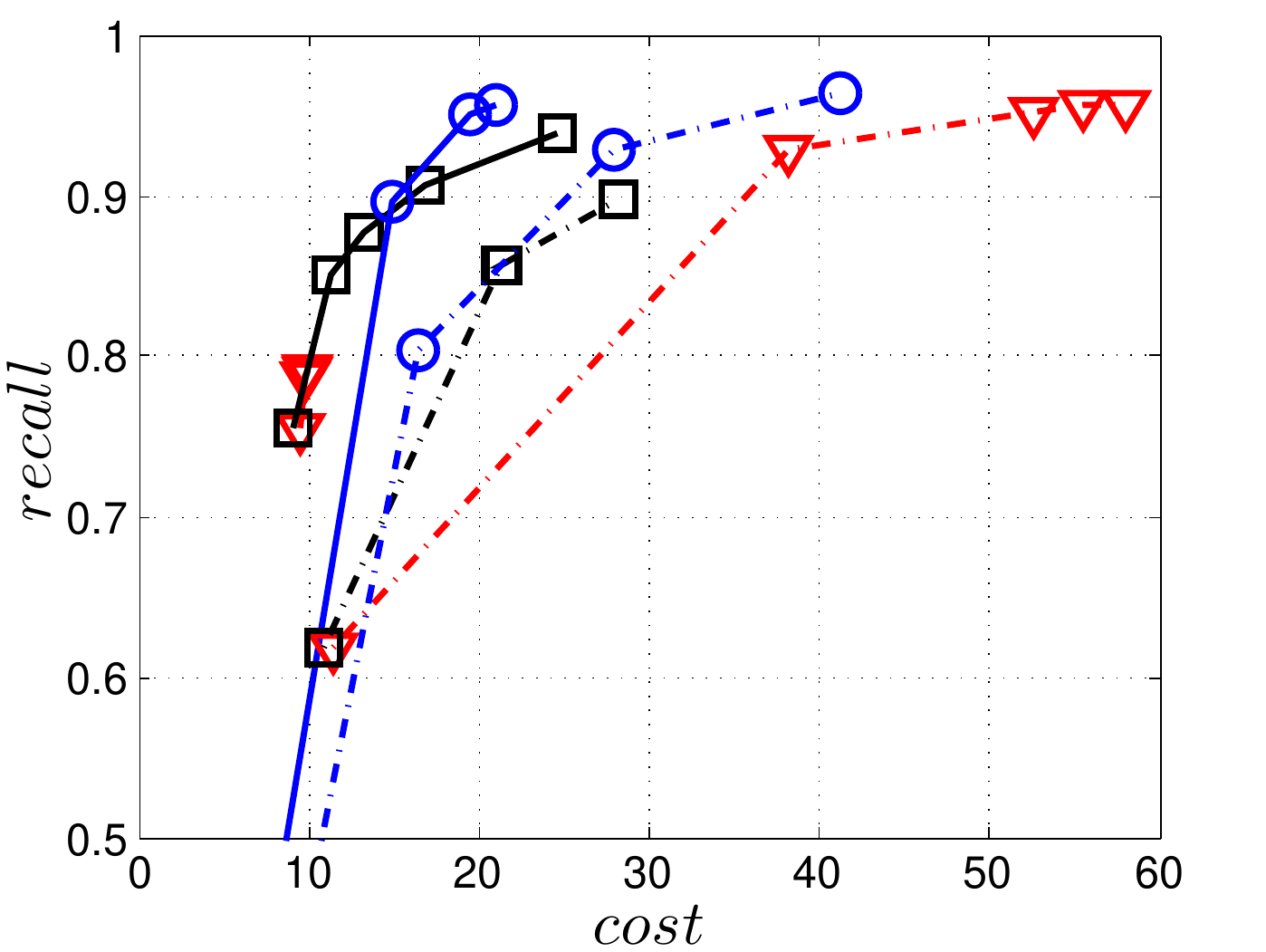}
\label{fig:kinflu_nuscm}
}
\subfigure[\textbf{Nusbow}]
{
\includegraphics[width=0.22\textwidth]{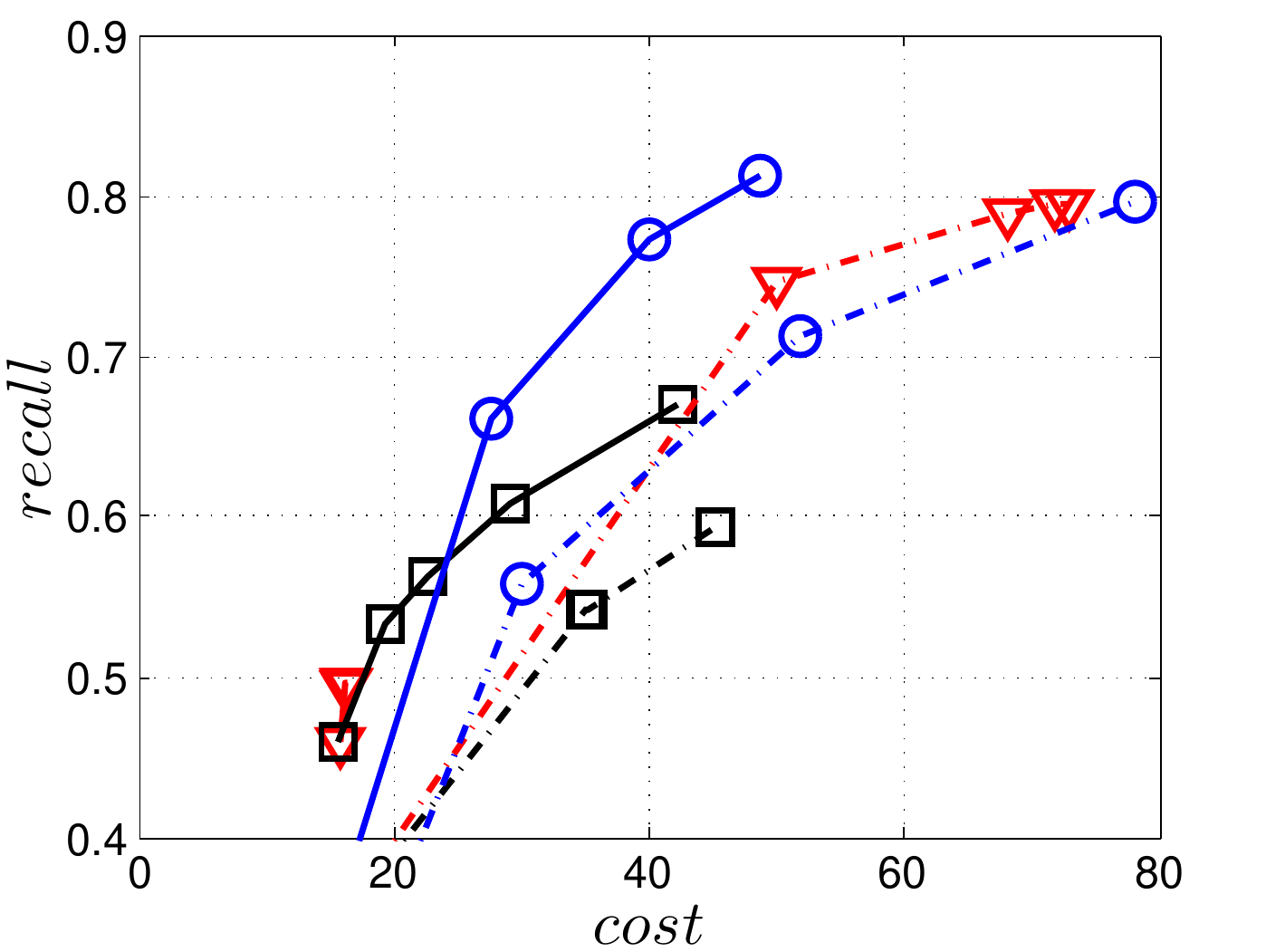}
\label{fig:kinflu_nusbow}
}
\subfigure[\textbf{Gist}]
{
\includegraphics[width=0.22\textwidth]{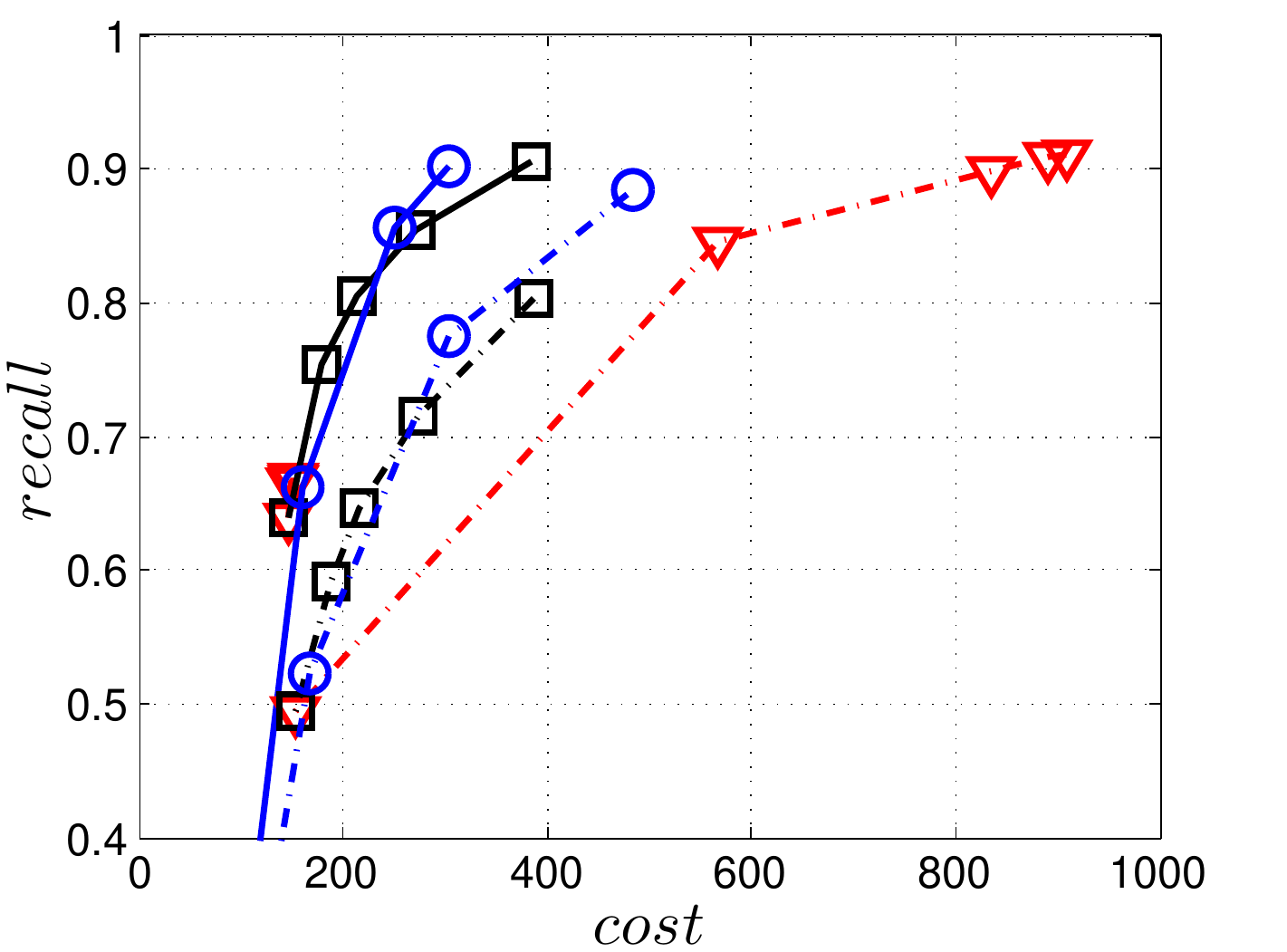}
\label{fig:kinflu_gist}
}
\subfigure[\textbf{Glove}]
{
\includegraphics[width=0.22\textwidth]{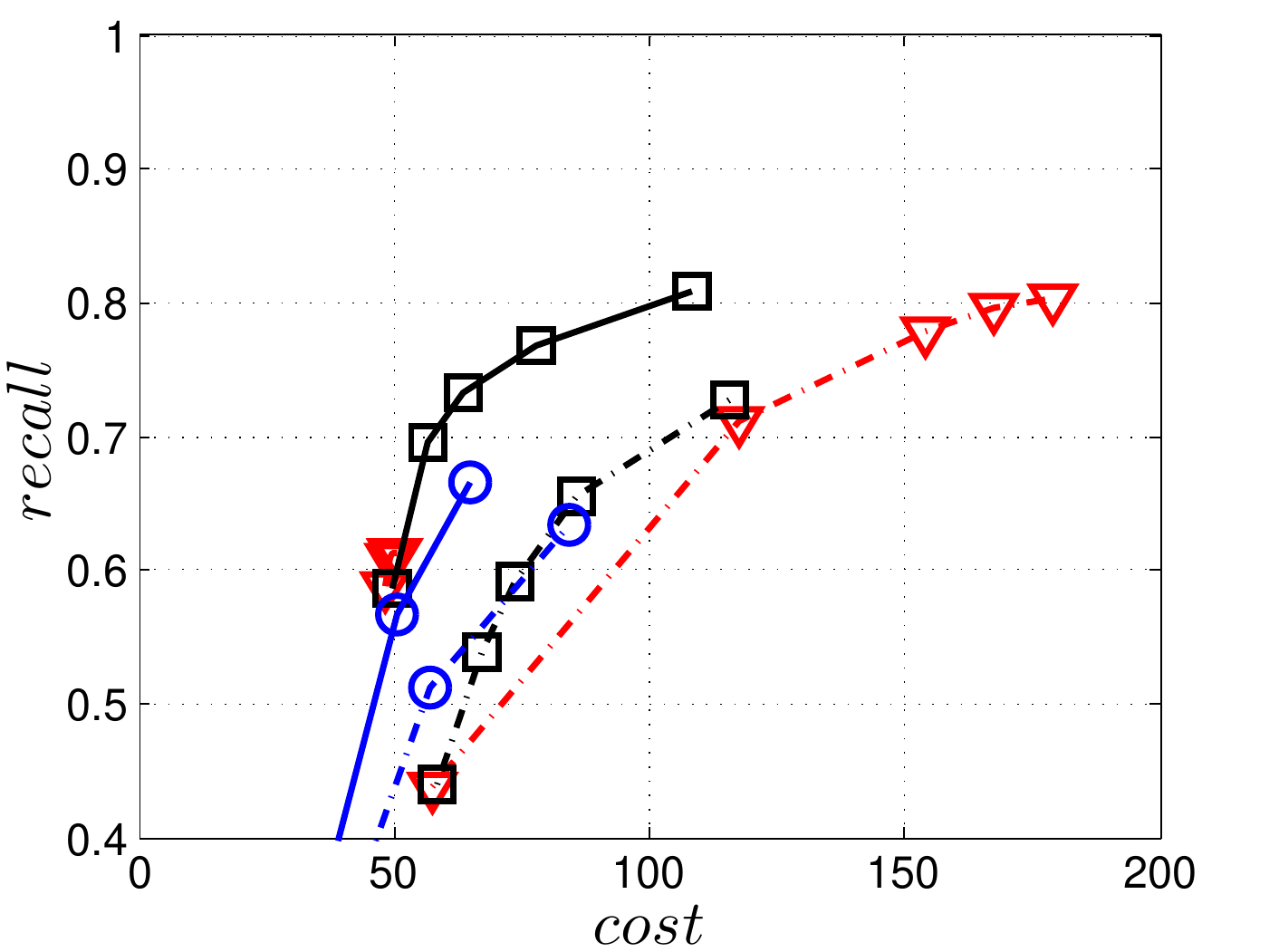}
\label{fig:kinflu_glove}
}
\subfigure[\textbf{Msdrp}]
{
\includegraphics[width=0.22\textwidth]{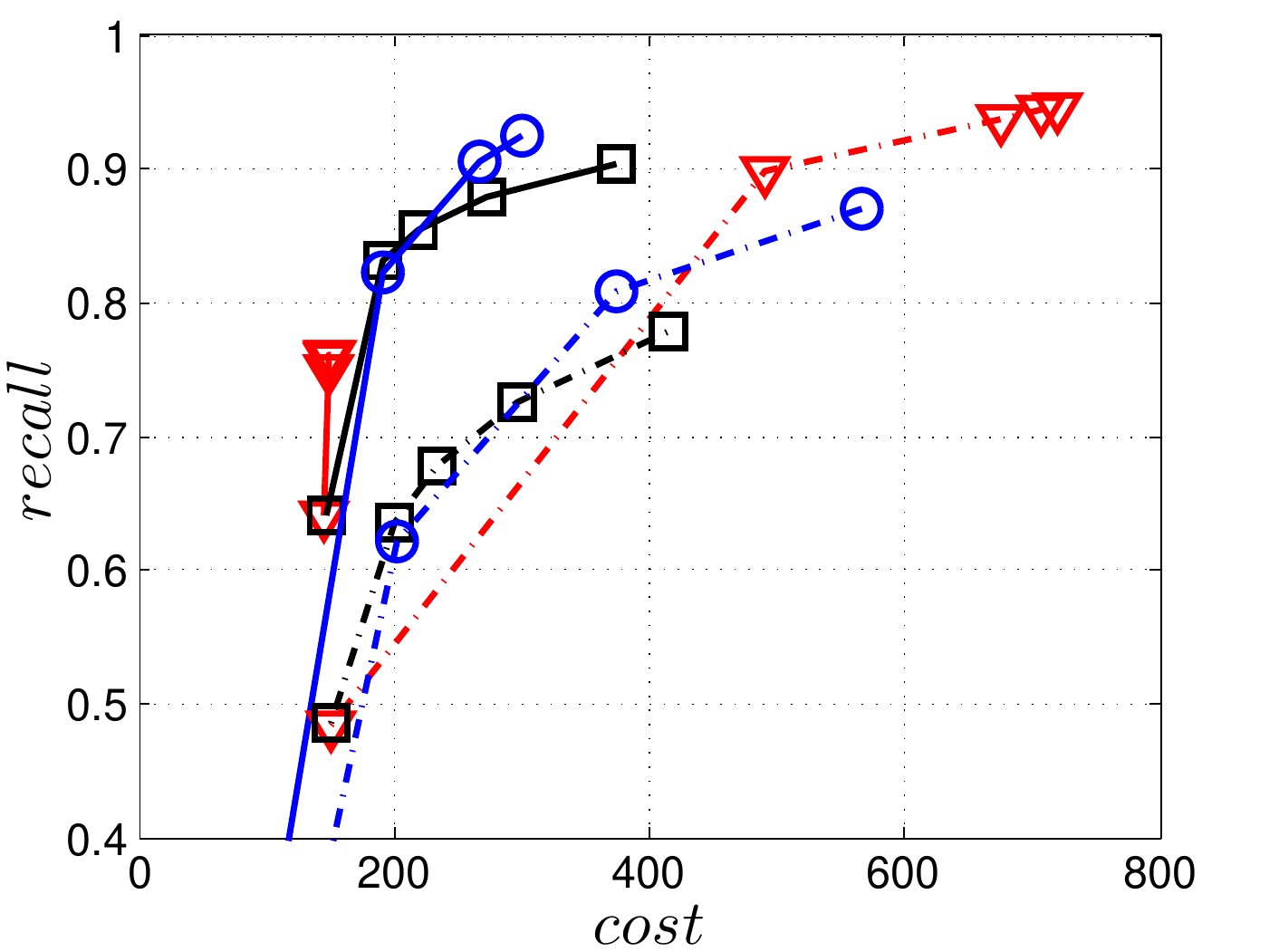}
\label{fig:kinflu_msdrp}
}
\subfigure[\textbf{Msdtrh}]
{
\includegraphics[width=0.22\textwidth]{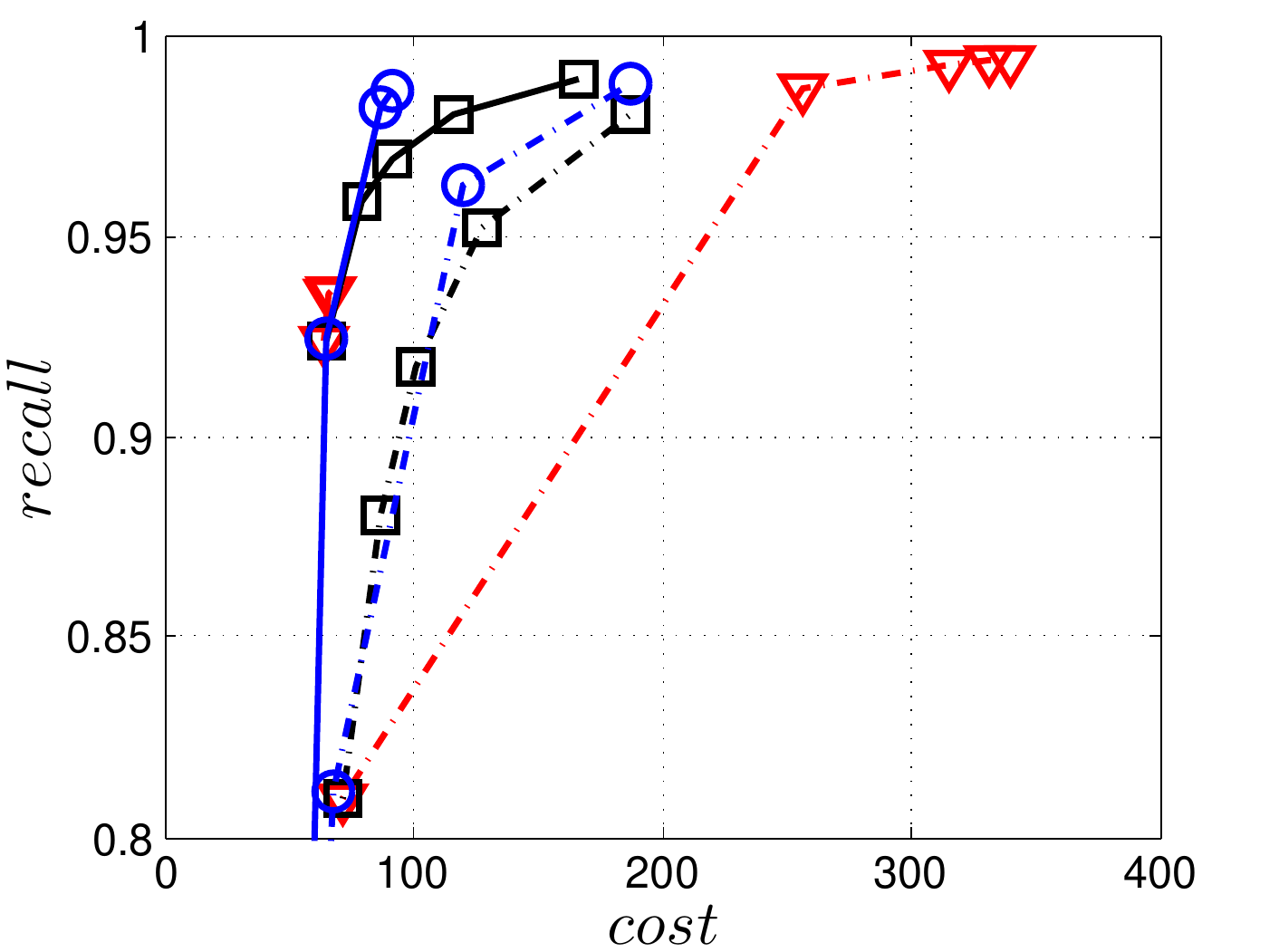}
\label{fig:kinflu_msdtrh}
}
\subfigure[\textbf{Uqv}]
{
\includegraphics[width=0.22\textwidth]{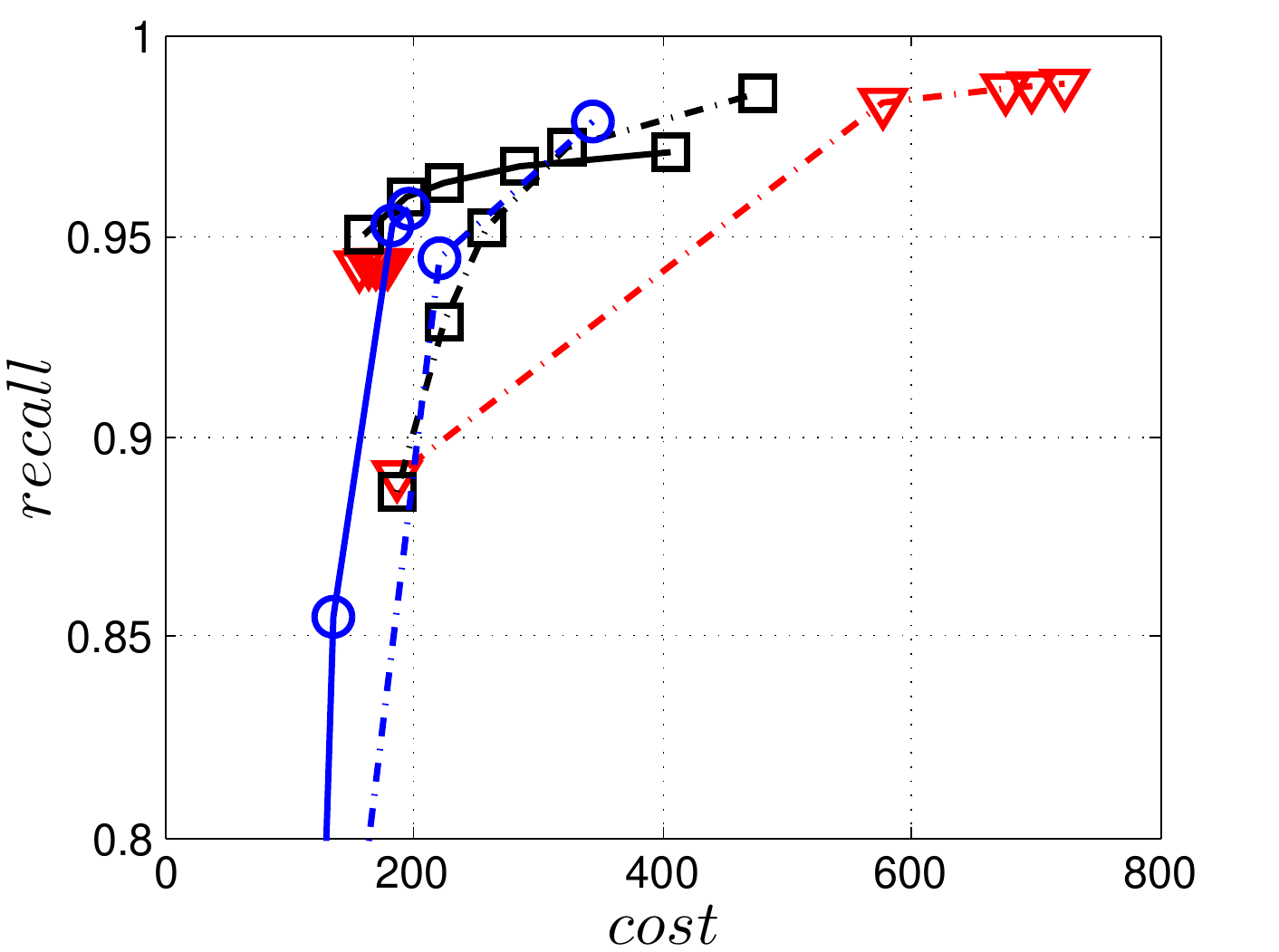}
\label{fig:kinflu_uqv}
}
\end{center}\vspace{-3ex}
\caption{The influences of $k$ on \texttt{NBPG} methods. $cost$ is measured in seconds.}
\label{fig:kinflu} \vspace{-2ex}
\end{figure*}

\eat{
\begin{figure*}[t]
\begin{center}
\includegraphics[width=0.75\textwidth, trim=40 392 45 195,clip]{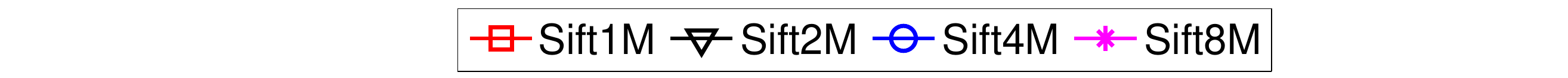}
\vspace{-1ex}
\\
\subfigure[\texttt{Uni HNSW}]
{
\includegraphics[width=0.22\textwidth]{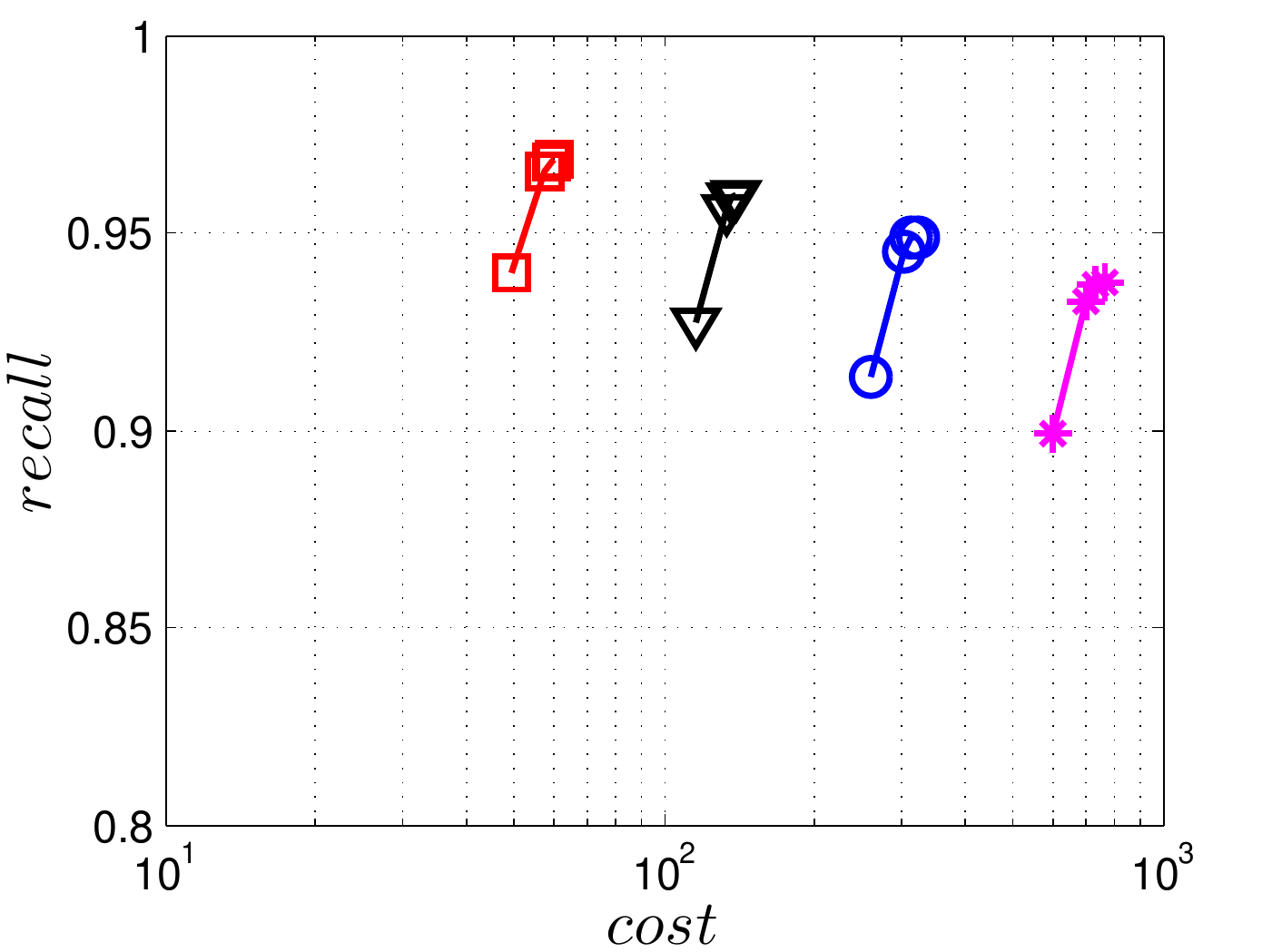}
\label{fig:scale_sift_dirhnsw}
}
\subfigure[\texttt{KGraph}]
{
\includegraphics[width=0.22\textwidth]{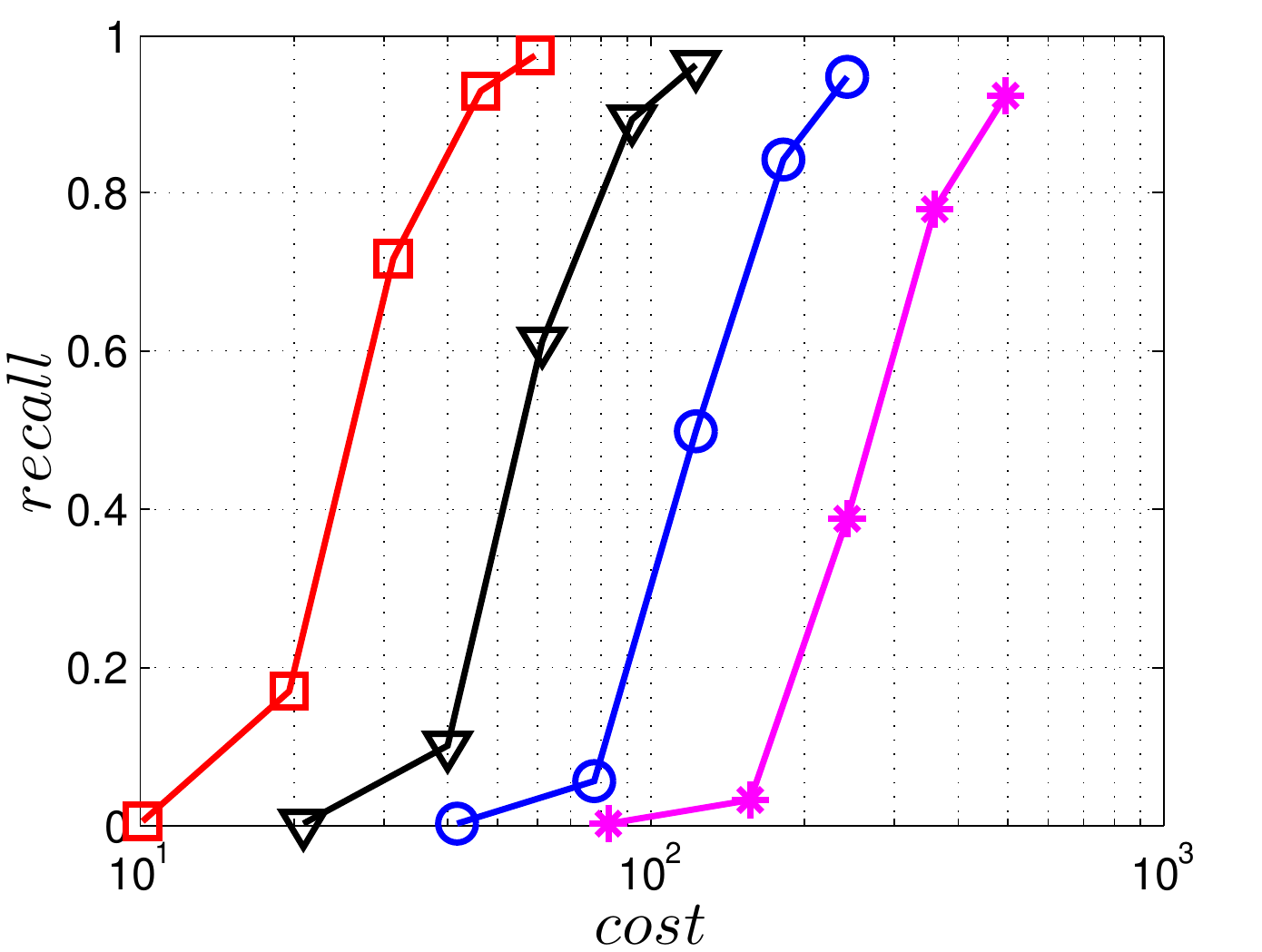}
\label{fig:scale_sift_kgraph}
}\subfigure[\texttt{Deep HNSW}]
{
\includegraphics[width=0.22\textwidth]{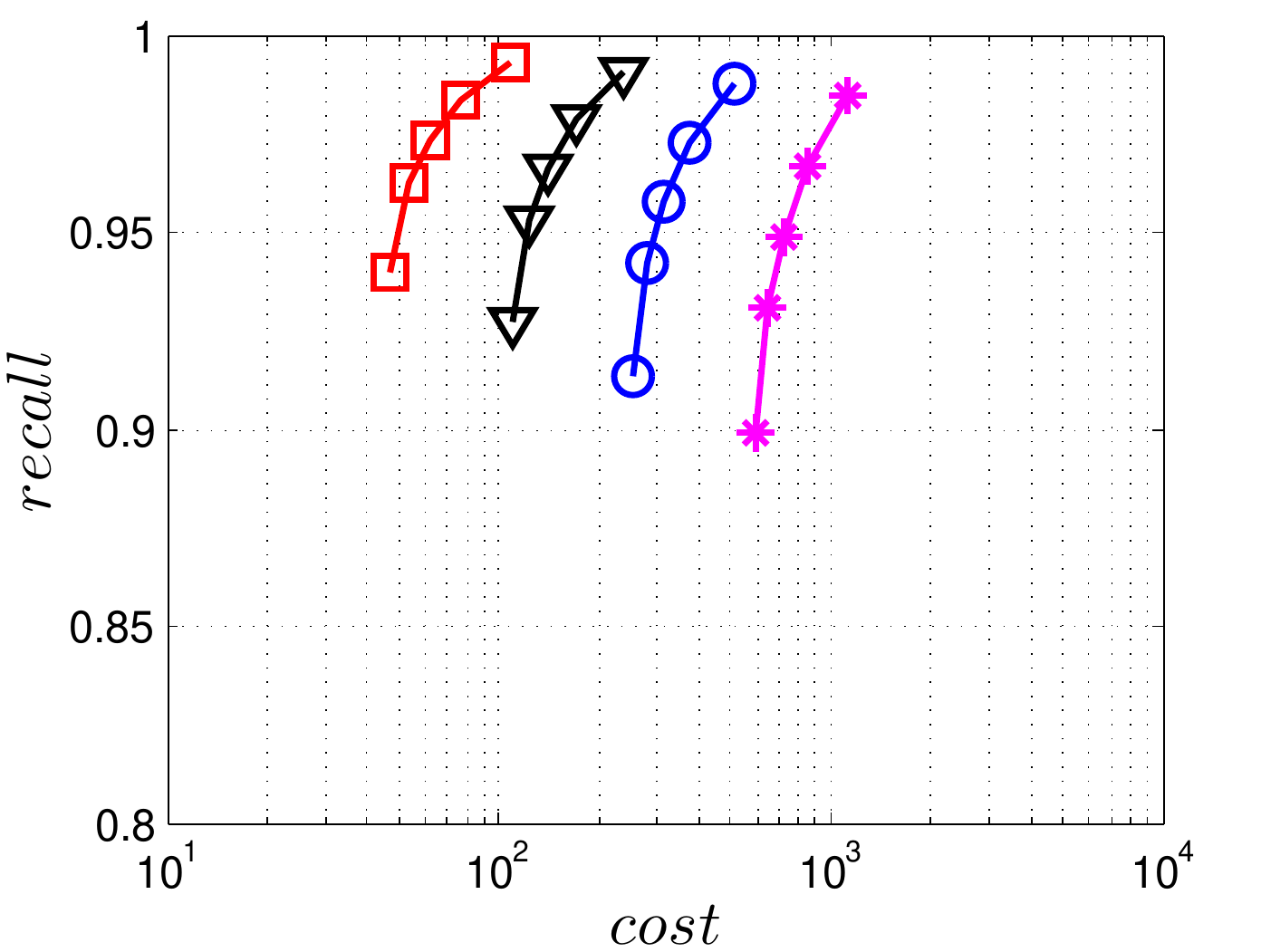}
\label{fig:scale_sift_deephnsw}
}
\end{center}\vspace{-3ex}
\caption{The Scalability of \texttt{NBPG} Methods. $cost$ is measured in seconds.}
\label{fig:scale}
\end{figure*}
}

We can see that \texttt{KGraph} with a good-quality initial KNNG does not improve its performance.  Even though \texttt{HnswKGraph} seems better than \texttt{KGraph} for a small or moderate $recall$, it is close to or even worse than \texttt{KGraph} when $recall$ reaches a high value.  Moreover, \texttt{MdKGraph} and \texttt{LvKGraph} are even worse than \texttt{KGraph}.  This is caused by the fact that determining $m^t (u)$ in \texttt{KGraph} is sensitive to the updates of KNNs according to Rule 3 in Section~\ref{ssec:bi_prop}.  When \texttt{KGraph} starts with an accurate KNNG, fewer updates happen in each iteration and thus fewer new neighbors will join in the next iteration. Moreover, there may exist repeated distance computations between \texttt{INIT} and \texttt{KGraph}, since some similar pairs may be computed repeatedly in these two steps. 

\subsubsection{Influences of $k$}
\label{sssec:kinflu}

As a key parameter in \texttt{NBPG}, we study the influence of $k$ on three representative methods, i.e., \texttt{Uni HNSW}, \texttt{Deep HNSW} and \texttt{KGraph}.  Here we select two $k$ values, i.e., a small value 5 and a large value 100, for each method. We show the results in Figure~\ref{fig:kinflu}. We can see that $k$ has an obvious influence on \texttt{Uni HNSW}.  With a large $k$, its $cost$ and $recall$ increase significantly, while keeping almost unchanged with a small $k$.  This is because up to $k^2$ candidates will be found for each query in one iteration, which leads to larger $cost$ and better $recall$. In addition, \texttt{Uni HNSW} is worse than its competitors for a large $k$, except on Nusbow and Msdrp.  As to \texttt{Deep HNSW} and \texttt{KGraph}, they have to pay more cost to achieve the same $recall$ for a large $k$, but they are not significantly affected by $k$ like \texttt{Uni HNSW}, because the way they find candidates is independent of $k$. 

\eat{
\subsubsection{Scalability Test}
To evaluate the scalability of various methods, we created four random subsets from a widely-used large scale data set with 100 million 128-dimensional SIFT vectors, called Sift100M~\footnote{http://corpus-texmex.irisa.fr/}.  Those samples are denoted as Sift1M, Sift2M, Sift4M and Sift8M, with sizes 1 million, 2 million, 4 million and 8 million respectively.  We show the results of three representative methods in Figure~\ref{fig:scale}. With the same parameter setting, the costs of these three methods increase almost linearly as the data size increases, while their $recall$'s decrease slightly. Hence all the three \texttt{NBPG} methods scale well with the data size.
}

\subsubsection{Performance on Large Data}
\label{sssec:large}
To test the scalability of \texttt{NBPG} methods, we use two large-scale data sets, i.e., Sift100M~\footnote{\textit{http://corpus-texmex.irisa.fr/}} and Deep100M to evaluate the performance of three representative methods, i.e., \texttt{Uni HNSW}, \texttt{KGraph} and \texttt{Deep HNSW}. Sift100M contains 100 million 128-dimensional SIFT vectors, while Deep100M consists of 100 million 96-dimensional float vectors, which are randomly sampled from the learn set of Deep1B~\footnote{\textit{https://yadi.sk/d/11eDCm7Dsn9GA}}. To deal with such large data, we set the number of threads as 48 for all three methods.  We show the results on Figure~\ref{fig:large}. We can see that \texttt{KGraph} obviously outperforms \texttt{Uni HNSW} and \texttt{Deep HNSW}. This is because the construction of the HNSW graph costs quite a lot time, i.e., around 6,000 seconds for both data sets. Note that KGraph requires much more memory space than its competitors.

\subsubsection{Summary}
\label{sssec:summary_nbpg}


There is no dominator in all cases for KNNG construction. We do not reccommend \texttt{UniProp}, since it cannot achieve high accuracy due to its fast convergence and is pretty sensitive to $k$. If the memory resource is enough and a high-recall KNNG is urgent, \texttt{KGraph} is the first choice. Otherwise, \texttt{Deep HNSW} is our first recommendation due to its superior balance among efficiency, accuracy and memory requirement.

\begin{table*}[t]
\centering
\caption{Data Hubness and Converged Recalls}\vspace{-2ex}
\begin{tabular}{|c|c|c|c|c|c|c|c|c|}
\hline
\multirow{2}{*}{} & \multicolumn{3}{c|}{Low Hub} & \multicolumn{2}{c|}{Moderate Hub} & \multicolumn{3}{c|}{High Hub} \\
\cline{2-9}
~ & Sift & Uqv & Msdtrh & Nuscm & Gist & Glove & Nusbow & Msdrp \\
\hline
$H_{20}(0.001, \cdot)$ & 0.007 & 0.014 & 0.012 & 0.036 & 0.039 & 0.075 & 0.066 & 0.259 \\
\hline
$H_{20}(0.01, \cdot)$ & 0.047 & 0.063 & 0.07 & 0.157 & 0.164 & 0.243 & 0.269 & 0.407 \\
\hline
$H_{20}(0.1, \cdot)$ & 0.285 & 0.319 & 0.355 & 0.531 & 0.549 & 0.647 & 0.694 & 0.740\\
\hline
\hline
$recall_{converged}^{\texttt{LargeVis}}$ & 0.929 & 0.967 & 0.929 & 0.835 & 0.673 & 0.619 & 0.599 & 0.865 \\
\hline
$recall_{converged}^{\texttt{Deep HNSW}}$ & \textbf{0.992} & \textbf{0.982} & \textbf{0.991} & 0.942 & 0.874 & \textbf{0.786} & 0.661 & 0.879 \\
\hline
$recall_{converged}^{\texttt{KGraph}}$ & 0.986 & 0.972 & 0.990 & \textbf{0.968} & \textbf{0.943} & 0.779 & \textbf{0.867} & \textbf{0.926} \\
\hline
\end{tabular}\vspace{-3ex}
\label{tb:hubness}
\end{table*}

\section{Exploring Neighborhood Propagation}
\label{sec:explo}

We are interested in the reasons why \texttt{NBPG} is effective.  The principle (i.e., ``a neighbor of my neighbor is also likely to be my neighbor''~\cite{Dong2011WWW}) is intuitively correct and works effectively in practice, but it lacks convincing insights on its effectiveness in the literature.  This motivates us to explore the \texttt{NBPG} mechanism. To achieve this goal, we employ the \textbf{node hubness} defined in \cite{Radovanovic2010} to characterize each node. Further, we develop this concept and propose the \textbf{data hubness} to characterize a data set. With those two definitions, we have two insights. First, the data hubness has a significant effect on the performance of \texttt{NBPG}. Second, the node hubness of a node has an obvious effect on its accuracy during the \texttt{NBPG} process. To present the effect of node hubness on its accuracy, we explore the dynamic process of three representative \texttt{NBPG} methods, i.e., \texttt{LargeVis}, \texttt{KGraph} and \texttt{Deep HNSW}, respectively.

\subsection{Hubness and Accuracy of Reverse KNNs}
\label{ssec:explo_def}

According to \cite{Radovanovic2010}, a hub node appears in the KNNs of many nodes, making it a popular exact KNN of other nodes.  Accordingly, a hub node has a large in-degree in the exact KNNG.  We show its definition from~\cite{Radovanovic2010} as follows.

\begin{definition}
\label{def:hub_point}
\textbf{Node Hubness}. Given a node $v \in D$ and $k$, we define its hubness, denoted as $h_k(v)$, as the number of its exact reverse KNNs, i.e., $h_k(v) = |R_k^*(v)|$, where $R_k^*(v) = \{ u | v \in N_k^*(u) \}$ and $N_k^*(u)$ represents the exact KNNs of $u$.
\end{definition}

According to \cite{Radovanovic2010}, a hub node is usually close to the data center or the cluster center, and thus it is naturally attractive to other nodes as one of their exact KNNs.  Further, we extend the hubness concept to the whole data set as follows. 

\begin{definition}
\label{def:hub_data}
\textbf{Data Hubness}. Given $D$, $k$ and a percentage $x$, we define the data hubness of $D$ at $x$ in Equation~\ref{eq:def_hubdata}, where $\pi$ is a permutation of nodes in $D$ and $u_{\pi(i)}$ is the node with the $i$-th largest node hubness.  
\begin{equation}
\label{eq:def_hubdata}
H_k(x, D) = \frac{\sum_{i=1}^{\lceil x \times n \rceil} h_k(u_{\pi(i)})}{n \times k}, x \in [0, 1]
\end{equation}
\end{definition}

We can see that $H_k(x, D)$ is actually the normalized sum of the largest $\lceil x\times n \rceil$ $h_k$ values and thus $H_k(x, D)$ is between 0 and 1.  For the eight data sets, we divide them into three groups according to their data hubness, i.e., low hub, moderate hub and high hub respectively, as shown in Table~\ref{tb:hubness}. We can see that \textbf{Msdrp} has the largest data hubness, where the first $0.1\%$ (995) points have $25.9\%$ hubness (corresponding to over 5 million edges in the exact KNNG).



In addition, we define the \textbf{accuracy of reverse KNNs}.  For a KNNG $G$, we have $R_k(v) = \{ u | v \in N_k (u) \}$ as $v$'s reverse KNNs.  Then we define the accuracy of $v$'s reverse KNNs as $recall^R(v) = |R_k(v) \cap R_k^*(v)| / |R_k^*(v)|$, where $R_k^*(v)$ is defined as in Definition~\ref{def:hub_point}. $recall^R$ reflects the accuracy of a KNNG to some extent, since a node probably has a high $recall^R$ value on an accurate KNNG. 

\subsection{Effect of Data Hubness on NBPG Performance}
\label{ssec:hub_effect}

\begin{figure}
\begin{center}
\subfigure[\textbf{Sift \& Gist}]
{
\includegraphics[width=0.22\textwidth]{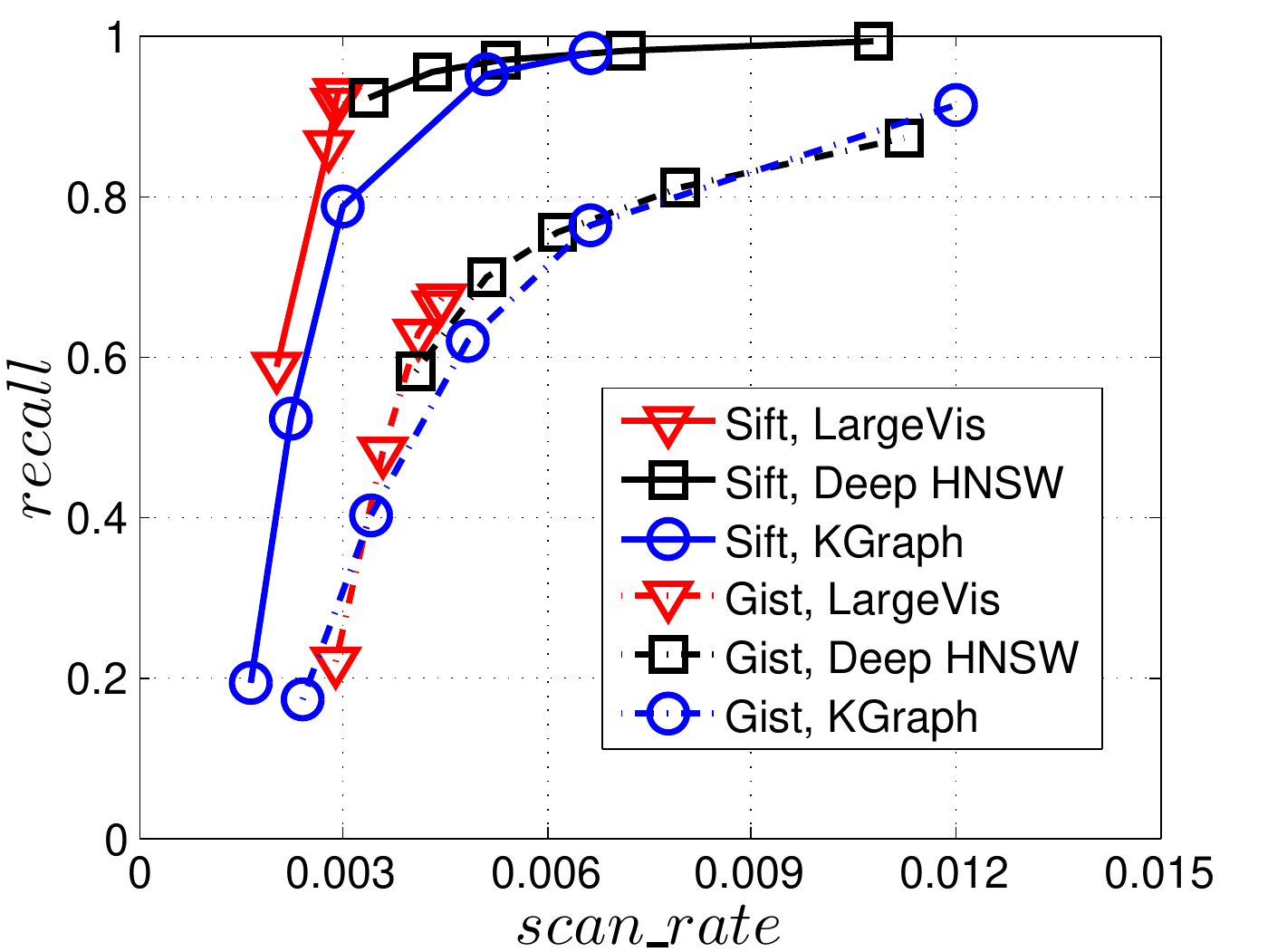}
\label{fig:hub_effect_1m}
}
\subfigure[\textbf{Nuscm \& Numbow}]
{
\includegraphics[width=0.22\textwidth]{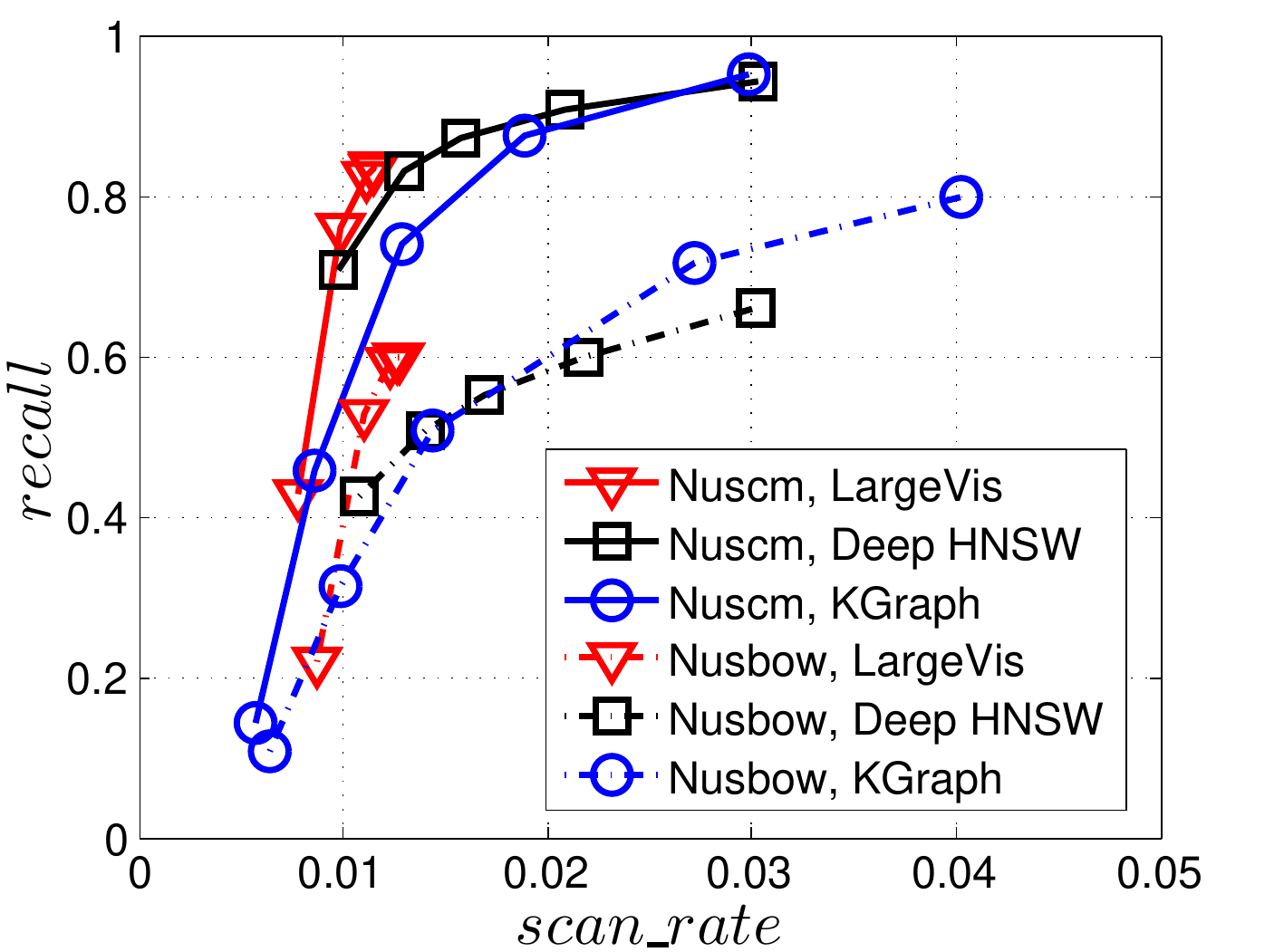}
\label{fig:hub_effect_}
}
\end{center} \vspace{-3ex}
\caption{Effect of Data Hubness on \texttt{NBPG}.} \vspace{-2ex}
\label{fig:hub_effect}
\end{figure}

In this part, we present the effect of data hubness on the performance of \texttt{NBPG}. We discuss the performance in two aspects, i.e., (1) computational cost vs accuracy and (2) the converged accuracy. We use $scan\_rate = \frac{\#total\_dist}{n*(n-1)/2}$ to measure the computational cost, where $\#total\_dist$ indicates the total number of distance computations during KNNG construction and $n*(n-1)/2$ is the number of all distance computations in the brute-force method. Unlike $cost$, $scan\_rate$ is independent of $d$.  Here, we show the results on data size in Figure~\ref{fig:hub_effect}. We select data sets with the same size in each subfigure, so that we can focus on the effects of data hubness. We can see that the data hubness has a significant effect on the performance of \texttt{NBPG} methods. With a larger data hubess, each \texttt{NBPG} method has to compute more distances to achieve the same $recall$. 

The converged $recall$ is such a value that once an \texttt{NBPG} method reaches it, paying more cost cannot improve the $recall$ value in a significant scale. We believe that the converged $recall$ reflects the characteristic of an \texttt{NBPG} method. Here we obtain this value of a method by setting its parameters large enough.  To be specific, the parameter settings are $nIter = 4$ in \texttt{LargeVis}, $efSearch = 160$ in \texttt{Deep HNSW} and $nIter = 16$ in \texttt{KGraph}. We show the results in Table~\ref{def:hub_data}. We can see that a lower hub data usually has a higher converged $recall$, while a higher hub data has a lower converged $recall$. This indicates that it is computationally expensive to get a high-recall KNNG for a high hub data.


\begin{figure*}[h]
\begin{center}
\subfigure[\textbf{Sift}]
{
\includegraphics[width=0.22\textwidth]{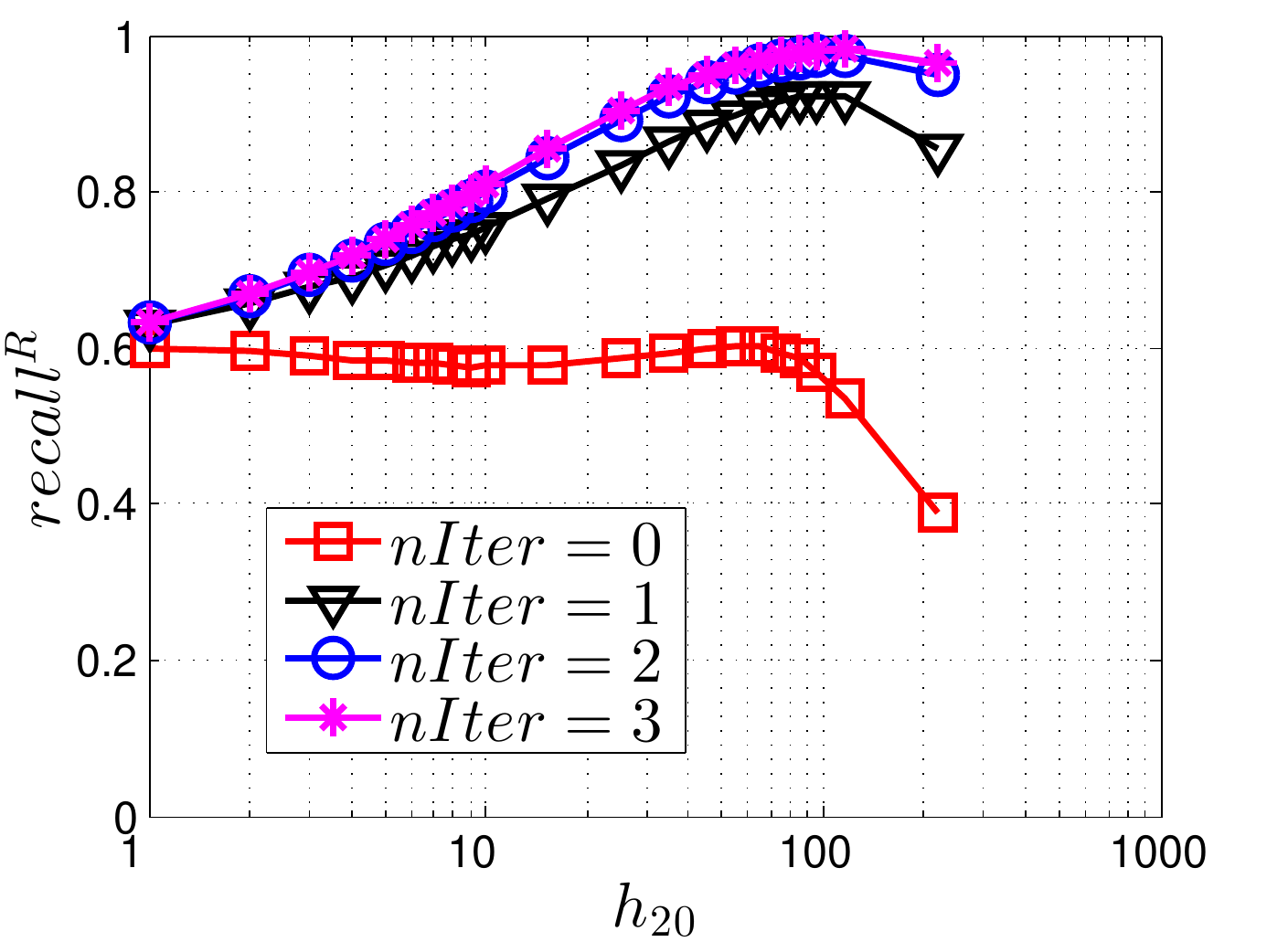}
\label{fig:lvg_tpr0_sift}
}
\subfigure[\textbf{Sift}]
{
\includegraphics[width=0.22\textwidth]{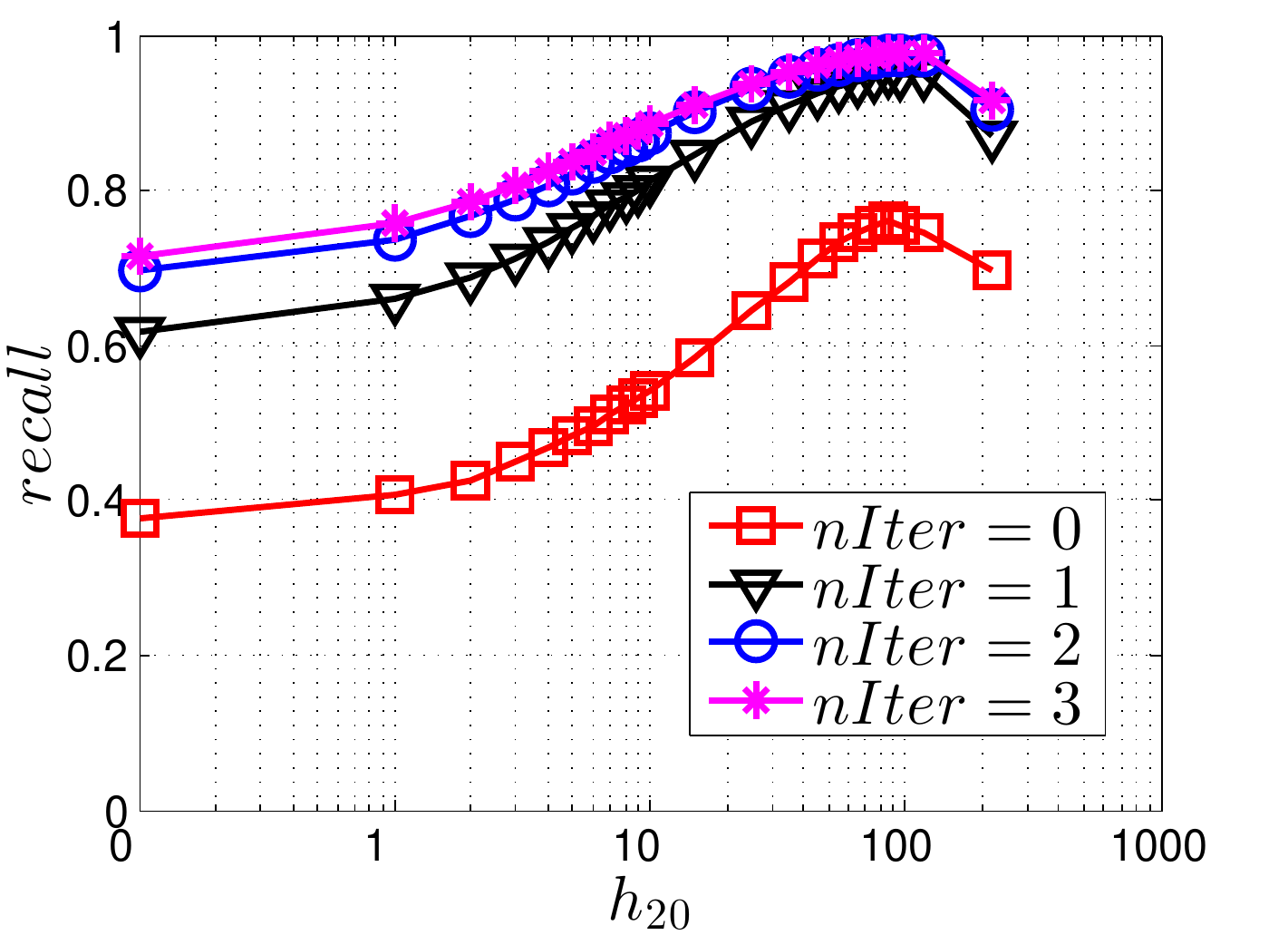}
\label{fig:lvg_tpr1_sift}
}
\subfigure[\textbf{Glove}]
{
\includegraphics[width=0.22\textwidth]{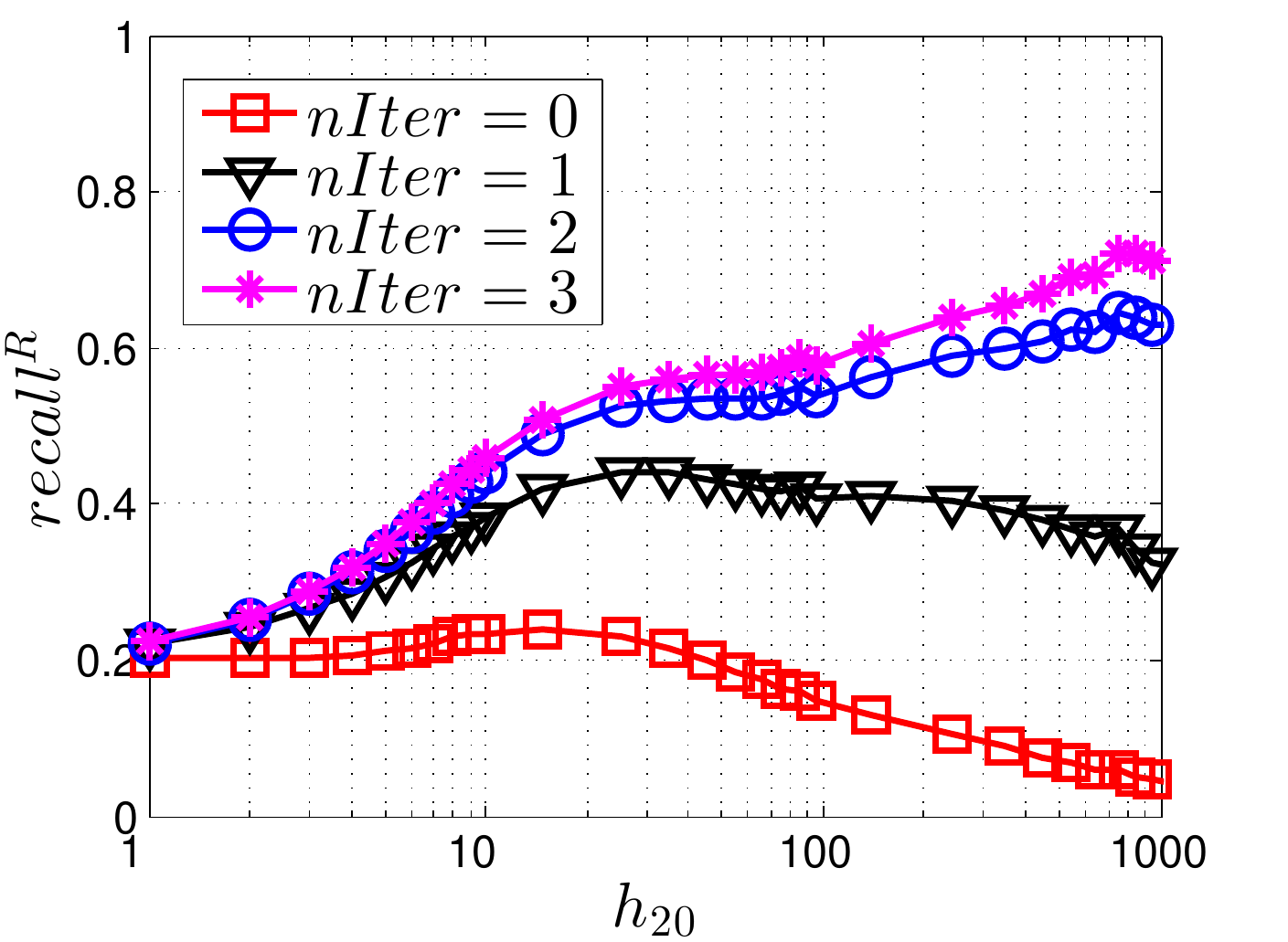}
\label{fig:lvg_tpr0_glove}
}
\subfigure[\textbf{Glove}]
{
\includegraphics[width=0.22\textwidth]{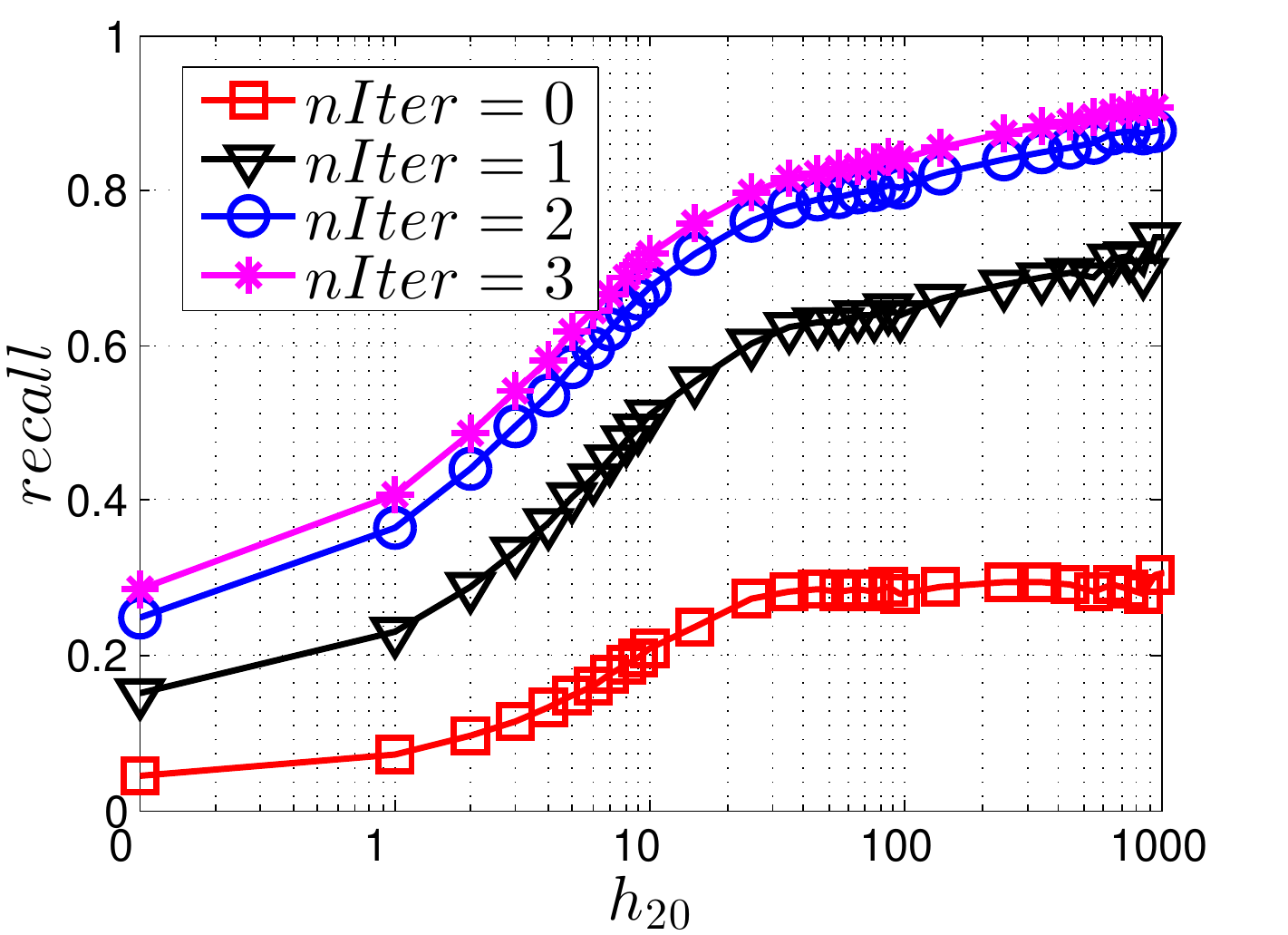}
\label{fig:lvg_tpr1_glove}
}
\end{center} \vspace{-3ex}
\caption{Exploring \texttt{LargeVis}.} \vspace{-2ex}
\label{fig:lvg_tpr}
\end{figure*}

\subsection{Exploring UniProp}
\label{ssec:explo_uni}

As a representative \texttt{UniProp} method, \texttt{LargeVis} starts from an initial KNNG with low or moderate accuracy and then conducts \texttt{UniProp} for a few iterations.  Here we vary the number $nIter$ of iterations from 0 to 3.  When $nIter = 0$, it corresponds to the initial KNNG. Due to the space limit, we only report the results on Sift and Glove in Figure~\ref{fig:lvg_tpr}.  The $x$-axis shows the node hubness of individual nodes and the $y$-axis shows the $recall$ or $recall^R$.  We can see the $recall$ values of all nodes increases substantially, especially in the first iteration. Notably, those newly found exact KNNs are mostly the hub nodes, as reflected by the significant increase in the $recall^R$ values of hub nodes.  

Even though the $recall^R$ values of high hub nodes are not high in the beginning, as the cardinality of $R^*_k(\cdot)$ for these high hub nodes is large, there are still a good number of them that have been correctly connected to by many nodes as exact KNNs.  Hence each node has a high chance to find those hub nodes via one of its KNNs in \texttt{UniProp}.  That is why we observe a significant increase of $recall$ in the first iteration. Overall, \texttt{UniProp} works well to find hub nodes as exact KNNs, but misses a significant part of non-hub nodes, which are less connected by other nodes and thus not easy to be found in \texttt{UniProp}. That is the reason why its converged recall (as shown in Table~\ref{tb:hubness}) is relatively low, especially on high hub data.  



\subsection{Exploring BiProp}
\label{ssec:explo_bi}

\begin{figure*}[t]
\begin{center}
\subfigure[\textbf{Msdtrh}]
{
\includegraphics[width=0.22\textwidth]{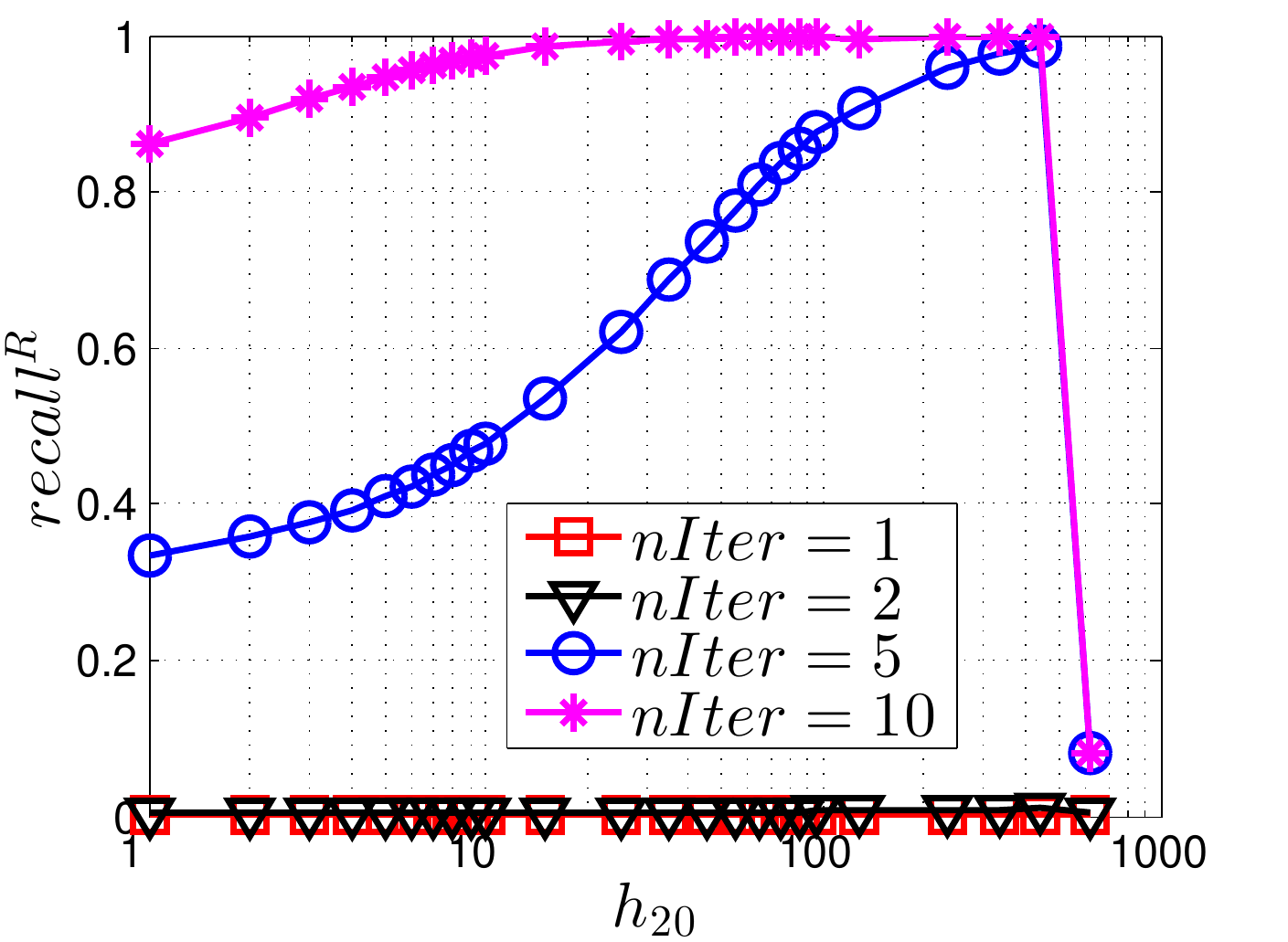}
\label{fig:kgraph_tpr0_msdtrh}
}
\subfigure[\textbf{Msdtrh}]
{
\includegraphics[width=0.22\textwidth]{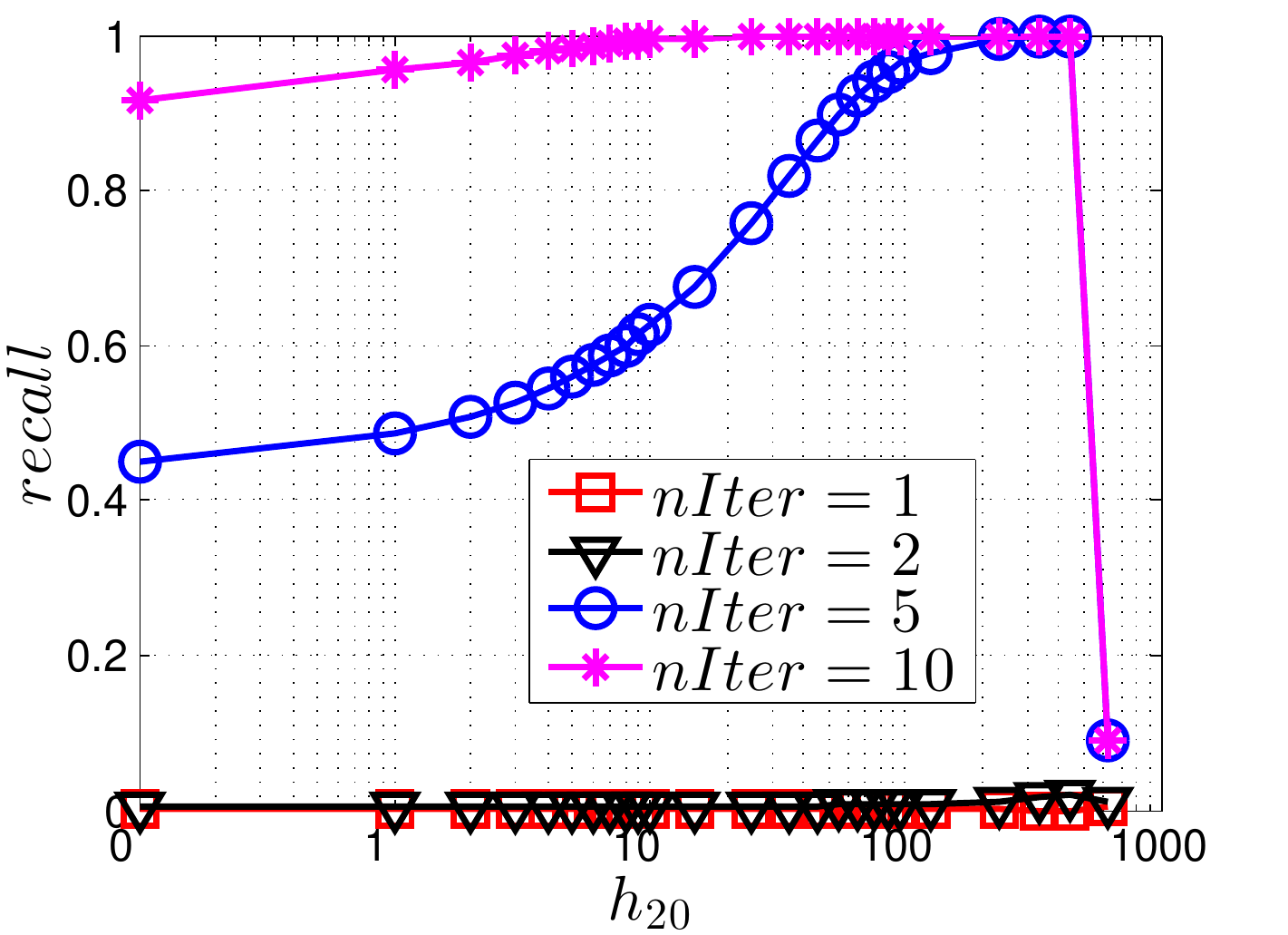}
\label{fig:kgraph_tpr1_msdtrh}
}
\subfigure[\textbf{Nusbow}]
{
\includegraphics[width=0.22\textwidth]{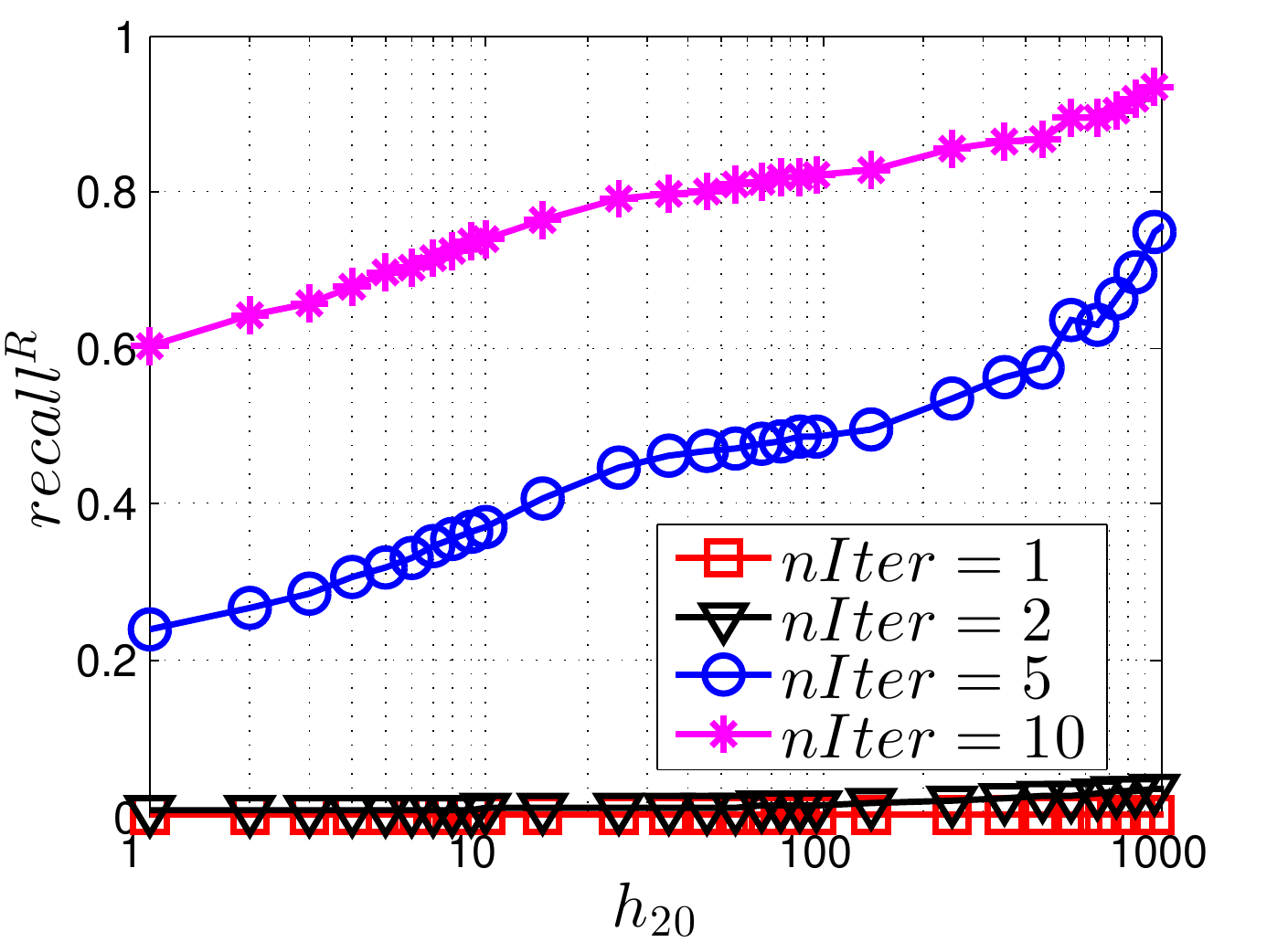}
\label{fig:kgraph_tpr0_nusbow}
}
\subfigure[\textbf{Nusbow}]
{
\includegraphics[width=0.22\textwidth]{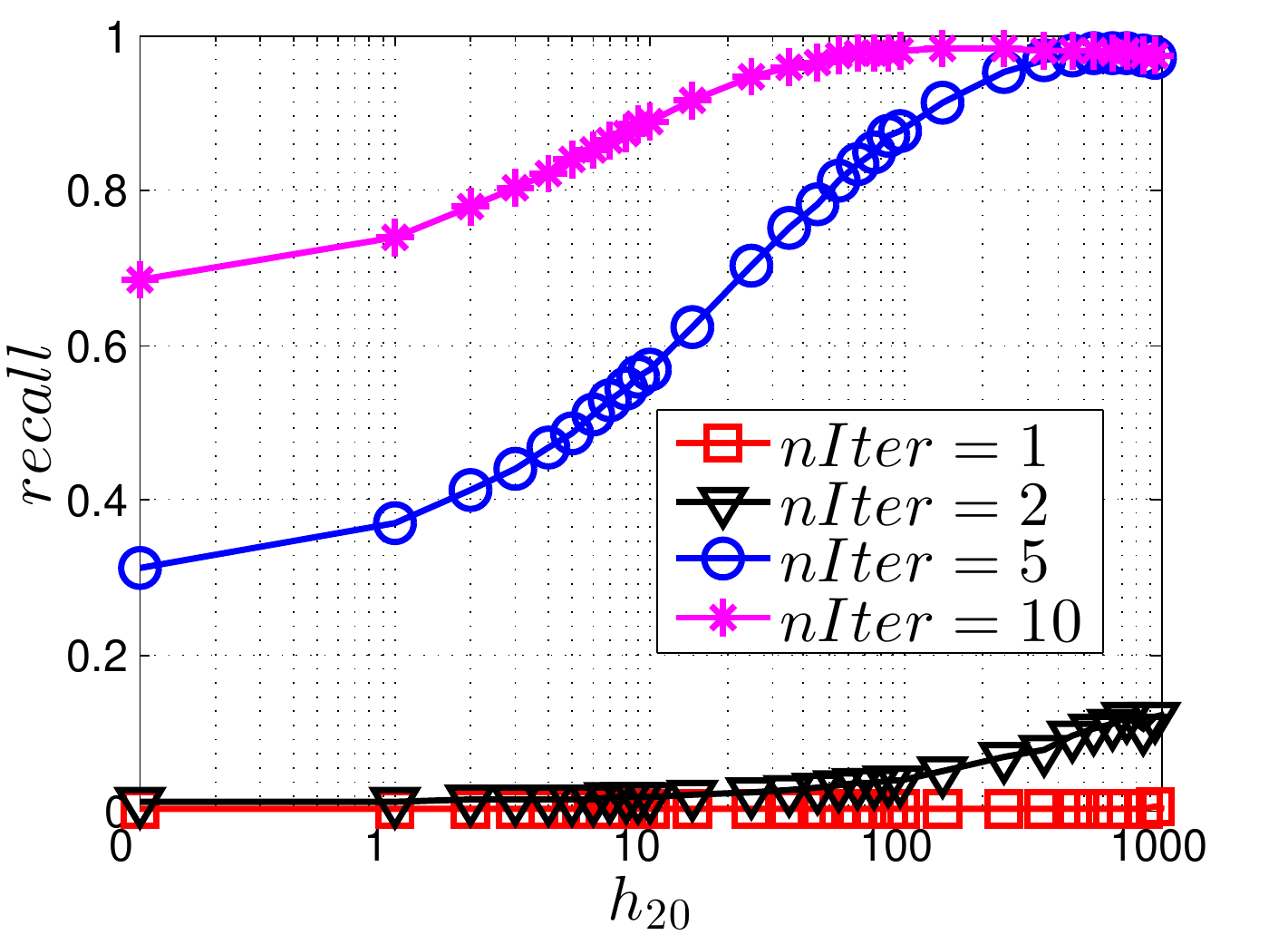}
\label{fig:kgraph_tpr1_nusbow}
}
\end{center} \vspace{-3ex}
\caption{Exploring \texttt{KGraph}.} \vspace{-2ex}
\label{fig:kgraph_tpr}
\end{figure*}

As a representative \texttt{BiProp} method, \texttt{KGraph} starts from a random initial graph and then conducts \texttt{NBPG} for a few iterations. Here we vary the number of iterations from 1 to 10. We show the results on Msdtrh and Nusbow in Figure~\ref{fig:kgraph_tpr}. We can see that both $recall$ and $recall^R$ are pretty low in the first two iterations, due to the random initial graph. As \texttt{KGraph} continues, both $recall$ and $recall^R$ of high hub nodes increase obviously and approach almost the converged values in the fifth iteration. Unlike \texttt{UniProp}, $recall$ and $recall^R$ of moderate and low hub nodes obviously grow in later iterations. This is because \texttt{KGraph} conducts the brute-force search in $B_m^t(\cdot)$, where each reverse neighbor (probably a non-hub node) will treat another reverse neighbor as a candidate.  In other words, \texttt{KGraph} enhances the communications among non-hub nodes.

Notably, on Msdtrh, $recall$ and $recall^R$ are pretty low for some high hub nodes, whose $h_{20}$ values are larger than 400. This is caused by the fact that \texttt{KGraph} imposes a threshold on the number of reverse neighbors in order to ensure the efficiency. Hence the reverse neighbors of high hub nodes will be reduced, which has a negative influence on $recall$ and $recall^R$ of high hub nodes.  But this loss could be made up by the increase in accuracy of non-hub nodes.

\begin{figure*}[t]
\begin{center}
\subfigure[\textbf{Nuscm}]
{
\includegraphics[width=0.22\textwidth]{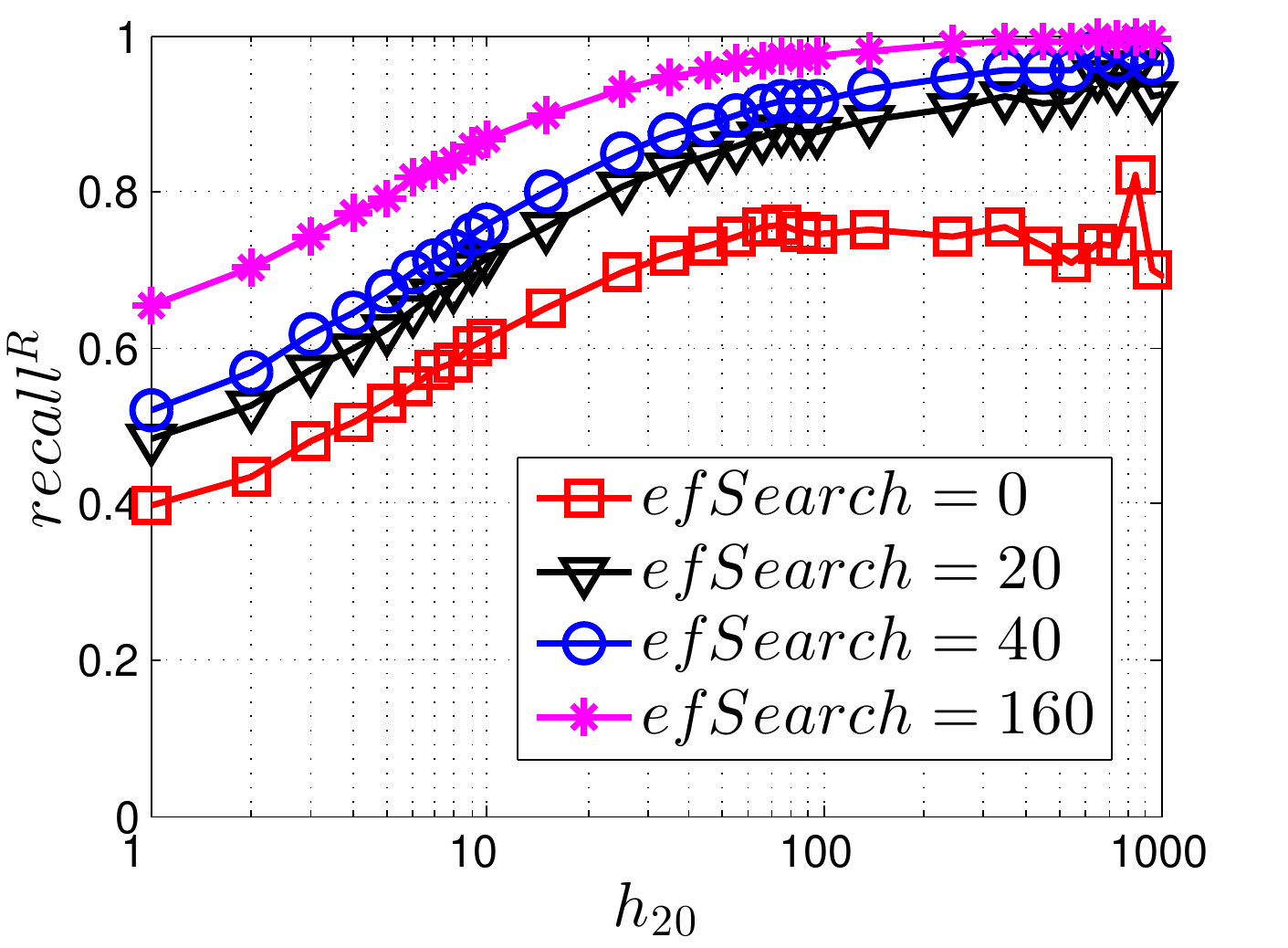}
\label{fig:deephnsw_tpr0_nuscm}
}
\subfigure[\textbf{Nuscm}]
{
\includegraphics[width=0.22\textwidth]{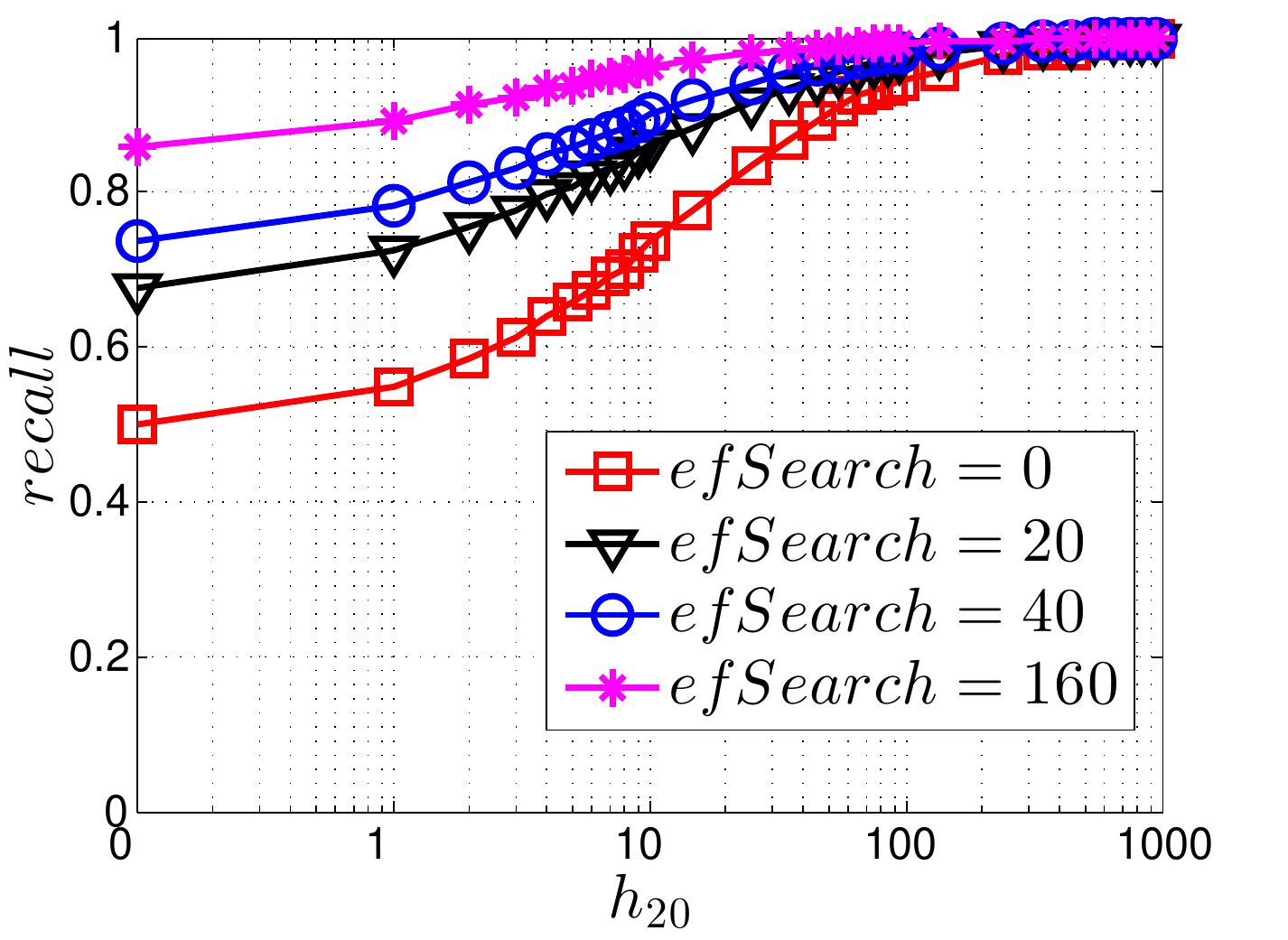}
\label{fig:deephnsw_tpr1_nuscm}
}
\subfigure[\textbf{Msdrp}]
{
\includegraphics[width=0.22\textwidth]{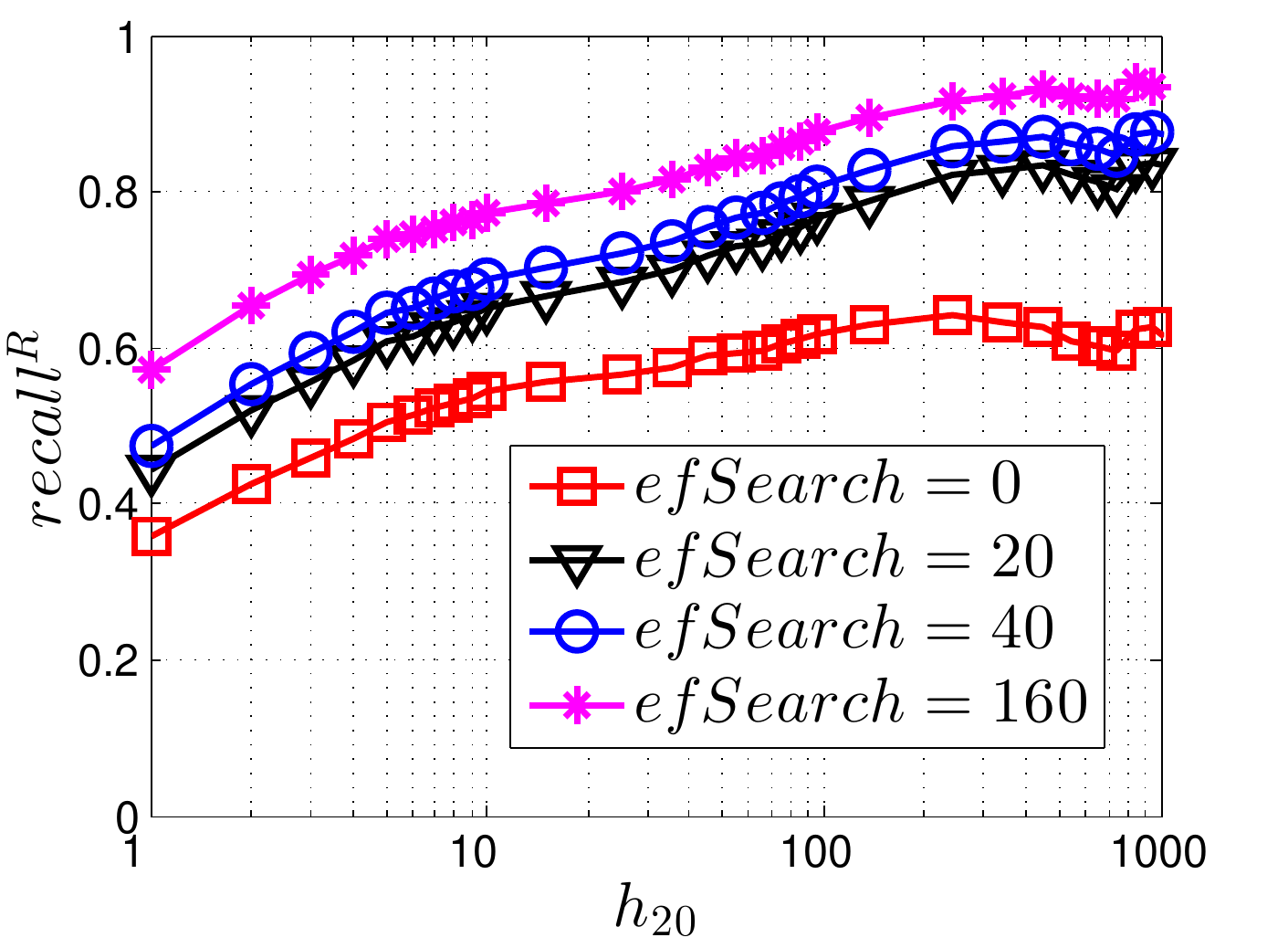}
\label{fig:deephnsw_tpr0_msdrp}
}
\subfigure[\textbf{Msdrp}]
{
\includegraphics[width=0.22\textwidth]{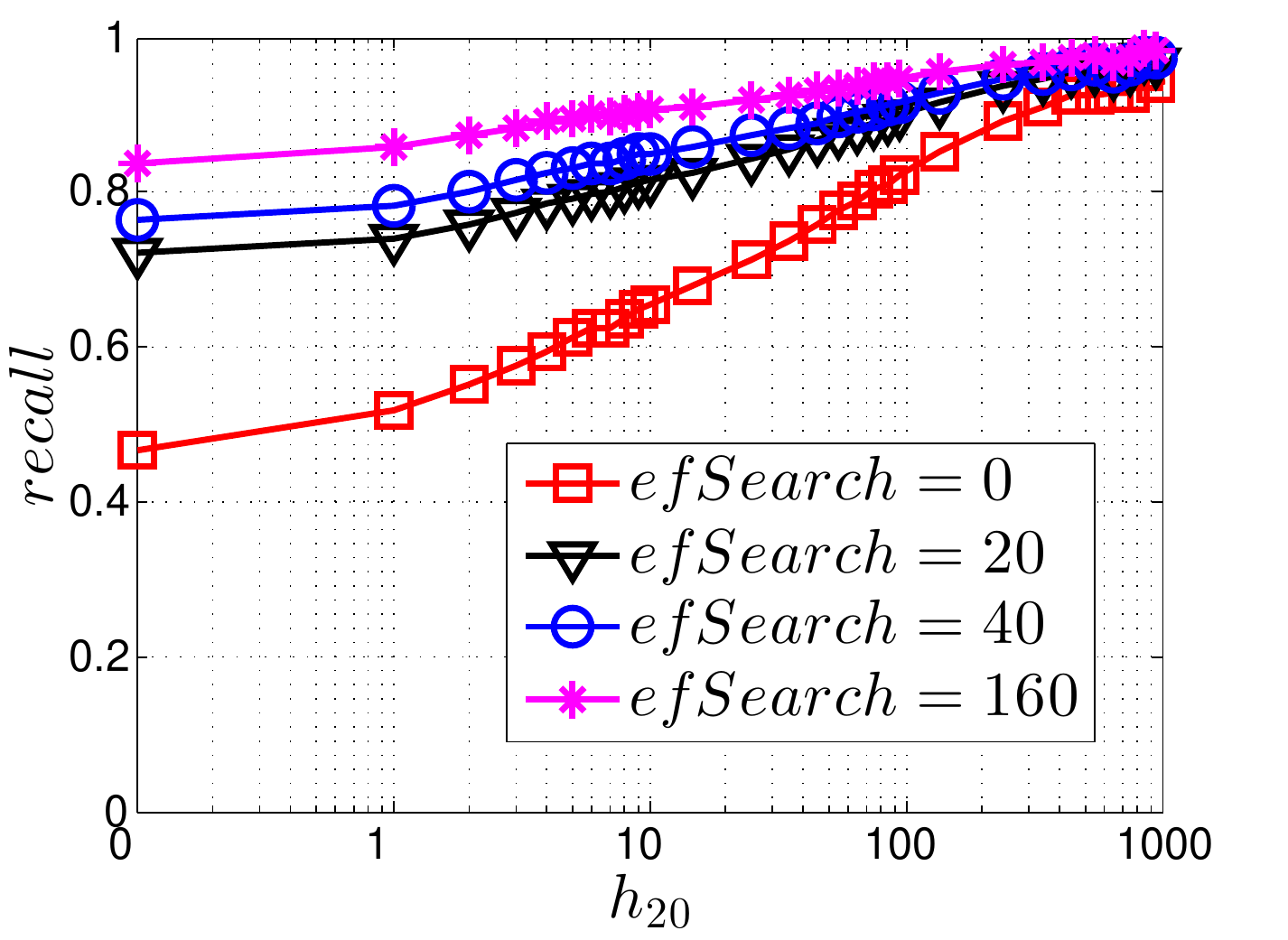}
\label{fig:deephnsw_tpr1_msdrp}
}
\end{center} \vspace{-3ex}
\caption{Exploring \texttt{Deep HNSW}.} \vspace{-2ex}
\label{fig:deephnsw_tpr}
\end{figure*}

\subsection{Exploring DeepSearch}
\label{ssec:explo_deep}

As a representative method of \texttt{DeepSearch}, the results of \texttt{Deep HNSW} are shown in Figure~\ref{fig:deephnsw_tpr}, where $efSearch$ varies from 0 and 160. When $efSearch = 0$, it corresponds to the initial KNNG, where both $recall$ and $recall^R$ of hub nodes are higher than those of non-hub nodes. Note that this is different from the initial KNNG generated by \texttt{LargeVis}, where hub nodes have even smaller $recall^R$ values than non-hub nodes. This is because the construction of HNSW graph also employs the similar idea of \texttt{NBPG} that establishes the advantages of hub nodes in accuracy.

Like \texttt{KGraph}, $recall$ and $recall^R$ of non-hub nodes increase progressively in \texttt{Deep HNSW} as $efSearch$ grows.  \texttt{DeepSearch} enhances its accuracy by enlarging the expanded neighborhood for each node. As a result, more candidates will be found with a larger $efSearch$, which increases the chance of finding non-hub nodes.

\subsection{Comparisons and Discussions}
\label{ssec:explo_comp}

\begin{figure*}
\begin{center}
\subfigure[\textbf{Uqv}]
{
\includegraphics[width=0.22\textwidth]{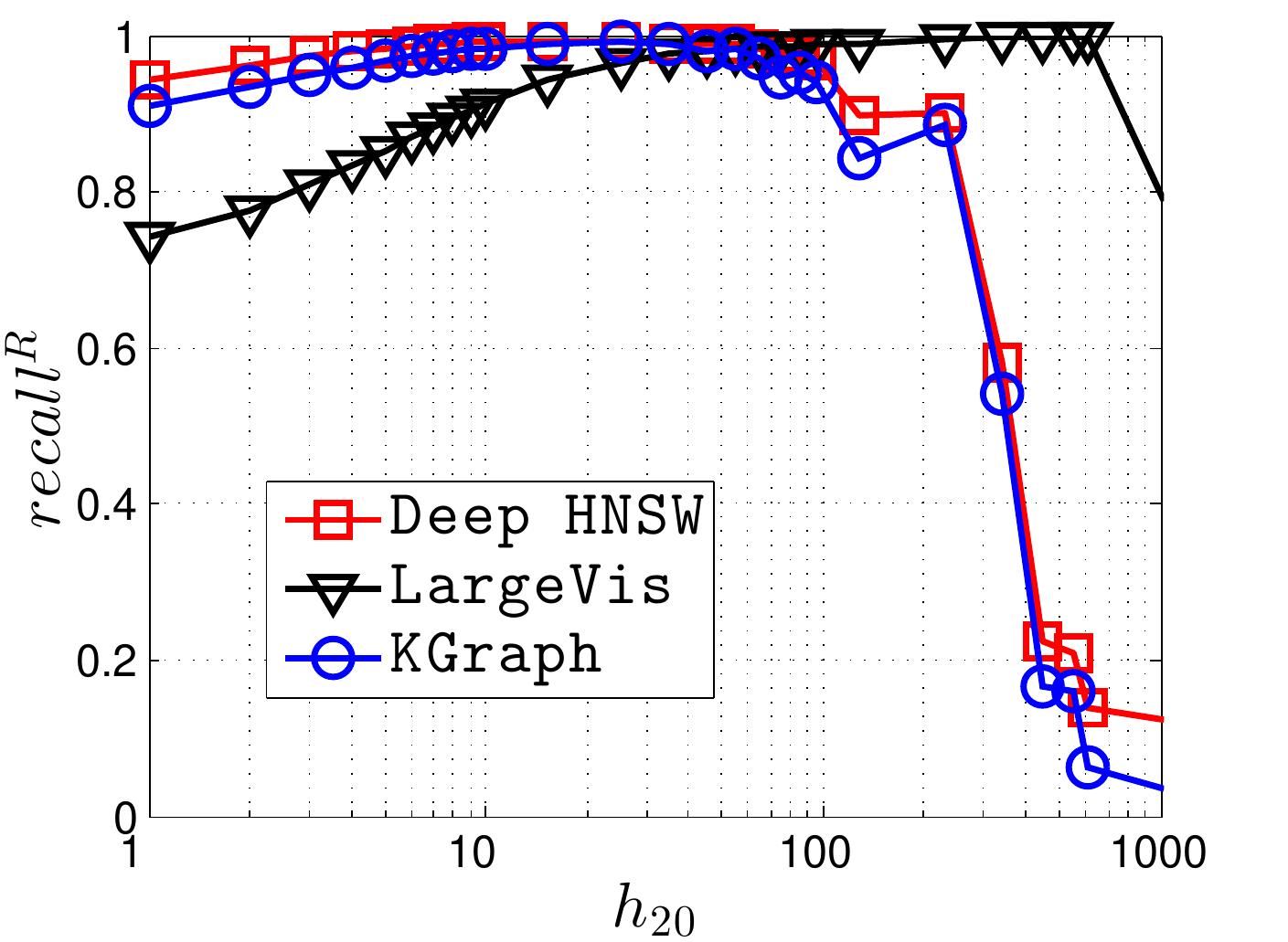}
\label{fig:tpr0_uqv}
}
\subfigure[\textbf{Uqv}]
{
\includegraphics[width=0.22\textwidth]{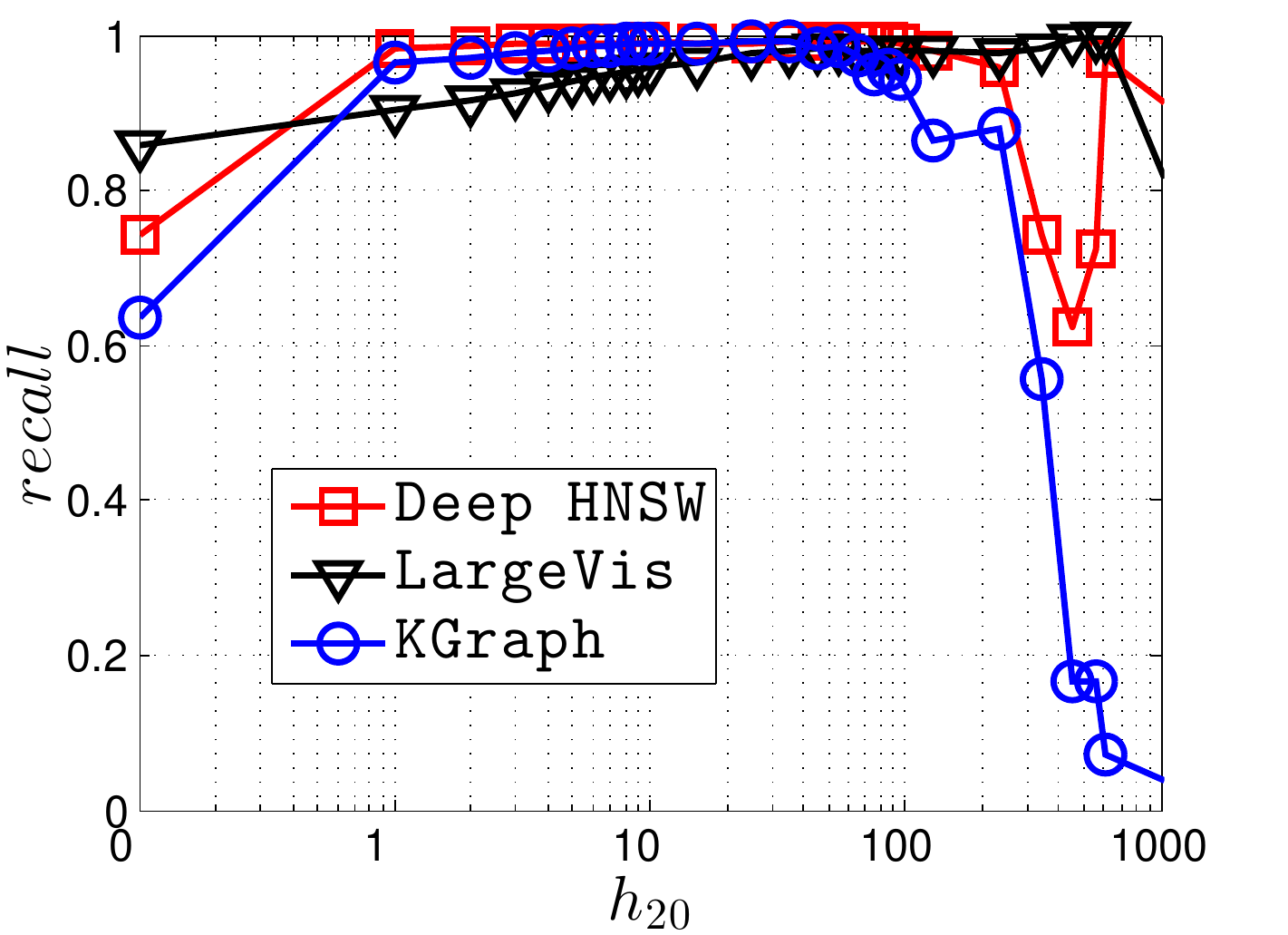}
\label{fig:tpr1_uqv}
}
\subfigure[\textbf{Gist}]
{
\includegraphics[width=0.22\textwidth]{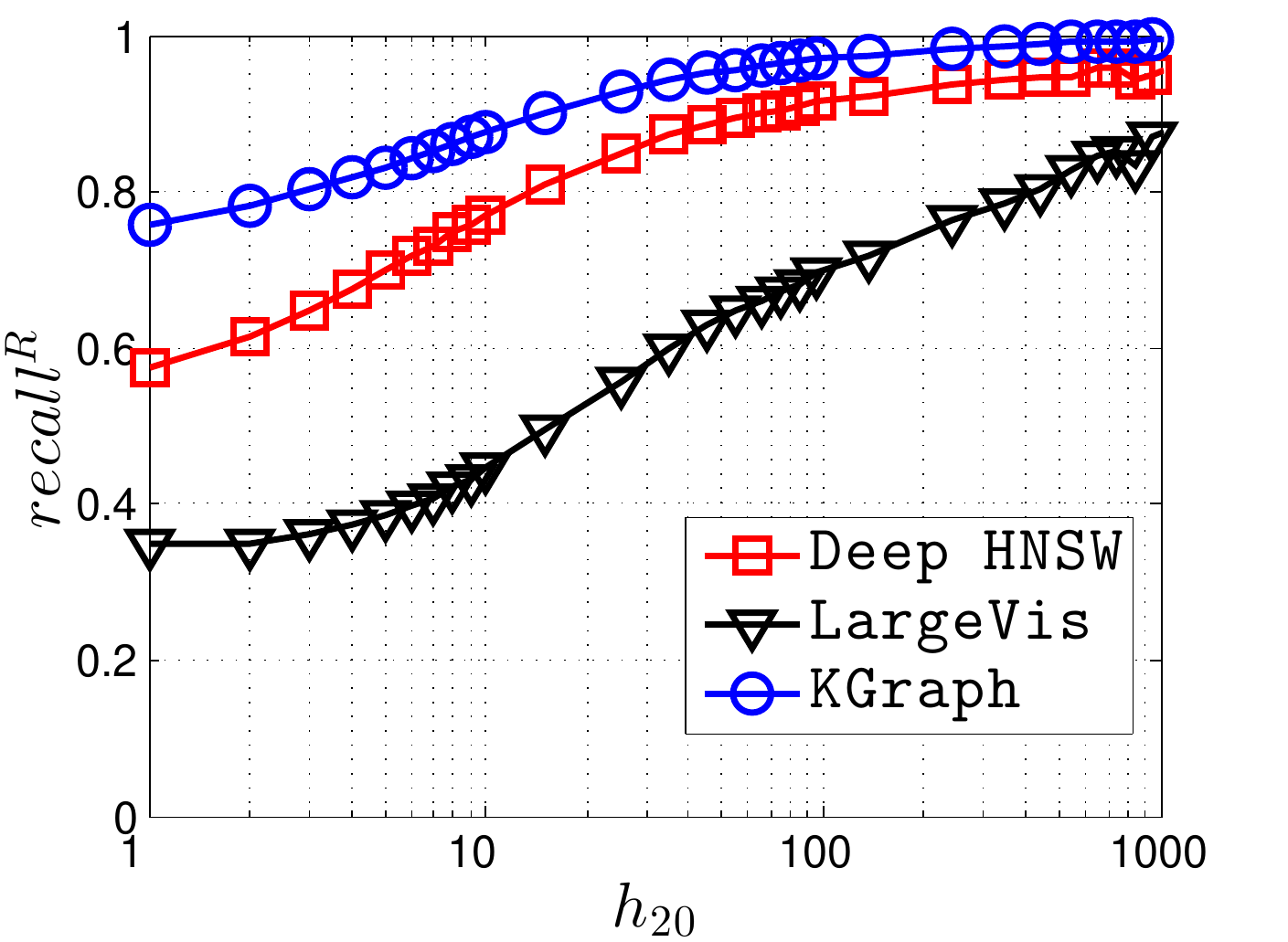}
\label{fig:tpr0_gist}
}
\subfigure[\textbf{Gist}]
{
\includegraphics[width=0.22\textwidth]{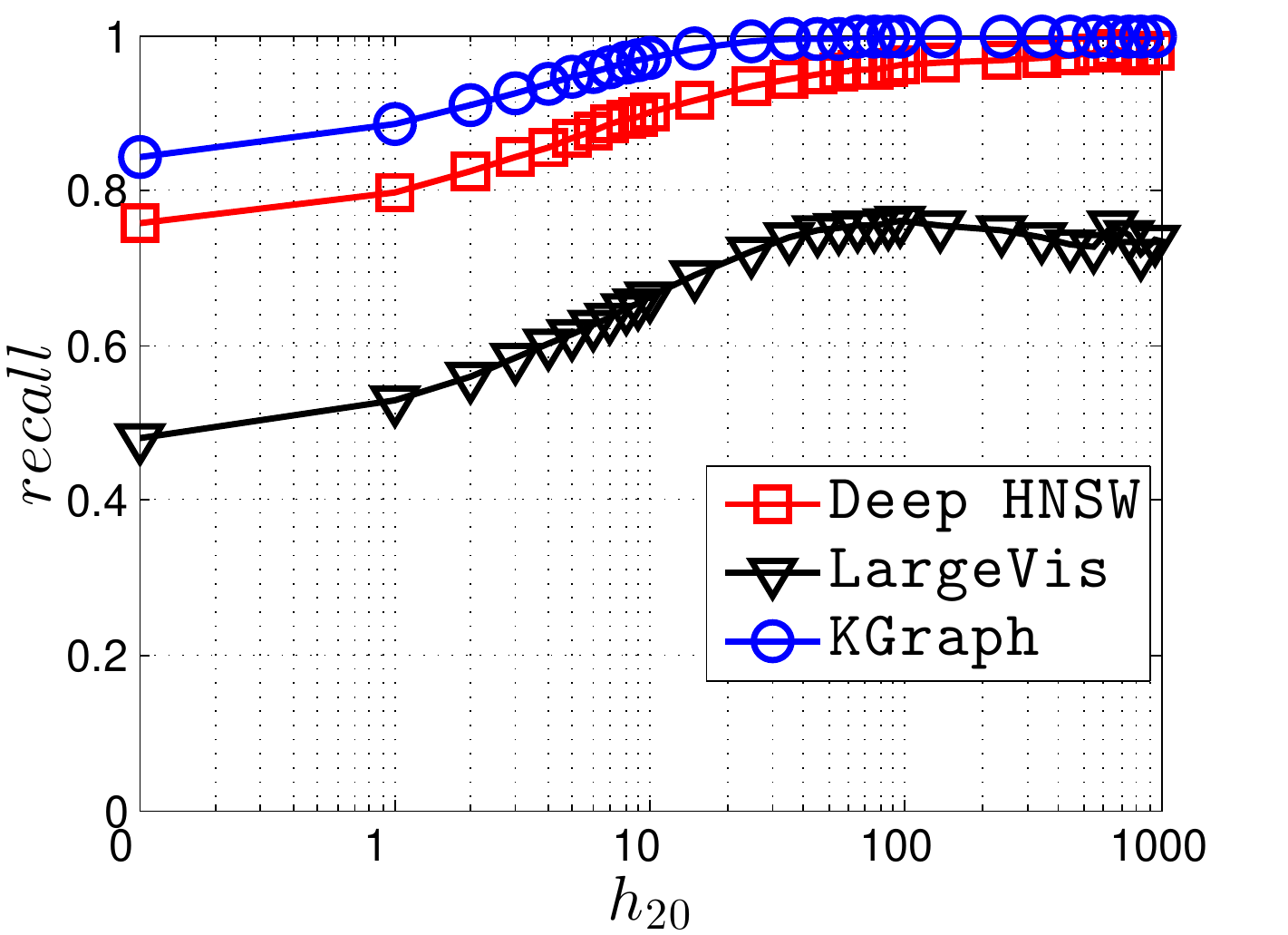}
\label{fig:tpr1_gist}
}
\end{center} \vspace{-3ex}
\caption{Comparing Converged Accuracy.} \vspace{-2ex}
\label{fig:tpr}
\end{figure*}

Now let us compare the converged accuracy of the three representative \texttt{NBPG} methods together. We show the results in Figure~\ref{fig:tpr}.  On the low hub data set Uqv, it is interesting that both \texttt{KGraph} and \texttt{Deep HNSW} have small $recall$ and $recall^R$ for high hub nodes (i.e., $h_k > 200$), while \texttt{LargeVis} has much larger values.  This is because both \texttt{KGraph} and \texttt{Deep HNSW} limit the neighborhood size for each node $u$.  To be more detailed, \texttt{KGraph} limits the size of $R_m^t(u)$, while the HNSW graph limits the number of $u$'s edges. Since low hub data has a small number of hub nodes, it has a small impact on the overall $recall$. To some extent, this phenomenon reflects the preference of \texttt{UniProp} on hub nodes. 

On the moderate hub data Gist, we can see the full advantages of \texttt{KGraph} in accuracy w.r.t.\ various hubness, followed by \texttt{Deep HNSW} and \texttt{LargeVis}. As indicated by the converged recall in Table~\ref{tb:hubness}, it will be more difficult to get high accuracy as the data hubness increases. The lower accuracy of \texttt{LargeVis} indicates that exploring the KNNs of each KNN does not work well for the low hub nodes of those difficult data, since \texttt{UniProp} obtains the candidates in the fixed neighborhood.  To find more exact KNNs, we have to enlarge the neighborhood. Compared with \texttt{LargeVis}, \texttt{Deep HNSW} explores more than $k$ neighbors' neighborhood on a fixed graph (i.e., the HNSW graph), while \texttt{KGraph} enlarges $B_m^t(u)$ for each promising point $u$ iteratively. Moreover, they both use the reverse neighbors in the neighborhood propagation, which enhances the accuracy of non-hub nodes. 

\vspace{-2ex}
\subsection{Summary}
\label{ssec:hub_summary}
We find that the hubness phenomenon has a significant effect on \texttt{NBPG} performance in two aspects. First, there is a strong correlation between the data hubness and the performance of an \texttt{NBPG} method. We can expect a good balance between efficiency and accuracy and a good converged recall on a low hub data, while probably the opposite situation on a high hub data. Second, the node hubness significantly affects its accuracy in both $recall$ and $recall^R$.  We can expect a high hub node has a high $recall$ and $recall^R$ value, while a low hub node the opposite situation.

\vspace{-2ex}
\section{Related Work}
\label{sec:rework}


Similarity search on high-dimensional data is another problem that is closely related to KNNG construction. Various index structures have been designed to accelerate similarity search. Tree structures were popular in early years and thus many trees~\cite{Guttman1984, Beckmann1990, Jagadish2005, Silpa2008} have been proposed. Locality sensitive hashing (LSH)~\cite{Indyk1998, Gionis1999} attracted a lot of attentions during the past decade, due to its good thorectical guarantee. Hence a few methods~\cite{Lv2007, Tao2009, Liu2014, Sun2015, HuangVLDB2016} based on LSH were created. Besides, there are also a few methods~\cite{Sivic2003, Philbin08, Jegou2011, Babenko2012, Baranchuk2018} based inverted index, which assigns similar points into the same inverted list and only accesses the most promising lists during the search procedure. Recently, proximity graphs~\cite{Dong2011WWW, Malkov2014IS, Malkov2018PAMI, Li2019TKDE, Fu2019VLDB} become more and more popular, due to their superior performance over other structures. 

\vspace{-2ex}
\section{Conclusion}
\label{sec:clu}

In this paper, we revisit existing studies of constructing KNNG on high-dimensional data.  We start from the framework adopted by existing methods, which contains two steps, i.e.,  \texttt{INIT} and \texttt{NBPG}.  We conduct a comprehensive experimental study on representative methods for each step respectively.  According to our experimental results, \texttt{KGraph} has the best performance to construct a high-recall KNNG, but requires much more memory space. A new combination called \texttt{Deep HSNW} presents better balance among efficiency, accuracy and memory requirement than its competitors. Notably, the popular \texttt{NBPG} technique, \texttt{UniProp}, is not recommended, since it cannot achieve high-recall KNNG and  is pretty sensitive to $k$.  Finally, we employ the hubness concepts to investigate the effectiveness of \texttt{NBPG} and have two interesting findings. First, the performance of \texttt{NBPG} is obviously affected by the data hubness. Second, the accuracy of each node is significantly influenced by its node hubness.


\vspace{-2ex}

\bibliographystyle{abbrv}
\bibliography{yfliuRef}

\vspace{-12ex}
\begin{IEEEbiography}[{\includegraphics[width=1.0in, height=1.25in,clip,keepaspectratio]{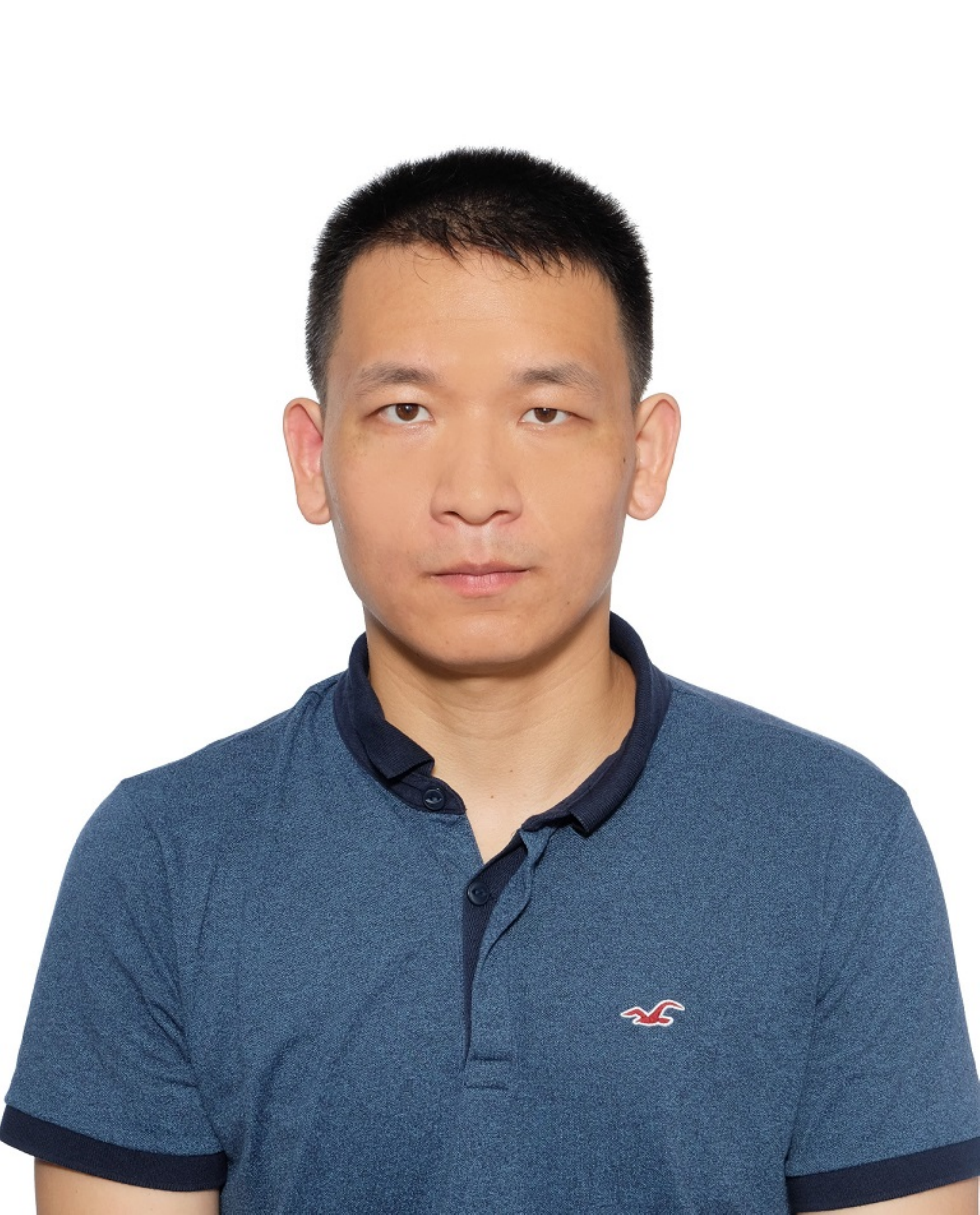}}]{Yingfan Liu} is currently a Lecturer in the School of Computer Science and Technology, Xidian University, China. He obtained his PhD degree in the department of Systems Engineering and Engineering Management of Chinese University of Hong Kong in 2019. His research interest contains the management of large-scale complex data and autonomous RDBMS.
\end{IEEEbiography}
\vspace{-12ex}

\begin{IEEEbiography}[{\includegraphics[width=1.0in, height=1.25in,clip,keepaspectratio]{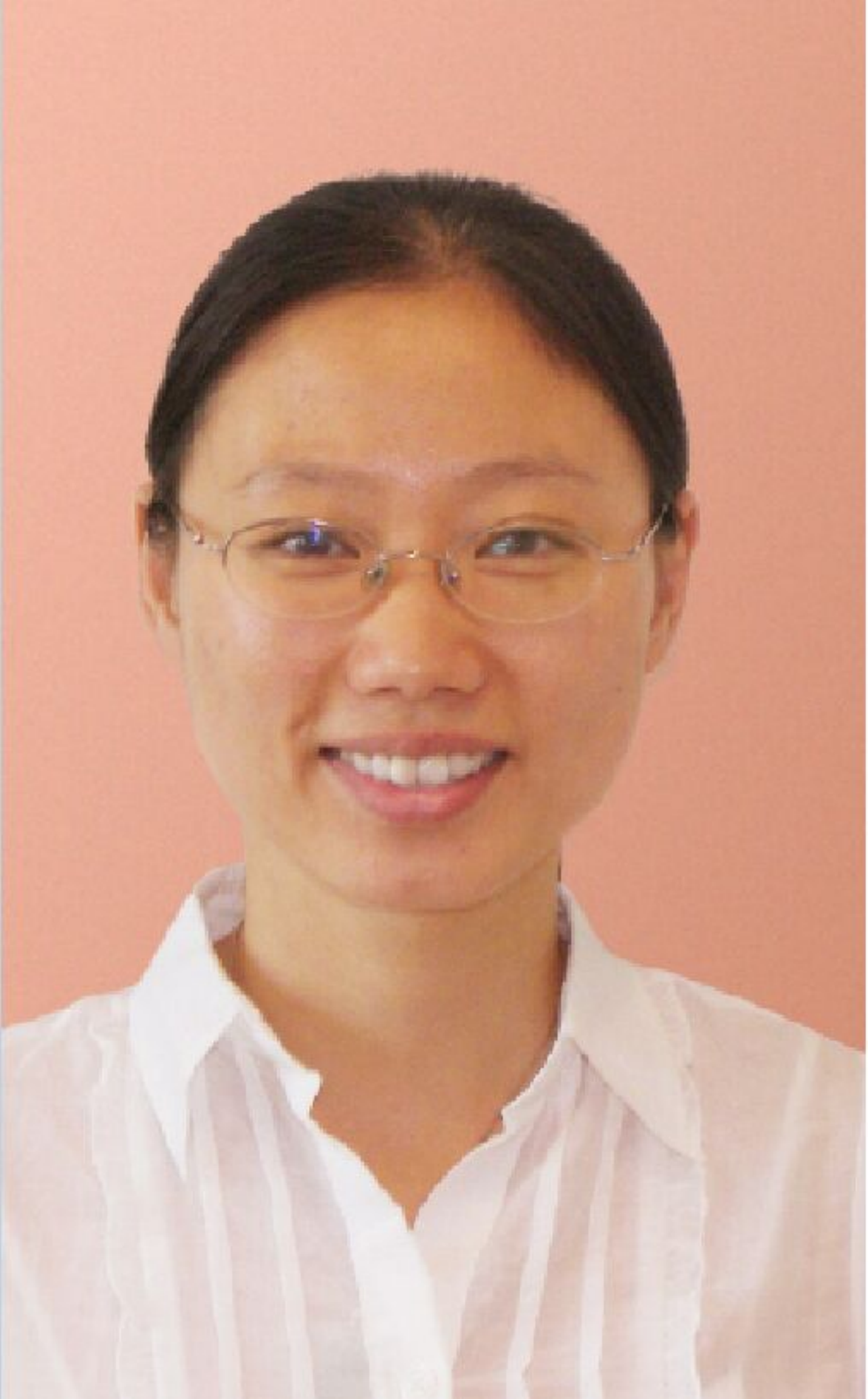}}]{Hong Cheng} is an Associate Professor in the Department of Systems Engineering and Engineering Management at the Chinese University of Hong Kong.  She received her Ph.D. degree from University of Illinois at Urbana-Champaign in 2008.  Her research interests include data mining, database systems, and machine learning.  She received research paper awards at ICDE'07, SIGKDD'06 and SIGKDD'05, and the certificate of recognition for the 2009 SIGKDD Doctoral Dissertation Award.  She is a recipient of the 2010 Vice-Chancellor's Exemplary Teaching Award at the Chinese University of Hong Kong.
\end{IEEEbiography}
\vspace{-12ex}

\begin{IEEEbiography}[{\includegraphics[width=1.0in, height=1.25in,clip,keepaspectratio]{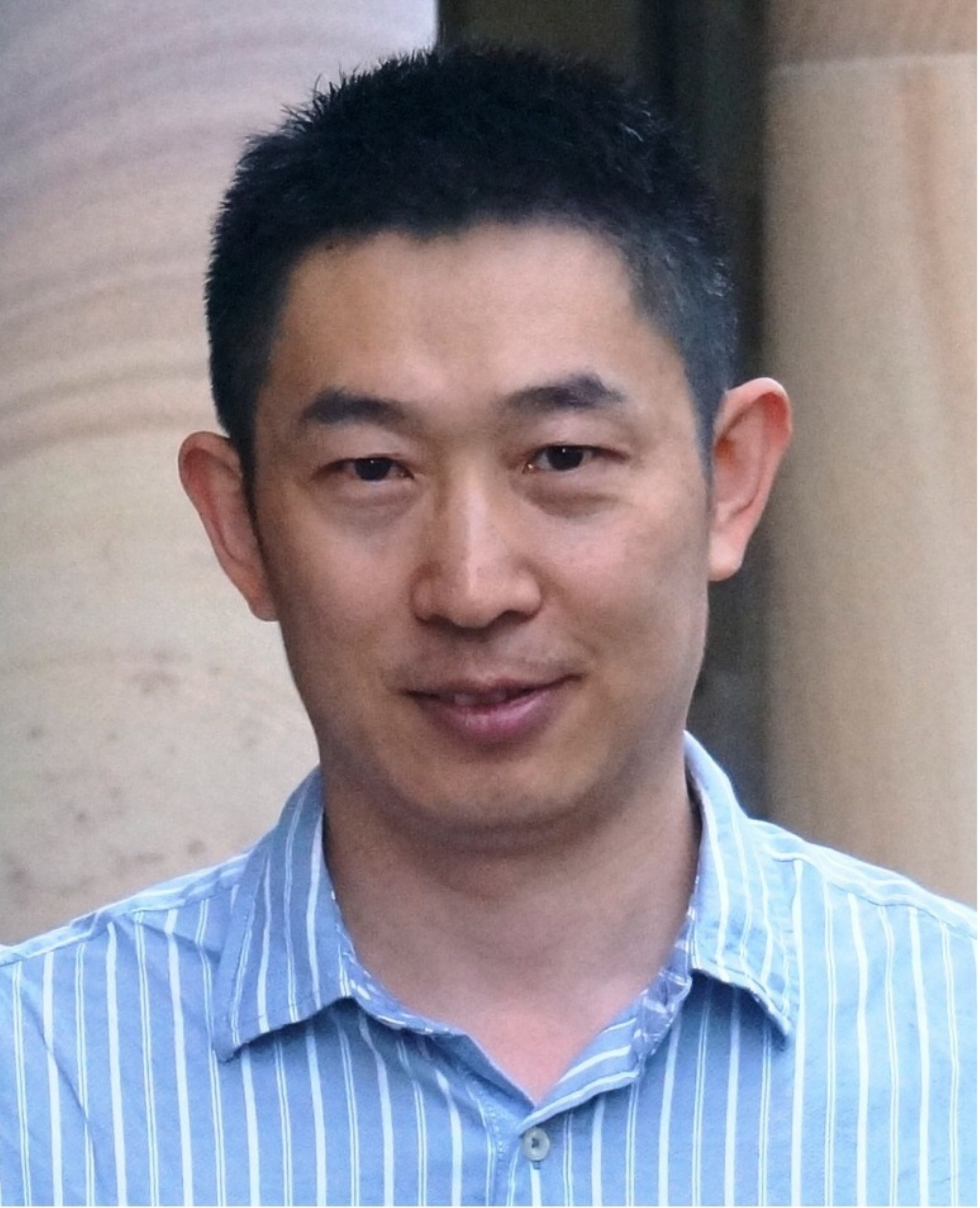}}]{Jiangtao Cui} received the MS and PhD degrees both in computer science from Xidian University, China, in 2001 and 2005, respectively. Between 2007 and 2008, he was with the Data and Knowledge Engineering group working on high-dimensional indexing for large scale image retrieval, in the University of Queensland, Australia. He is currently a professor in the School of Computer Science and Technology, Xidian University, China. His current research interests include data and knowledge engineering, data security, and high-dimensional indexing.
\end{IEEEbiography}

\appendices

\section{The cost model of \texttt{SW KNNG}}
\label{sec:append_cost_sw}

\begin{figure}[!htp]
\begin{center}
\includegraphics[width=0.22\textwidth]{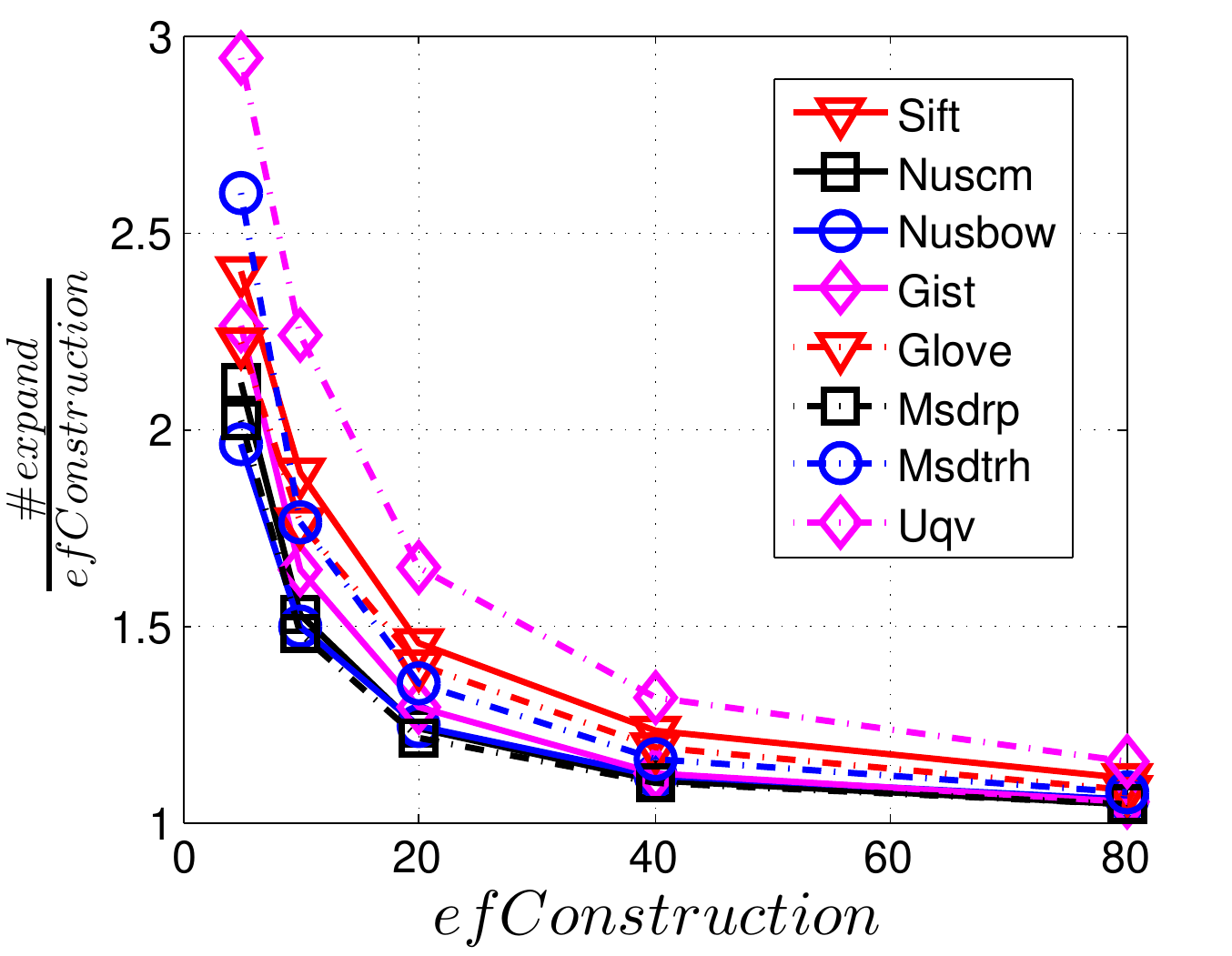}
\end{center} \vspace{-3ex}
\caption{$\#expand$ vs $efConstruction$ in \texttt{SW KNNG}.}
\label{fig:sw_efc_ratio}
\end{figure}

The main cost of \texttt{SW KNNG} is to conduct ANN search for each $u \in D$. For each query $u$, the search iteratively expands $u$'s close nodes until the termination condition is met. Hence we care about the number $\#expand$ of expanded nodes for each query. To investigate it, we conduct experiments on real data. We show the results in Figure~\ref{fig:sw_efc_ratio}. We can see that as $efConstruction$ increases, the ratio $\frac{\#expand}{efConstruction}$ decreases and becomes graudally close to 1. On average, we conclude that each query expands $O(efConstruction)$ nodes. Unfortunately, nodes in an SW graph have various number of neighbors. Hence, we cannot figure out the number of candidates accessed precisely. Let $M_{sw}^{max}$ the maximum neighborhood size in the SW graph. In the worst case, the time complexity of \texttt{SW KNNG} is $O(n * d * efConstruction * M_{sw}^{max})$.  

\section{The cost model of \texttt{HNSW KNNG}}
\label{sec:append_cost_hnsw}

\begin{figure}[!htp]
\begin{center}
\includegraphics[width=0.22\textwidth]{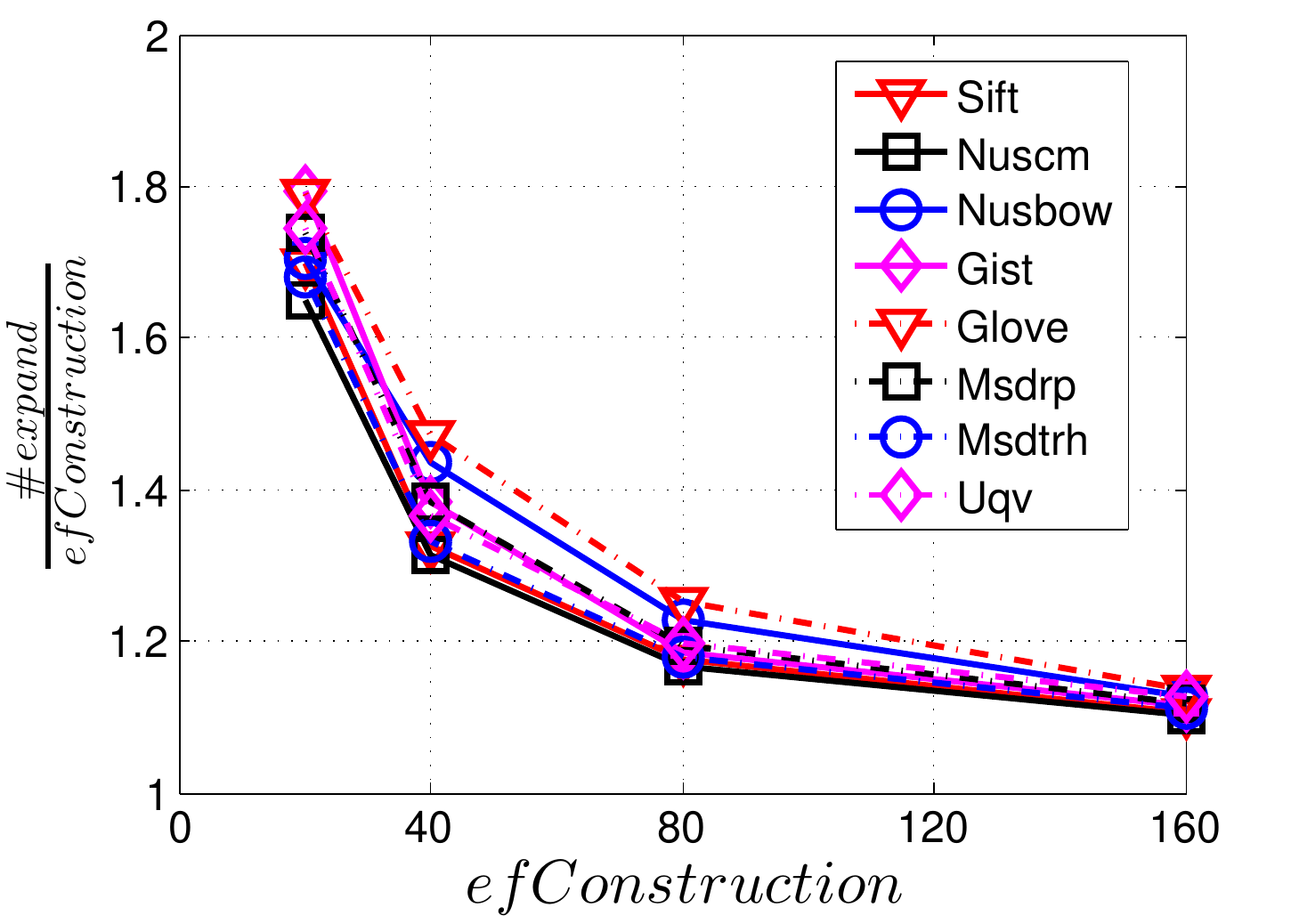}
\end{center}\vspace{-3ex}
\caption{$\#expand$ vs $efConstruction$ in \texttt{HNSW KNNG}.}
\label{fig:hnsw_efc_ratio}
\end{figure}

The main cost of \texttt{HNSW KNNG} contains two parts, i.e. ANN search for each node and pruning oversize nodes. Like \texttt{SW KNNG}, we find that the number $\#expand$ of expanded nodes for each query is pretty close to $efConstruction$. We show the relationship between the ratio $\frac{\#expand}{efConstruction}$ and $efConstruction$ in Figure~\ref{fig:hnsw_efc_ratio}. We can see that the ratio approaches 1 as $efConstruction$ increases. Hence, we can conclude that each query expands $O(efConstruction)$ nodes on average.  Unlike an SW graph, each node in an HNSW graph has at most $M_{hnsw}$ neighbors. Moreover, after each search, HNSW selects $M_{hnsw}$ neighbors from $efConstruction$ ones, following the link diversification strategy like the pruning operations. Each such selection costs $O(efConstruction * M_{hnsw} * d)$. In total, the part of ANN search for $n$ nodes costs $O(n * efConstruction * M_{hnsw} * d)$. 

\begin{figure}[!htp]
\begin{center}
\includegraphics[width=0.22\textwidth]{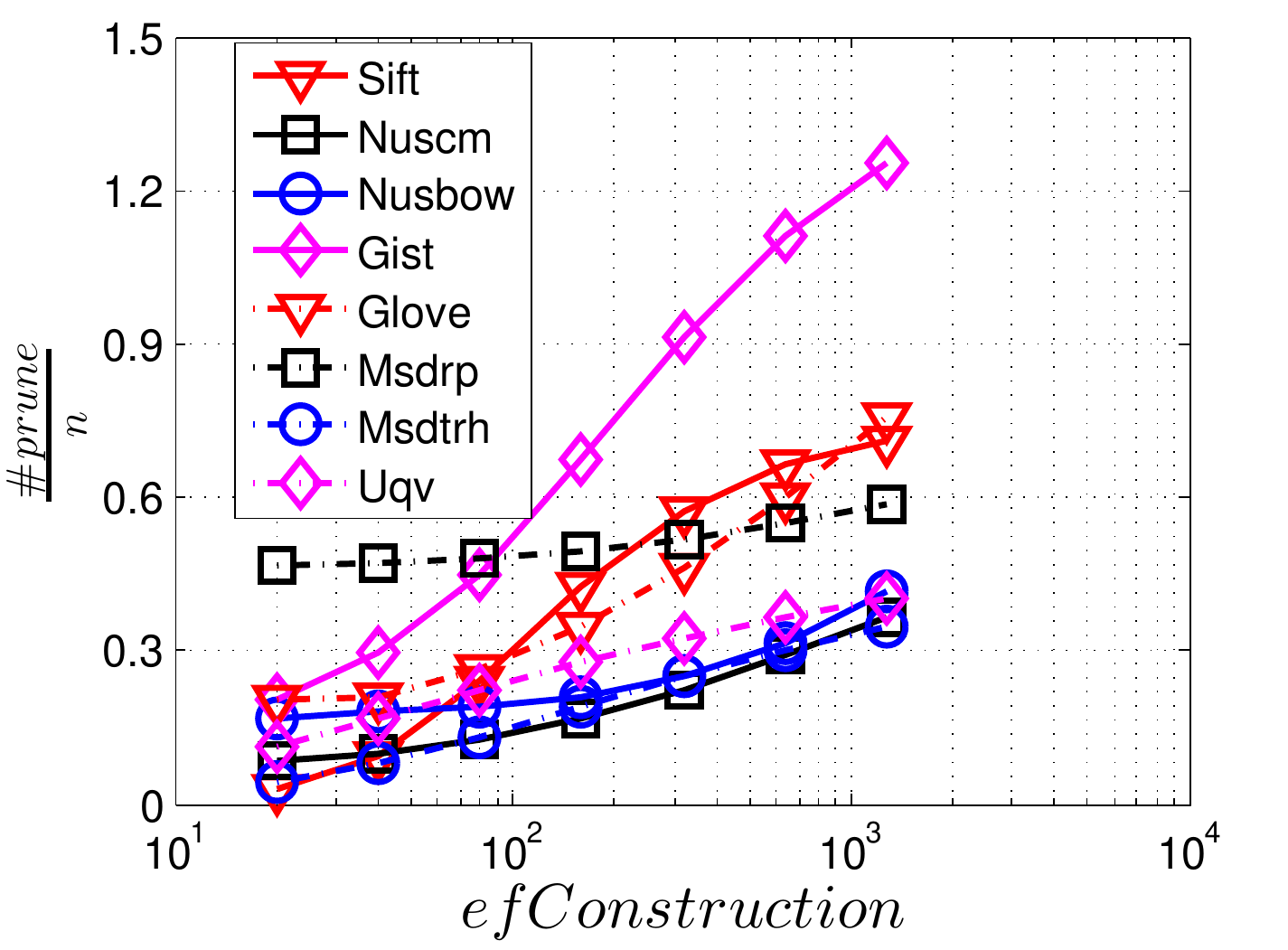}
\end{center}\vspace{-3ex}
\caption{The effect of $efConstruction$ on $\#prune$.}
\label{fig:hnsw_efc_prune}
\end{figure}

Like $\#expand$, we investigate the number $\#prune$ of pruning operations by experiments. We use the heuristic 2 (i.e., the default setting) in nmslib library as the link diversificaiton strategy, which generates fewer pruning operations. We show the experimental results in Figure~\ref{fig:hnsw_efc_prune}. We can see that $\#prune$ increases obviously as $efConstruction$ rises. However, we set $efConstruction$ as a small enough value (e.g., 80) in this paper. This is because we do not expect a pretty accurate KNNG for an \texttt{INIT} method. Hence, the practical $\#prune$ will be sufficiently small. 

\begin{figure}[!htp]
\begin{center}
\includegraphics[width=0.22\textwidth]{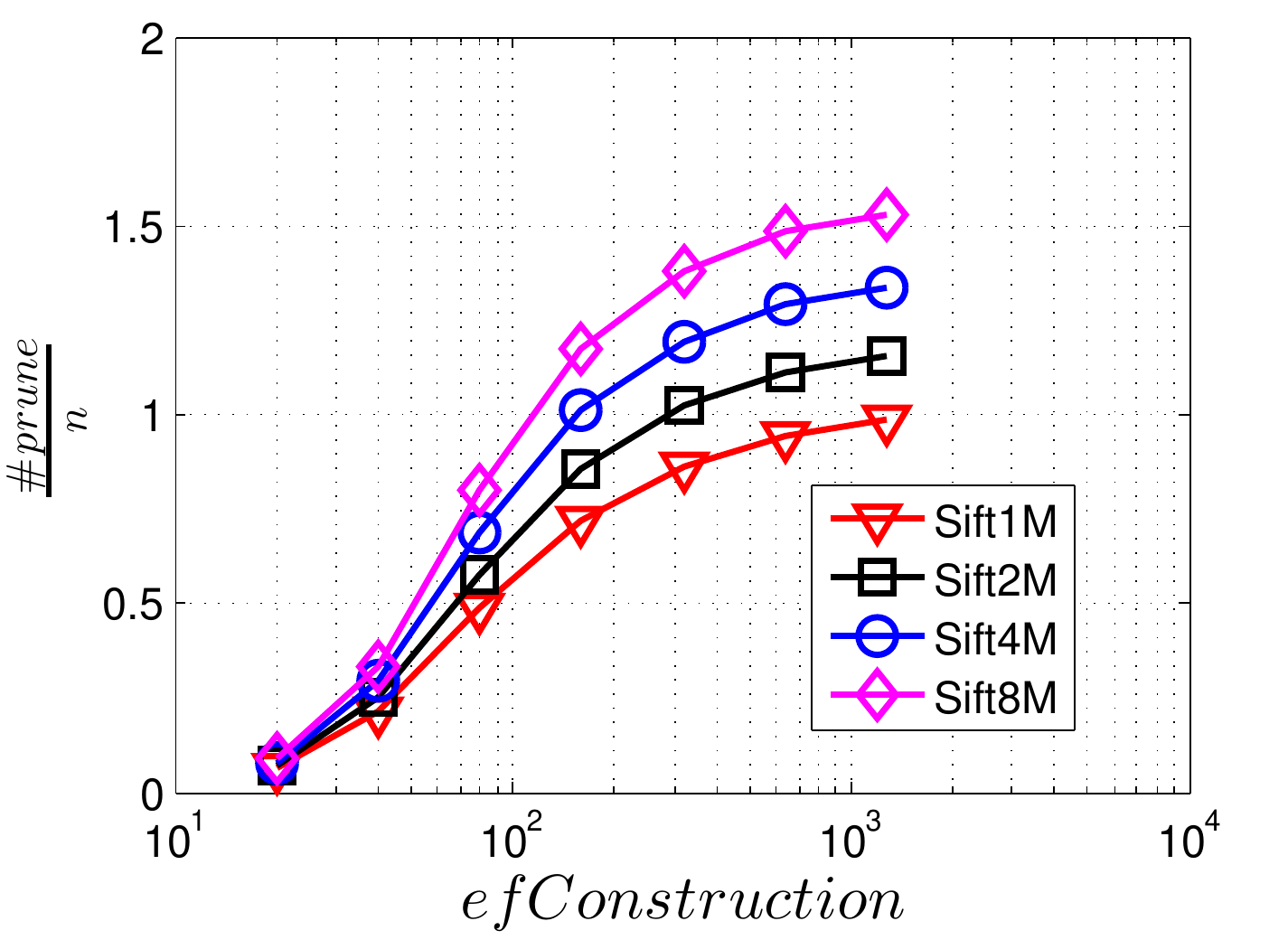}
\end{center}\vspace{-3ex}
\caption{The effect of data size on $\#prune$.} 
\label{fig:hnsw_n_prune}
\end{figure}

In addition, the data hubness (defined in Section~\ref{ssec:explo_def}) also has a positive influence on $\#prune$.  Let us take Sift and Gist as an example. With the same $efConstruction$ value, Gist has an obvious $\#prune$ than Sift. Similar phenomena could be found when comparing the pair of Nuscm and Nusbow and the pair of Msdrp and Msdtrh respectively. Moreover, the data size also positively affect $\#prune$. In Figure~\ref{fig:hnsw_n_prune}, we show the results on four subsets with sizes 1 million, 2 million, 4 million and 8 million of Sift100M, denoted as Sift1M, Sift2M, Sift4M and Sift8M respectively. The details of Sift100M could be found in Section~\ref{sssec:large}.

Even affected by a few factors, $\#prune$ is pretty close to $O(n)$ in practice, especially considering a small $efConstruction$ value. Hence, the practical values of $\#prune$ are far smaller than the worst case $n*M_{hnsw}$, where $M_{hnsw} = 20$ in our experiments. 

\section{The cost model of \texttt{DeepSearch}}
\label{sec:append_cost_deep}

\begin{figure}[!htp]
\begin{center}
\subfigure[\texttt{DeepMdiv}]
{
\includegraphics[width=0.22\textwidth]{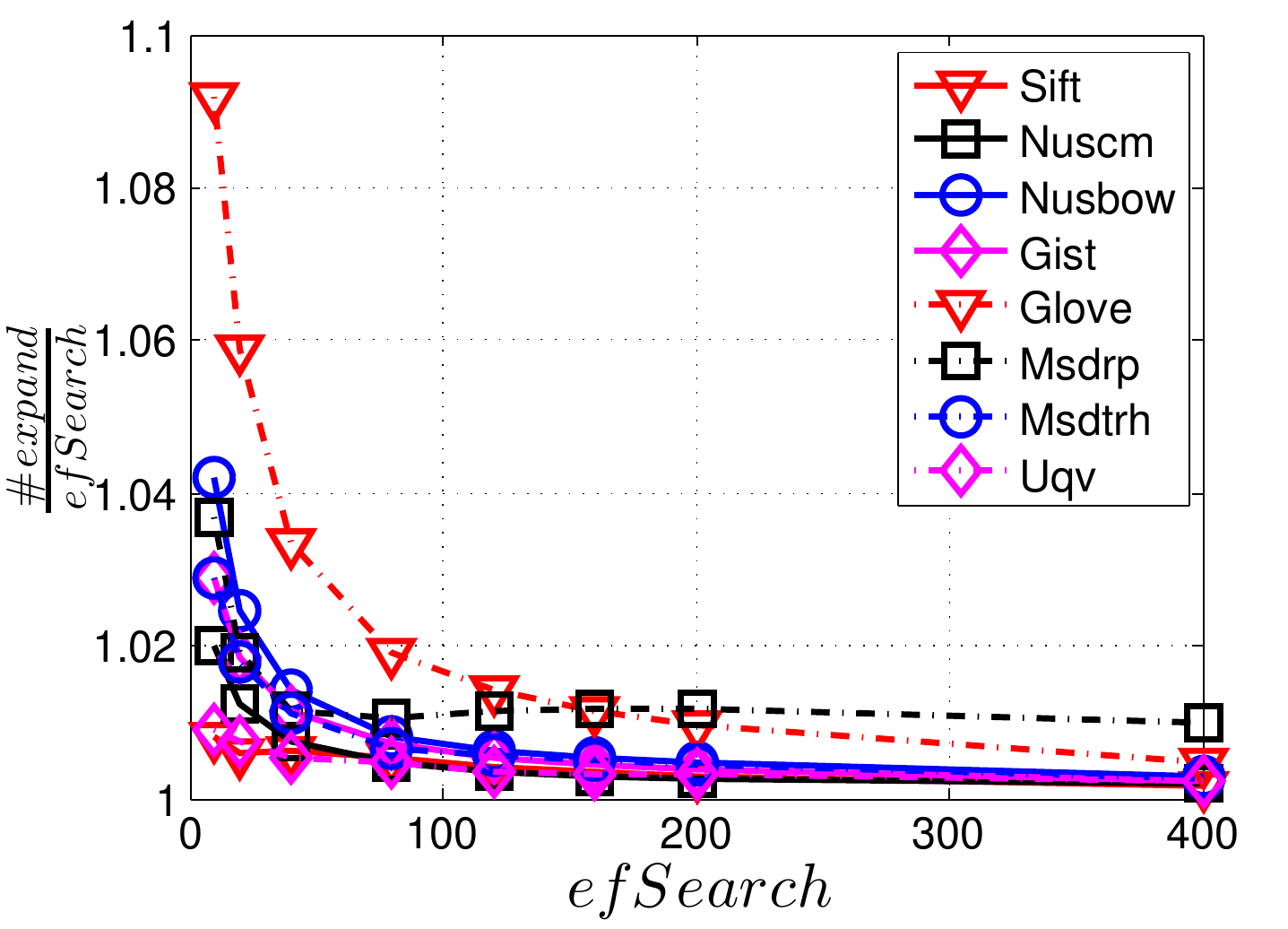}
\label{fig:mdiv_efs_ratio}
}
\subfigure[\texttt{Deep HNSW}]
{
\includegraphics[width=0.22\textwidth]{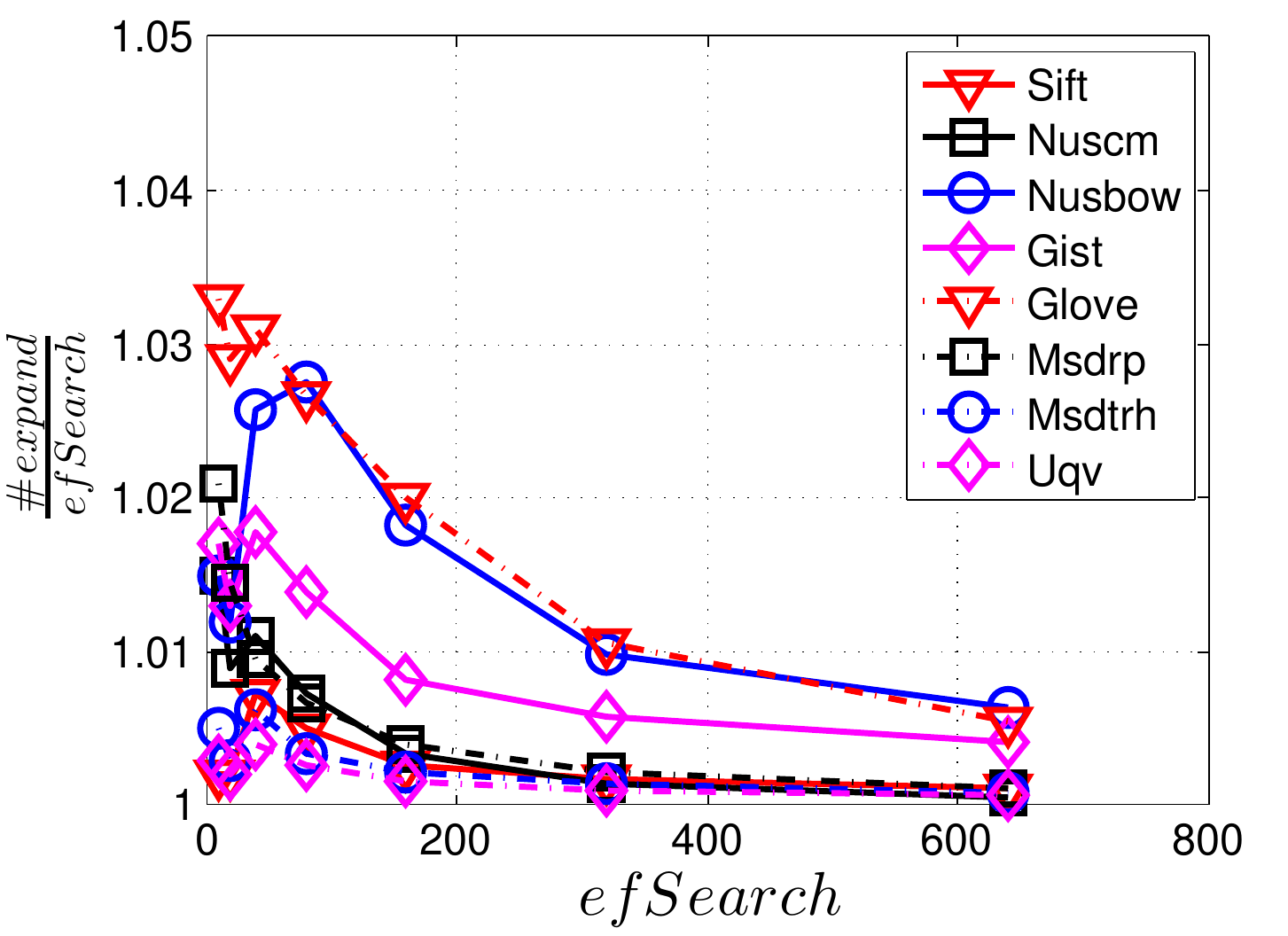}
\label{fig:hnsw_efs_ratio}
}
\end{center}\vspace{-3ex}
\caption{$\#expand$ vs $efSearch$ in \texttt{DeepSearch}.}
\label{fig:deep_efs}
\end{figure}

\texttt{DeepSearch} conducts ANN search on a online-built proximity graph $H$ for each query $u \in D$. The search process is controlled by the parameter $efSearch$. The main cost of \texttt{DeepSearch} is to expand similar nodes to each query in $H$. Hence, the number $\#expand$ is key to the cost model of \texttt{DeepSearch}. Here, we investigate two variants of \texttt{DeepSearch}, i.e. \texttt{DeepMdiv} and \texttt{Deep HNSW}. We show relationship between $\frac{\#expand}{efSearch}$ and $efSearch$ in Figure~\ref{fig:deep_efs}. 

We can see that the ratio gradually approaches 1 as $efSearch$ increases. We can say that \texttt{DeepMdiv} and \texttt{Deep HNSW} expand $O(efSearch)$ nodes for each query. \texttt{DeepMdiv} takes the initial KNNG as the proximity graph, where each node has exactly $k$ neighbors. As a result, the cost of \texttt{DeepMidv} is $O(n * d * efSearch * k)$. \texttt{Deep HNSW} uses the HNSW graph as the proximity graph, where each node has up to $M_{hnsw}$ neighbors. Hence, the cost of \texttt{Deep HNSW} is $O(n * d * efSearch * M_{hnsw})$.

\end{document}